\documentclass[revtex4]{emulateapj}
\usepackage{subfigure}
\usepackage{natbib}
\usepackage{epsfig}
\usepackage{graphicx}
\usepackage{multirow}
\usepackage{epstopdf}
\usepackage{adjustbox}
\usepackage{hyperref}

\slugcomment{Accepted for Publication in AJ}

\shorttitle{Bolometric Flux Estimation for Cool Evolved Stars}

\shortauthors{van Belle, Creech-Eakman \& Ruiz-Velasco}

\begin{document}

\title{Bolometric Flux Estimation for Cool Evolved Stars}

\author{Gerard T. van Belle\altaffilmark{1}, Michelle Creech-Eakman\altaffilmark{2}, Alma Ruiz-Velasco\altaffilmark{1}}

\altaffiltext{1}{Lowell Observatory, 1400 W. Mars Hill Rd., Flagstaff, Arizona, USA; gerard@lowell.edu, alma@lowell.edu}
\altaffiltext{2}{New Mexico Institute of Mining \& Technology, 801 Leroy Pl., Socorro, NM 87801; mce@kestrel.nmt.edu}
%----------------------------------------------------------------------

\begin{abstract}
Estimation of bolometric fluxes $(F_{\rm BOL})$ is an essential component of stellar effective temperature determination with optical and near-infrared interferometry.  Reliable estimation of $F_{\rm BOL}$ simply from broad-band $K$-band photometry data is a useful tool in those cases were contemporaneous and/or wide-range photometry is unavailable for a detailed spectral energy distribution (SED) fit, as was demonstrated in \citet{Dyck1974ApJ...189...89D}.  Recalibrating the intrinsic $F_{\rm BOL}$ versus observed $F_{{\rm 2.2} \mu {\rm m}}$ relationship of that study with modern SED fitting routines, which incorporate the significantly non-blackbody, empirical spectral templates of the INGS spectral library \citep[an update of the library in][]{Pickles1998PASP..110..863P} and estimation of reddening, serves to greatly improve the accuracy and observational utility of this relationship.  We find that $F_{\rm BOL}$ values predicted are roughly 11\% less than the corresponding values predicted in \citet{Dyck1974ApJ...189...89D}, indicating the effects of SED absorption features across bolometric flux curves.
\end{abstract}

%----------------------------------------------------------------------
%
% see http://www.journals.uchicago.edu/page/apj/instruct.key.html


%----------------------------------------------------------------------

Draft: \today

\section{Introduction}\label{sec_introduction}

A substantial number of angular sizes of cool, evolved stars have been measured over the past two decades with increasingly sophisticated optical interferometry facilities operating in both the near-infrared (taken here to be 1.0-2.5 $\mu{\rm m}$) and the optical (0.3-1.0 $\mu{\rm m}$).  However, without ancillary measurements, these apparent sizes $(\theta)$ are of limited utility.  When combined with distance $(d)$, a linear size may be obtained; when combined with a measurement of bolometric flux $(F_{\rm BOL})$, the effective temperature $(T_{\rm EFF})$ may be measured.  The latter comes from the definition of luminosity $(L)$, relating it to stellar radius $R$ and $T_{\rm EFF}$ (and the usual mathematical / physical constants):
\begin{equation}
L = 4 \pi \sigma R^2 T_{\rm EFF}^4
\end{equation}
Dividing both sides by $d$ reduces this to the relationship:
\begin{equation}\label{eqn_teff}
T_{\rm EFF} = 2341 \left( {F_{\rm BOL} \over \theta_R} \right)^{1/4}
\end{equation}
where the Rosseland angular size $\theta_R$ is in units of milliarcseconds (mas), and $F_{\rm BOL}$ is given as $10^{-8}$~ergs~cm$^{-2}$~sec$^{-1}$; the resultant $T_{\rm EFF}$ is in Kelvin (K).  The Rosseland mean angular size is the wavelength-averaged angular size incorporating the wavelength-weighted Rosseland mean opacity -- effectively, the mean radiating surface of the star \citep[definitional discussions can be found in][]{Seaton1994MNRAS.266..805S,Seaton2004MNRAS.354..457S}.  The value for $F_{\rm BOL}$ is to be taken as the flux arriving at Earth from the observed star, assuming no intervening extinction between it and the observer.

Unfortunately, many interferometric size determinations are not done in direct conjunction with matching measures of $d$ or $F_{\rm BOL}$.  In the case of distance, this (relatively) invariant parameter can be measured long after (or before) the corresponding measure of $\theta$ takes place, with no significant effect upon the ultimately determined linear size, $R$.

For bolometric flux, the situation is more challenging.  Complete characterizations of the total flux output from the shortest to longest wavelengths are rarely done for most objects, and even simple photometry is rarely carried out at the time of the interferometric observations.  Many of the coolest objects that are of particular interest to interferometry vary in brightness substantially on time periods short compared to the lag between flux and size measurements.  For the bandpasses where they are brightest -- in the near-infrared -- instrumentation tends to be available only on larger telescopes, meaning these objects saturate detectors and fall squarely in a surprising `blind spot' of modern astronomy.  To complicate matters further, many of these objects are distant, which means a correction for interstellar reddening must be taken into account as well.  An important piece of the solution is the fact that interferometers typically collect measurements of the incoherent flux levels\footnote{The light at the end of the tunnel, as it were.} at their wavelengths of operations -- typically $K$-band  -- in addition to measures of interference during their observations.

Historically, effective use of these data was directly affected by the advent of useful infrared detectors in the 1960's, which resulted in the seminal Two Micron Sky Survey \citep{Neugebauer1969tmss.book.....N} and the dissertation work of \citet{Wing1967PhDT.........5W}.  The ensuing work by \citet[][hereafter DLC74]{Dyck1974ApJ...189...89D} demonstrated relationships between band-to-bolometric flux ratios and spectral type, at 0.55, 1.04, and $2.2 \mu$m (see Figure 6 of DLC74 in particular), for `relatively normal' Mira variables and giants.  Spectral types in the DLC74 study were determined contemporaneously to account for this parameter appearing to change with variable star phase.  For the $F_{1.04 \mu{\rm m}} / F_{\rm BOL}$ ratio, the relationship was essentially flat for spectral types M0 through $\sim$M6; for $F_{2.2 \mu{\rm m}} / F_{\rm BOL}$, the ratio changes slowly and in a linear fashion from M0 all the way out to M10.

However, with a number of recent advances, the specific results of DLC74 bear a revisit and recalibration.  The spectral template library of \citet{Pickles1998PASP..110..863P} allows for a much more fine-grained fitting of spectral energy distributions to the the photometry present in DLC74 - e.g. see Figures 1-3 in that paper, which effectively does a simple Riemann sum as a function of wavelength to determine their values for $F_{\rm BOL}$.  Such a sum approximates a blackbody curve, and does not account for absorption features present between photometric bands -- which, particularly for the later-type stars, can be considerable (non-)contributions of the stellar spectral energy distribution (SED) curve.  Current sophisticated SED fitting codes can make use of the unreddened Pickles templates, constrain them by the photometry in DLC74 -- which remain high-quality measurements -- and also incorporate reddening corrections.

%-------------------

\section{Photometry and Data Reduction}\label{sec_observations}

\subsection{Targets and Photometric Data}\label{sec_21_targetsanddata}

The target data come from 84 observations of 60 objects found in DLC74, all of which are cool evolved stars at spectral types M4.0III and later.  Two objects have measurements at 4 separate epochs, three objects have 3 epochs, and 11 have 2 epochs.  These 60 objects include contemporaneous flux measurements in $\log F_\lambda$ (in W cm$^{-2}$ $\mu$m$^{-1}$) across up to 12 bands from 0.55$\mu$m to 10.2$\mu$m, along with epoch-specific spectral type determinations for 70 of the 84 observations, and 34 $F_{\rm BOL}$ determinations.  Broad-band filters representing the $V, J, H, K, L, M,$ and $N$ passbands were used, along with narrowband filters at 0.78, 0.87, 0.88, 1.04 and 1.05~$\mu$m with filter-specific properties listed in Table \ref{table-LWphotometric}.  Further details on the narrowband system can be found in DLC94, along with \citet{Lockwood1971ApJ...169...63L,Lockwood1972ApJS...24..375L,Lockwood1973ApJ...180..845L,Wing1973ApJ...184..873W}.

\begin{table}[h]
\caption{Properties of the \citet{Lockwood1971ApJ...169...63L} photometric system.}\label{table-LWphotometric}
\begin{tabular}{llll}
\hline
Designation & \begin{tabular}[c]{@{}l@{}}Central\\ wavelength\\ (\AA)\end{tabular} & \begin{tabular}[c]{@{}l@{}}Width at\\  half power\\ (\AA)\end{tabular} & \begin{tabular}[c]{@{}l@{}}Feature\\ Measured\end{tabular} \\ \hline
78          & 7817                                                               & 90                                                                  & TiO                                                        \\
87          & 8778                                                               & 82                                                                  & Continuum                                                       \\
&&& (TiO after M5)  \\
88          & 8884                                                               & 114                                                                 & TiO                                                        \\
104         & 10351                                                              & 125                                                                 & Continuum                                                    \\
105         & 10506                                                              & 100                                                                 & VO                                                       \\ \hline
\end{tabular}
\end{table}

For the purposes of this study, we have taken the data in DLC74 and converted it into Janskys (Jy), which can be found in Table \ref{table-DLC74-Jy}.
Spectral types listed in this table for each of the stars were determined contemporaneously by DLC74 using the TiO and VO band strengths in the narrowband photometry, using the $T_1, T_2, V_1, D$ color indices calibration and technique found in \citet{Lockwood1972ApJS...24..375L}.  The color indices found in \citet{Lockwood1972ApJS...24..375L} were used to type stars in integer subtypes between M1 and M3, half-integer subtypes between M4.0 and M8.0, and quarter-integer subtypes between M8.0 and M10.

DLC74 cite their photometry errors as ``2-3 percent'' for their narrowband data, ``5 percent or less'' for 1-4$\mu$m, and ``10 percent or slightly greater'' for their 5 and 10.2 $\mu$m data points; no error is given for $V$-band data, except to note that their stars are ``extremely faint'' (and imply that larger errors result).  From the numerical modeling presented in the next section, this appears to correspond to a median measurement values of $\lesssim 10^{-9}$ W cm$^{-2}$ $\mu$m$^{-1}$. In practice, we found that inflating these errors by a factor of $2\times$ was necessary to achieve average reduced $\chi^2$ values approaching 1.0 in the next section.  We additionally found that for the latest type stars ($>$M7.5III), data from the 3 shortest wavelength filters was also ``extremely faint'' (also $\lesssim 10^{-9}$ W cm$^{-2}$ $\mu$m$^{-1}$) and that 2-3\% errors for these data were underestimates.

\subsection{Spectral Energy Distribution Fitting with {\tt sedFit}}\label{sec-sedFitting}

For SED fitting, we used the {\tt sedFit} code that we have employed on previous occasions \citep[e.g.][]{vanbelle2007ApJ...657.1058V,vanBelle2008ApJS..176..276V,vanBelle2009ApJ...694.1085V} for determination of bolometric flux and other relevant stellar parameters.  The flux data from DLC74 were matched against a grid of M-giant template spectra; for comparison, the epoch-specific spectral types determined in DLC74 were also specifically evaluated.

The template spectra were based upon the INGS (\underline{I}UE / \underline{NG}SL / \underline{S}peX) library of Pickles\footnote{http://lcogt.net/user/apickles/dev/INGS/}, an update of the earlier catalog of \citet{Pickles1998PASP..110..863P}.  The INGS combines IUE data \citep{Heck1984ESASP1052.....H}, NGSL data \citep{Heap2011ASPC..448..887H}, and SpeX data \citep{Rayner2009ApJS..185..289R} to produce a compendium of 143 stellar spectra, and significantly improves the late-type spectra relative to \citet{Pickles1998PASP..110..863P}.  This is primarily due to the later spectra being now observationally based, in contrast to \citet{Pickles1998PASP..110..863P} which used the theoretical spectra from \citet{Fluks1994A&AS..105..311F} for the late M giant spectra.  A search grid was developed from the 11 subtype templates found in the INGS between M0III and M10III, with half-subtypes generated by interpolating between those 11 templates.
This grid of these spectra were searched by chi-squared minimization by fitting, for each template, the bolometric flux level $F_{\rm BOL}$ and wavelength-dependent reddening.  For all but 3 of the 84 flux data sets, a slightly different spectral subtype from the grid search fit better than the indicated DLC74 spectral type; typically this was roughly one subtype bluer than DLC74.

In combination with an effective temperature estimate that assigned to each spectral type \citep[based upon the values in -- interpolated or extraopolated where necessary -- ][]{vanBelle1999AJ....117..521V}, an estimate of the angular sizes of these objects was produced as a potentially useful secondary data product.  Reddening corrections were based upon the empirical reddening determination by \citet{Cardelli1989ApJ...345..245C}, which differs little from van de Hulst's theoretical reddening curve No. 15 \citep{Johnson1968nim..book..167J,Dyck1996AJ....111.1705D}.  Zero-magnitude flux density calibrations for broadband photometry data points were based upon the values given in \citet{Fukugita1995PASP..107..945F} and \citet{cox00}.  In contrast to similar earlier SED efforts of ours, no literature photometry \citep[such as Stromgren $ubvy\beta$ or 2MASS data in][]{ruf76, Cutri2003tmc..book.....C} was utilized herein, given the non-contemporaneous nature of those data sets.

The $V$-band data were not used in the SED fitting for a number of reasons.  Principally, the more variable stars (e.g. the Miras) have variations in $T_{\rm EFF}$ and $R$, and dust production which causes up to 8 magnitudes of variation in $V$, rendering data in this bandpass useless for SED fitting.  Additionally, the weighting on the data points was unclear given the non-numerical quantification of the signal-to-noise in DLC74.  As such, the $V$ data had a disproportionate effect in determination of reddening, even when given low statistical weight.  Determination of reddening was still well-characterized for these stars given the inclusion of the shorter-wavelength narrow-band filters (LW $78$, $87$, $88$).  For determination of the intrinsic bolometric flux ($F_{\rm BOL}$) below, it is important to be as unbiased as possible in characterizing reddening.  Contributions longwards ($\lambda > 2.5 \mu$m) of the data presented in the template spectra were estimated via a Rayleigh-Jeans tail and typically were a minor ($<10$\%) contribution.

The results from the SED fitting can be seen in Table~\ref{table-FBOLresults}, including estimates for reddening, angular size, and bolometric flux.  The full set of summary {\tt sedFit} output plots can be seen in the Appendix in Figures~A1-A14.
%for objects selected from either end of the spectral types in question.\footnote{For objects of specific interest to the reader, detailed {\tt sedFit} output plot packages are available upon request.}
Six objects had abnormally large $\chi^2$ per degree-of-freedom values ($\chi^2_\nu > 6$) and were not considered in the ensuing analyses.  Unsurprisingly, four of these six (IRC-10236, IRC+40485, IRC+50096, IRC+50357) are those objects noted in DLC74 as carbon stars;
the remaining two (IRC+40004, IRC+40448) are M-type in DLC74 but had sufficiently spurious fits that they were set aside.
Interestingly, all objects designed in DLC74 as S-type (R And, R Cam, R Cyg) had SED fits with reasonable $\chi^2_\nu$ values indistinguishable from the M-type stars.

\begin{figure}
\centering
    \includegraphics[width=0.44\textwidth,angle=0]{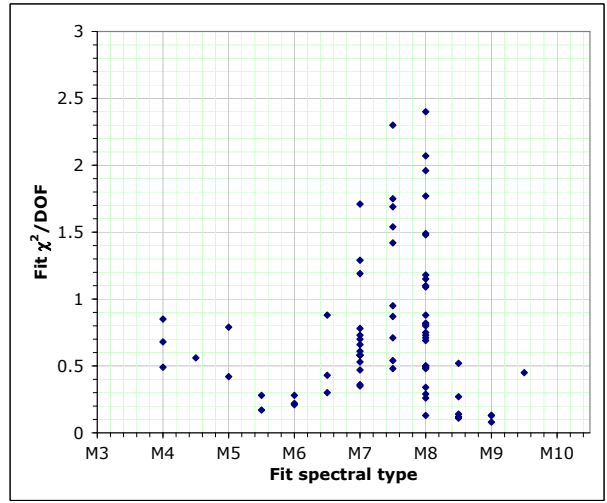}
    \caption{The reduced $\chi^2$ ($\chi^2_\nu$) of each SED fit versus spectral type.}\label{fig-chi2vsST}
\end{figure}

\begin{figure}
\centering
    \includegraphics[width=0.44\textwidth,angle=0]{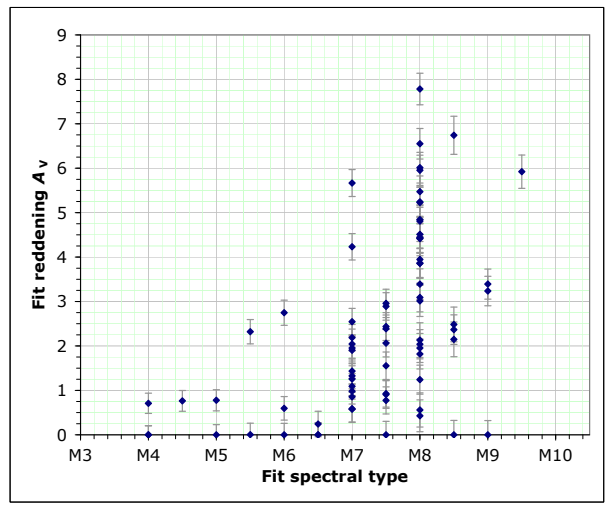}
    \caption{The reddening $A_{\rm V}$ of each SED fit versus spectral type.}\label{fig-AVvsST}
\end{figure}

\begin{figure}
\centering
    \includegraphics[width=0.44\textwidth,angle=0]{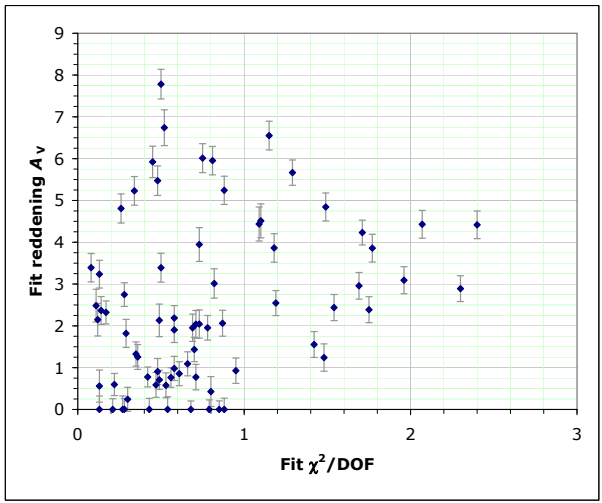}
    \caption{The reddening $A_{\rm V}$ of each SED fit versus reduced $\chi^2$ ($\chi^2_\nu$).}\label{fig-AVvschi2}
\end{figure}

For the remaining objects the $\chi^2_\nu$ values for the SED fits have a median value of $\simeq$0.64, with a maximum of 2.4.\footnote{An earlier investigation utilizing the \citet{Pickles1998PASP..110..863P} templates in like manner to the INGS templates had median values of $\chi^2_\nu \simeq$4, with a significant increasing trend in $\chi^2_\nu$ versus spectral subtype; it is on this basis that we find that the INGS templates are significantly better for the later subtypes.}  All $\chi^2_\nu$ values are based upon employing the error bars discussed at the end of \S \ref{sec_21_targetsanddata}.  There is a correlation between these $\chi^2_\nu$ values, seen in Figure~\ref{fig-chi2vsST}, with a significant increase in $\chi^2_\nu$ for the redder stars, beginning at M7.5III.  Similarly, we see an increase in the range of reddening fit values in Figure~\ref{fig-AVvsST} for the later spectral types, also beginning at M7.5III, and possibly one-half subtype earlier.  However, considering $A_{\rm V}$ versus $\chi^2_\nu$ (Figure~\ref{fig-AVvschi2}), there is quite a bit of scatter between these parameters.

Taken together, Figures \ref{fig-chi2vsST}, \ref{fig-AVvsST}, and \ref{fig-AVvschi2} lead us to the following conclusions: first, the spectral type templates of the Pickles INGS are somewhat problematic for spectral types around M8III, for those bandpasses found in the narrowband filters of \citet{Lockwood1971ApJ...169...63L}.  Second, increasing values of $A_{\rm V}$ for the later spectral types are associated with increasing amounts of circumstellar, rather than interstellar, dust, given the known increase in mass loss rates with increasing spectral type \citep{Dupree1986ARA&A..24..377D,Guandalini2010A&A...513A...4G}.  Third, the slight trend in $A_{\rm V}$ versus $\chi^2_\nu$ towards increasing  $A_{\rm V}$ for $\chi^2_\nu>1$ indicates that poorness of fit -- potentially due to a lack of sufficient granularity in the INGS spectral template grid -- might result a increased value for $A_{\rm V}$.

%{\it Alternate spectral types.}  For completeness, we also explored fitting the photometry data for each star against a complete grid

{\it Objects with Anomalous Results.}  As noted above, a number of objects explicitly identified as carbon stars were presented in DLC74 (IRC-10236, IRC+40485, IRC+50357, IRC+50096), and fits using M10III templates were attempted, under the notion that these latest spectral types would be most analogous to the carbon stars.  However, as seen in the summary plots in the Appendix, this approach failed: SEDs for these object exhibit a distinctive, almost linear increase in flux with increasing wavelength.   Additionally, stars IRC+40004, and IRC+40448 were assigned M-type subtypes in DLC74 (M9III, and M7III, respectively) but exhibited photometry reminiscent of the carbon star phenomenon.
None of these objects that exhibited carbon behavior were included in Figures \ref{fig-chi2vsST}-\ref{fig-AVvschi2} or utilized in the discussion of \S \ref{sec_discussion}.
A cursory attempt was made to compare the data to model SED templates for the carbon stars \citep[e.g.][]{Aringer2009A_A...503..913A}, but this was not successful; fits to carbon star SEDs remain an interesting and involved challenge \citep[e.g. see extensive discussion in the recent work by][]{Rau2015arXiv150603978R}.

%The object R And was identified in DLC74 as `S-type'; however, a M10III spectral template was provided reasonable fits to both epochs of this star's data  ($\chi^2_\nu$ = 3.25, 2.09).  As such, it {\it is} included in Figures \ref{fig-chi2vsST}-\ref{fig-AVvschi2} and discussion.  The first epoch of R Tri data has a suspiciously S-type appearance, with $<$1$\mu$m excesses, but was kept given the goodness-of-fit in the ensuing 2 epochs.  U Ari's second epoch had a poor fit ($\chi^2_\nu$ = 11.46) but appears to eye to be a reasonable fit; given that and the earlier epoch's better fit ($\chi^2_\nu$ = 4.41), these data were also kept.

\section{Discussion}\label{sec_discussion}

\subsection{Bolometric Flux Relationships}

\begin{figure}
\centering
\includegraphics[width=0.44\textwidth,angle=0]{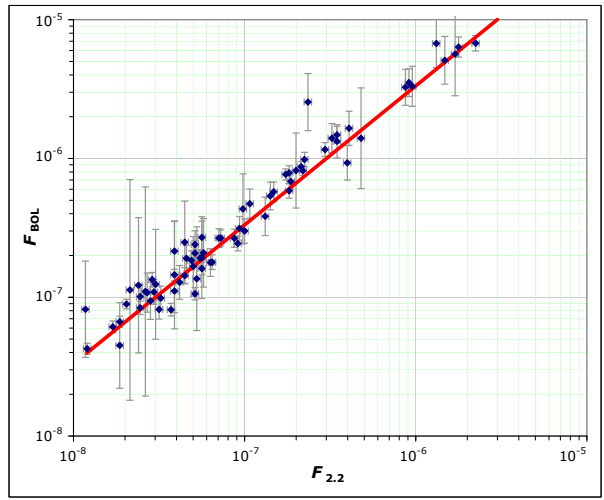}
\caption{Intrinsic bolometric flux ($F_{\rm BOL}$) versus observed 2.2 $\mu$m flux ($F_{2.2 \mu m}$), in ergs cm$^{-2}$ s$^{-1}$. The functional form of the red fit line is presented in Equation~\ref{eqn-fbol-f22}.}\label{fig-FBOLvsF2.2}
\end{figure}

As inspired by DLC74 (particularly Figure~6 in that article), we examined the intrinsic bolometric flux ($F_{\rm BOL}$) versus observed 2.2 $\mu$m flux ($F_{2.2 \mu m}$) in Figure~\ref{fig-FBOLvsF2.2}, and found the simple relationship fit:
\begin{equation}\label{eqn-fbol-f22}
F_{\rm BOL} = F_{2.2 \mu m} \times (3.340 \pm 0.101) - (0.04 \pm 0.33)\times 10^{-8}
\end{equation}
in erg cm$^{-2}$ s$^{-2}$, with a reduced $\chi^2$ of 1.66.
Direct comparison of this fit for $F_{\rm BOL}$ against the values from the SED fitting indicates a median intrinsic scatter of this relationship, across the range of bolometric fluxes in question, of 17\%.  It is important to emphasize that this is the relationship between {\it observed} 2.2 $\mu$m flux versus {\it intrinsic} bolometric flux -- the influence of reddening is considered in transforming from $F_{2.2 \mu m}$ to $F_{\rm BOL}$ using this relationship.  As such, this is a relationship of significant observational utility, but of limited value for understanding the underlying astrophysics.

We can leverage the estimate for $A_{\rm V}$ found in Table~\ref{table-FBOLresults} to ``correct'' the observed 2.2$\mu$m flux for the intrinsic 2.2$\mu$m flux $(F_{2.2corr, \mu m})$, which modifies the previous relationship to:
\begin{equation}
F_{\rm BOL} = F_{2.2corr, \mu m} \times (2.996 \pm 0.093) + (0.46 \pm 0.34)\times 10^{-8}
\end{equation}
in erg cm$^{-2}$ s$^{-2}$, with a reduced $\chi^2$ of 1.55; as expected, this decreases the slope of the linear fit -- the amount of intrinsic 2.2$\mu$m flux relative to the overall intrinsic bolometric flux is greater than the reddening-decreased apparent 2.2$\mu$m flux.

\subsection{Comparison to DLC74}\label{sec-DLC74comparison}

The intent of this investigation is to expand and update the intrinsic $F_{\rm BOL}$ versus $F_{{\rm 2.2} \mu {\rm m}}$ relationship presented in DLC74.  In particular, the use of specific Pickles INGS spectral templates, with the inclusion of significant absorption bands, along with an estimate of reddening, is expected to refine the values obtained for $F_{\rm BOL}$.  Inspection of how the fits compare to data in DLC74 is warranted before examining that derived quantity.  The most direct comparison is that of spectral type.  DLC74 determined their spectral types by referencing the Lockwood \& Wing narrowband fluxes against the color calibrations of \citet{Lockwood1972ApJS...24..375L}; spectral types in this investigation were fixed to those values.

%For completeness, we also explored fitting the photometry data for each star against a complete grid of spectral types between M3 and M10 from \citet{Pickles1998PASP..110..863P}, finding which template optimally fit the flux data when adjusting simultaneously for $F_{\rm BOL}$ and $A_{\rm V}$.  We found there was very good agreement of $\pm$0.5 subtypes between M4.0III and M7.0III, but then a systematic offset of roughly +1 subtypes for objects later than M7.5III.  Our inference from this result is that it was merely an observational bias, resulting from increasingly small flux values for the \citet{Lockwood1971ApJ...169...63L} filters at 0.78, 0.87, and 0.88 $\mu$m: for a small flux value with a underestimated error bar (noted above in \S \ref{sec_21_targetsanddata}), the {\tt sedFit} code would take an earlier subtype and simply increase the reddening of the fit.
For completeness, we also explored fitting the photometry data for each star against subtypes fixed to those types found in DLC74.  We found significantly poorer $\chi^2_\nu$ values (as seen in Table 3) for these types which were, on average, 1 subtype later than the template found from searching a grid of INGS templates.   Interestingly, the resultant $F_{\rm BOL}$ values were still comparable between fits using a DLC74-fixed subtype, versus a grid fitted subtype value, differing only at the $\sim$6\% level.

\begin{figure}
\centering
    \includegraphics[width=0.44\textwidth,angle=0]{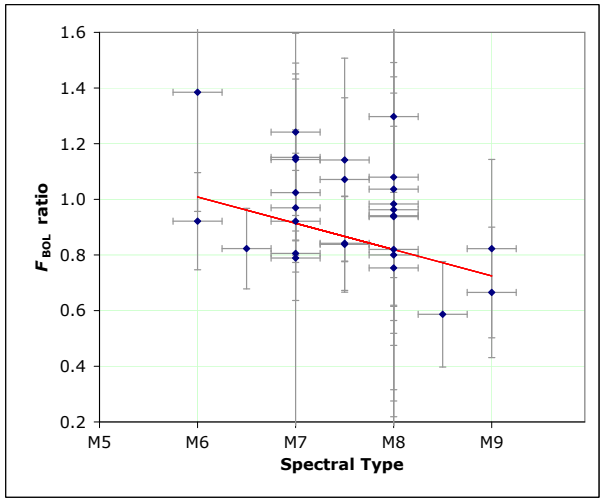}
    \caption{Ratio of $F_{\rm BOL}$ values for those values given in DLC74, versus the $F_{\rm BOL}$ values determined herein, as a function of M-class spectral subtype.}\label{fig-FBOLratios}
\end{figure}

Overall, the inclusion of reddening in our SED fitting has the effect of increasing the indicated intrinsic $F_{\rm BOL}$ on a star-by-star basis, when the extinguishing effect of that phenomenon is taken into account.  The competing effect of major near-infrared `gaps' (due to absorption features in the spectra) in the SED leading to a lower integrated $F_{\rm BOL}$ relative to a $F_{\rm BOL}$ obtained from a simple Riemann sum (cf. consider the areas under the curve of our summary {\tt sedFit} figures in the Appendix, and Figures~1, 2, or 3 of DLC74) will diminish the indicated $F_{\rm BOL}$.  This second effect should increase for increasingly redder spectral types, and this is borne out by examination of the ratio of DLC74 $F_{\rm BOL}$ values versus those determined herein (Figure~\ref{fig-FBOLratios}).  A fit to these data points gives
\begin{equation}
r = (-0.095 \pm 0.062) \times ST + (1.58 \pm 0.47)
\end{equation}
with a reduced $\chi^2=0.30$; where $r$ is the $F_{\rm BOL}$ ratio, $r = F_{\rm BOL, DLC74} / F_{\rm BOL, {\tt sedFit}}$, and ST the M-class spectral subtype value; not every star in the previous figures is seen in this plot, because DLC74 did not provide a $F_{\rm BOL, DLC74}$ for every one of their stars.

%Fitting the DLC74 values:
%\begin{equation}
%F_{\rm BOL} = F_{2.2 \mu m} \times (3.882 \pm 0.113) + (-1.25 \pm 0.56)\times 10^{-8}
%\end{equation}

\begin{figure}
\centering
    \includegraphics[width=0.44\textwidth,angle=0]{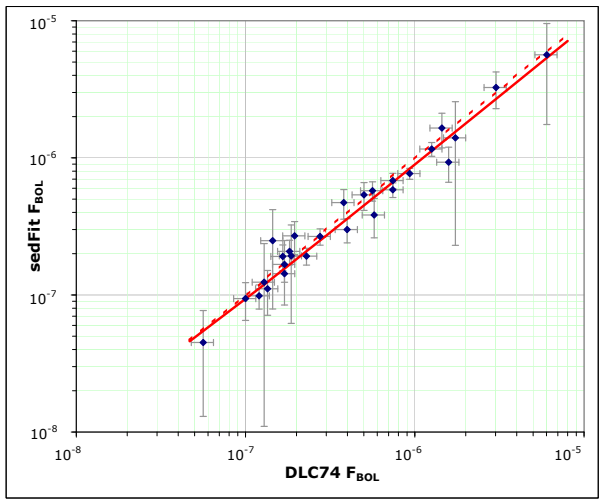}
    \caption{$F_{\rm BOL}$ as determined in this work by SED fitting, versus the values in DLC74.  The red dotted line is the 1-to-1 correspondence line, while the red solid line is the line fit (Equation~\ref{eqn-fbolvsfbol}).}\label{fig-FBOLvsFBOL}
\end{figure}

A separate way of looking at the effects of reddening and near-infrared absorption is to compare directly the derived $F_{\rm BOL}$ values from DLC74 and our work (Figure~\ref{fig-FBOLvsFBOL}).  Fitting for the relationship between the two sets of values gives:
\begin{equation}\label{eqn-fbolvsfbol}
F_{{\rm BOL}, {\tt sedFit}} = (0.89 \pm 0.08) \times F_{{\rm BOL}, {\rm DLC74}} + (0.44 \pm 1.69)\times 10^{-8}
\end{equation}
in units of ergs cm$^{-2}$ s$^{-1}$ with a $\chi^2_\nu$ of 0.39.  On average, our $F_{\rm BOL}$ values are $\simeq$11\% lower than those found in DLC74.%, though this ranges from $\simeq$30\% less on the dim end (at $F_{\rm BOL} \simeq 10^{-7}$ erg cm$^{-2}$ s$^{-1}$) to $\simeq$11\% less on the bright end (at $F_{\rm BOL} \simeq 10^{-5}$ erg cm$^{-2}$ s$^{-1}$). The overall trend for our $F_{\rm BOL}$ values to be lower at all values is indicative of accounting for near-infrared absorption with the SED templates employed in our fitting, captured in the less-than-unit slope of this relationship.

\subsection{Further Improvement upon Empirical Determination of $F_{\rm BOL}$}

As seen in DLC74, a second-order correction to the relationship between observed 2.2$\mu$m flux and overall intrinsic $F_{\rm BOL}$ can be considered.  For DLC74, they present this as a weak spectral-type dependency (see their Figure 6 and discussion in \S V.a).  Unfortunately, in the notional observational scenario we are considering -- angular size measurements contemporaneous with only $K$-band, and possibly $V$-band, measurements -- contemporaneous spectral typing is not available.  As such, development of a modern refinement of such a relationship seems of limited utility, particularly given the complications of spectral type discussed in \S \ref{sec-DLC74comparison}.

Instead of spectral type, a more directly quantitative index such as $\log F_{0.55 \mu{\rm m}} / F_{2.2 \mu{\rm m}}$ can be considered. However, a cursory examination of the ratio of $\log F_{\rm BOL} / F_{2.2 \mu{\rm m}}$ to $\log F_{0.55 \mu{\rm m}} / F_{2.2 \mu{\rm m}}$ shows no significant improvement in determination of $F_{\rm BOL}$ over the 17\% uncertainty found for Equation~\ref{eqn-fbol-f22} -- which provides no motivation for the complication introduced by now levying a requirement for contemporaneous $V$-band data in conjunction with the $K$-band data.

Incorporating our SED fitting $A_{\rm V}$ fit values into this analysis, to index with unreddened values, i.e. $\log F_{\rm BOL} / F_{2.2 \mu{\rm m, corr}}$ versus $\log F_{0.55 \mu{\rm m, corr}} / F_{2.2 \mu{\rm m, corr}}$, shows a reduction in the predictive uncertainty for $F_{\rm BOL}$ decreasing to $\simeq$9\%.  While this is a significant improvement, it is a rather unrealistic one, presuming some knowledge of $A_{\rm V}$ which is not obtained without the complication of (and requirements for contemporaneous observational data over a wide bandpass) a full SED fit.

\subsection{Empirical (non-)Determination of $T_{\rm EFF}$}

While this investigation has emphasized determination of $F_{\rm BOL}$, we do not see it of direct utility in determining effective temperature $(T_{\rm EFF})$.  As noted in \S \ref{sec-sedFitting}, a $(T_{\rm EFF})$ is assigned to each spectral template, based upon the earlier calibration found in \citet{vanBelle1999AJ....117..521V}; this enabled an estimate of angular size to be derived from this SED fitting exercise.  A recalibration of the $(T_{\rm EFF})$ versus spectral type (or other index, such as $V$-$K$ color) would be useful and is in fact part of a larger effort on the part of the authors.  However, the necessary new information -- namely direct measures of angular sizes for highly evolved stars -- will be part of a forthcoming publication; as such, any $(T_{\rm EFF})$ recalibration is deferred until that investigation.  This investigation directly enables that work by providing the photometrically-based determination of $F_{\rm BOL}$ for Equation \ref{eqn_teff}.

%Discussion of errors
%DLC74
%V band not used, disproportionate effect on reddening determination
%Scaling of errors
%Achieve red-chi$^2$ on average of 1.0
%FBOL fits
%discussion of AVs
%[bump forward by half a spectral type if AV=0 ?]
%
%* FBOL_new versus FBOL_old
%line fit with errors in 2D
%source of difference?
%
%* FBOL versus 2.2um fitting
%
%[* other FBOL versus xxx fits?]
%
%Effect that recalibrated FBOL relationship has on resultant TEFF
%
%Recalibration of photometric system against STIS?

\section{Conclusions}

Evaluation of the photometric data of DLC74 with current computing techniques, against empirical spectral templates, makes considerable utility of this still-impressive dataset.  Refinement of the intrinsic $F_{\rm BOL}$ versus observed $F_{{\rm 2.2} \mu {\rm m}}$ relationship, first noted in DLC74, provides observers with a useful tool in deducing the fundamental stellar parameter of $T_{\rm EFF}$ of these stars, even with limited observational data.  This approach has considerable merit in further application against other early but high-quality photoelectric photometry, such as the extensive \citet{Wing1967PhDT.........5W} data set.

\acknowledgements

{\it Acknowledgements.} This work would not have been possible without the {\tt sedFit} code written by Andy Boden, and his generous technical support on the use thereof.  We have made extensive use of the SIMBAD database and the VizieR catalogue access tool, operated by the CDS in Strasbourg, France \citep{Ochsenbein2000A&AS..143...23O}.  This research has made use of the AFOEV database, operated at CDS, France, and the GCPD database at the University of Lausanne, Switzerland \citep{Mermilliod1997A&AS..124..349M} . This research has made use of NASA's Astrophysics Data System.  %Portions of this work were performed at the Jet Propulsion Laboratory, California Institute of Technology under contract with the National Aeronautics and Space Administration.

Helpful input for this article resulted from discussions with and feedback from Wes Lockwood and Kaspar von~Braun.  Funding for this research has been generously provided in part by Lowell Observatory and CONACyT, Mexico.  This material is based upon work supported by the National Science Foundation under Grant No. AST-1212203, and NASA Grant No. NNX13AF01G.

%----------------------------------------------------------------------

%\bibliography{ms}

%----------------------------------------------------------------------

\clearpage

\appendix

%\LongTables
\begin{table}[h]
\caption{\small Observational data for the program stars derived from DLC74, including spectral types, bolometric flux (log [W cm$^{-2}$]), and filter fluxes (Jy).}\label{table-DLC74-Jy}
\begin{adjustbox}{width=0.95\textwidth}
\small
\begin{tabular}{lcclrrrrrrrrrrrr}
Star      & \begin{tabular}[c]{@{}l@{}}JD\\ 2440000+\end{tabular} & \begin{tabular}[c]{@{}l@{}}Spectral\\ Type\end{tabular} & $\log F_{\rm BOL}$ & 0.55$\mu$m   & 78     & 87     & 88     & 104     & 105     & 1.25$\mu$m   & 1.65$\mu$m    & 2.2$\mu$m      & 3.4$\mu$m     & 5.0$\mu$m    & 10.2$\mu$m   \\
IRC-30217 & 1050                                                  & M9.8                                                    & -14      & 0.0019  & 0.301  & 1.51   & 1.86   & 12.67   & 6.54    &         & 39.61    & 45.47     & 46.33    &         & 34.90   \\
IRC-30217 & 1107                                                  & M9.3                                                    &          & 0.0124  & 1.043  & 4.67   & 5.50   & 25.87   & 16.06   &         &          & 51.02     & 58.32    & 49.07   & 54.06   \\
IRC-20293 & 1051                                                  & M7.9                                                    & -13.74   & 0.0580  & 1.813  & 8.91   & 8.71   & 29.71   & 28.56   &         & 99.51    & 82.74     & 59.68    &         & 10.07   \\
IRC-10236 & 1049                                                  & c                                                       &          & 0.0171  & 0.600  & 1.02   & 1.10   & 2.20    & 2.43    &         & 30.05    & 111.62    & 376.56   &         & 516.27  \\
IRC+00028 & 1215                                                  & M7.9                                                    & -13.56   & 0.1752  & 3.702  & 18.18  & 16.60  & 52.83   & 50.79   & 94.78   & 137.36   & 116.88    & 76.88    & 40.81   & 31.11   \\
IRC+00266 & 1050                                                  & M9.6                                                    & -14.25   & 0.0006  & 0.069  & 0.47   & 0.58   & 4.29    & 2.43    &         & 24.43    & 30.04     & 30.61    &         & 19.18   \\
IRC+00266 & 1107                                                  & M9.9                                                    &          & 0.0003  & 0.045  & 0.35   & 0.44   & 3.49    & 1.80    &         &          & 18.96     & 20.22    &         &         \\
IRC+00281 & 1049                                                  & M7.8                                                    &          & 0.1158  & 2.446  & 11.21  & 10.47  & 32.57   & 32.04   &         & 97.24    & 79.02     & 47.41    &         &         \\
IRC+10011 & 1251                                                  &                                                         &          &         &        & 0.02   &        & 0.40    & 0.22    & 1.94    & 15.06    & 84.67     & 367.99   &         &         \\
IRC+10050 & 1277                                                  & M9.5                                                    &          & 0.0450  & 7.053  & 39.78  & 50.13  & 373.98  & 221.69  & 655.69  & 1252.70  & 1540.72   & 2117.56  &         &         \\
IRC+10313 & 1215                                                  & M8.4                                                    &          & 0.0206  & 0.756  & 3.80   & 3.72   & 16.32   & 13.67   & 32.11   & 53.44    & 47.61     & 33.56    &         &         \\
IRC+10523 & 1215                                                  & M9.1                                                    & -13.4    & 0.0298  & 2.082  & 10.46  & 10.97  & 56.60   & 37.65   & 103.92  & 157.71   & 161.33    & 153.40   & 138.30  & 220.23  \\
IRC+10523 & 1251                                                  & M9.2                                                    &          & 0.0211  & 1.543  & 8.12   & 8.91   & 54.06   & 34.34   & 94.78   & 150.61   & 150.57    & 143.16   &         &         \\
IRC+10525 & 1251                                                  & M8.5                                                    &          & 0.0061  & 0.810  & 4.89   & 5.13   & 21.52   & 17.61   & 37.73   & 65.74    & 82.74     & 111.13   &         & 148.89  \\
IRC+20052 & 1215                                                  & M8.1                                                    & -13.42   & 0.0820  & 2.873  & 14.44  & 13.81  & 54.06   & 49.63   & 116.60  & 189.60   & 172.87    & 124.69   & 76.00   & 41.01   \\
IRC+20281 & 1109                                                  & M9.3                                                    &          & 0.0006  & 0.043  & 0.28   & 0.34   & 2.53    & 1.61    &         &          & 42.43     & 78.67    &         & 87.68   \\
IRC+20328 & 1106                                                  & M9.8                                                    &          & 0.0030  & 0.406  & 2.24   & 2.75   & 19.63   & 10.61   &         &          & 92.84     & 116.37   &         & 87.68   \\
IRC+30021 & 1106                                                  & M8.9                                                    &          & 0.0016  & 0.287  & 1.78   & 1.95   & 11.56   & 8.43    &         &          & 62.77     & 75.13    & 67.74   & 138.96  \\
IRC+30021 & 1214                                                  & M9.4                                                    & -13.89   &         & 0.125  & 0.69   & 0.87   & 6.07    & 3.76    & 16.47   & 39.61    & 48.72     & 58.32    & 89.29   & 126.73  \\
IRC+30021 & 1251                                                  & M9.6                                                    &          & 0.0004  & 0.085  & 0.58   & 0.69   & 5.16    & 2.99    & 13.70   & 30.75    & 43.42     & 53.19    &         &         \\
IRC+30021 & 1277                                                  & M9.7                                                    &          &         & 0.063  & 0.41   & 0.52   & 4.20    & 2.38    & 12.21   & 29.37    & 38.70     & 47.41    &         &         \\
IRC+30055 & 1215                                                  & M7.9                                                    & -13.78   & 0.0401  & 1.440  & 7.24   & 6.61   & 24.15   & 23.75   & 48.61   & 80.88    & 73.74     & 50.80    & 43.73   & 24.71   \\
IRC+30292 & 1074                                                  & M8.6                                                    & -13.87   &         & 0.154  & 0.93   & 1.05   & 6.50    & 5.32    &         & 39.61    & 62.77     & 90.33    & 63.21   & 121.03  \\
IRC+30292 & 1215                                                  & M9.6                                                    & -14.06   &         & 0.039  & 0.26   & 0.34   & 2.47    & 1.43    & 7.19    & 20.79    & 34.49     & 48.51    & 66.19   & 89.72   \\
IRC+30515 & 1215                                                  & M7.9                                                    & -13.25   & 0.2475  & 6.583  & 32.33  & 28.85  & 96.13   & 92.42   & 184.80  & 280.44   & 238.63    & 156.98   & 109.85  & 59.28   \\
IRC+40004 & 1215                                                  &                                                         & -13.51   &         &        & 0.13   & 0.15   & 2.10    & 1.61    & 14.34   & 52.22    & 99.48     & 176.13   & 295.68  & 460.13  \\
IRC+40004 & 1251                                                  & M9.1                                                    &          & 0.0006  & 0.011  & 0.13   & 0.15   & 1.67    & 1.19    & 10.88   & 43.44    & 84.67     & 149.91   &         & 290.32  \\
IRC+40442 & 1251                                                  & M9.3                                                    &          & 0.0045  & 0.548  & 3.39   & 4.07   & 27.09   & 16.82   & 71.90   & 147.18   & 157.66    & 133.61   &         & 49.30   \\
IRC+40448 & 1215                                                  & M7.0                                                    & -12.63   &         & 0.059  & 0.50   & 0.48   & 3.18    & 3.60    & 14.34   & 82.76    & 361.18    & 1367.21  & 2635.23 & 6499.48 \\
IRC+40485 & 1215                                                  & c                                                       & -13.47   & 0.0175  & 0.791  & 1.55   & 1.74   & 3.83    & 4.22    & 11.66   & 41.48    & 106.59    & 211.76   & 363.76  & 304.00  \\
IRC+40485 & 1251                                                  & c                                                       &          & 0.0253  & 1.043  & 1.95   & 2.29   & 4.93    & 5.32    & 14.34   & 48.74    & 122.38    & 232.19   &         & 215.22  \\
IRC+50096 & 1215                                                  & c                                                       & -13.13   & 0.0899  & 2.180  & 3.80   & 4.37   & 8.37    & 9.03    & 22.74   & 68.84    & 193.97    & 485.11   & 814.36  & 918.08  \\
IRC+50096 & 1251                                                  & c                                                       &          & 0.1056  & 2.282  & 4.17   & 4.68   & 9.18    & 10.37   & 26.10   & 80.88    & 222.70    & 556.98   &         &         \\
IRC+50260 & 1054                                                  & M7.0                                                    &          & 0.2257  & 2.180  & 9.11   & 8.13   &         & 21.17   &         & 52.22    & 39.60     & 24.31    &         &         \\
IRC+50261 & 1078                                                  & M6.0                                                    &          & 0.1712  & 2.682  & 9.54   & 8.71   & 16.70   & 17.61   &         & 34.50    & 27.40     & 19.31    &         &         \\
IRC+50357 & 1215                                                  & c                                                       & -13.83   &         &        & 0.05   &        & 0.15    &         & 0.84    & 5.34     & 22.79     & 88.27    & 209.32  & 200.85  \\
IRC+60015 & 1251                                                  & M7.8                                                    &          & 0.1392  & 3.224  & 16.21  & 14.79  & 44.96   & 45.26   & 84.47   & 131.17   & 114.22    & 75.13    &         & 18.74   \\
IRC+60052 & 1251                                                  & M6.0                                                    &          & 0.0985  & 2.940  & 12.29  & 11.75  & 27.72   & 29.91   & 52.08   & 92.86    & 82.74     & 58.32    &         & 21.03   \\
IRC+60092 & 1215                                                  & M8.6                                                    & -13.24   &         & 0.477  & 3.55   & 3.80   & 23.60   & 19.31   & 59.80   & 143.83   & 212.68    & 299.11   & 408.15  & 799.61  \\
IRC+60169 & 1054                                                  & M9.3                                                    &          & 0.0127  & 1.543  & 7.58   & 8.91   & 52.83   & 32.79   &         & 274.06   & 321.90    & 343.43   &         & 460.13  \\
IRC+60184 & 1054                                                  & M8.8                                                    &          & 0.0175  & 0.930  & 4.46   & 4.68   & 22.02   & 16.43   &         & 49.87    & 59.94     & 47.41    &         &         \\
IRC+60288 & 1106                                                  & M9.3                                                    &          & 0.0040  & 0.435  & 3.09   & 3.47   & 23.06   & 14.31   &         &          & 67.25     & 62.49    & 47.95   & 33.33   \\
IRC+60288 & 1214                                                  & M8.2                                                    & -13.64   &         & 2.873  & 15.48  & 14.79  & 42.94   & 36.79   & 62.62   & 95.03    & 88.66     & 82.38    & 66.19   & 68.06   \\
IRC+60288 & 1251                                                  & M8.2                                                    &          & 0.0430  & 2.446  & 13.79  & 13.49  & 42.94   & 36.79   & 65.57   & 111.65   & 104.17    & 92.44    &         &         \\
IRC+60289 & 1215                                                  & M7.9                                                    &          & 0.0221  & 0.773  & 4.36   & 3.98   & 13.89   & 13.36   & 28.62   & 44.45    & 39.60     & 27.91    & 24.59   &         \\
IRC+60316 & 1215                                                  & M6.0                                                    & -13.71   & 0.0608  & 2.336  & 10.96  & 10.24  & 27.09   & 28.56   & 55.81   & 97.24    & 90.72     & 63.95    & 36.38   & 24.71   \\
IRC+60334 & 1251                                                  & M8.9                                                    &          & 0.0076  & 0.629  & 3.31   & 3.72   & 19.63   & 13.99   & 39.51   & 79.04    & 101.79    & 111.13   &         & 78.14   \\
IRC+70102 & 1052                                                  & M9.0                                                    & -13.92   & 0.0111  & 0.791  & 3.71   & 4.07   & 18.74   & 12.47   &         & 54.68    & 52.21     & 50.80    &         & 39.16   \\
IRC+70171 & 1215                                                  & M9.2                                                    & -13.73   & 0.0023  & 0.338  & 1.90   & 2.34   & 13.27   & 8.43    & 34.41   & 75.48    & 90.72     & 90.33    & 70.93   & 74.62   \\
IRC+70171 & 1251                                                  & M9.3                                                    & -13.84   & 0.0016  & 0.262  & 1.44   & 1.82   & 10.54   & 6.54    & 23.81   & 58.59    & 72.06     & 71.75    &         & 62.07   \\
IRC+80005 & 1050                                                  & M9.5                                                    &          & 0.0011  & 0.181  & 1.44   & 1.78   & 12.38   & 7.34    & 32.86   & 61.35    & 62.77     & 51.98    &         &         \\
IRC+80005 & 1215                                                  & M9.2                                                    & -13.77   & 0.0007  & 0.294  & 2.24   & 2.34   & 15.59   & 9.90    & 45.36   & 79.04    & 80.86     & 66.96    & 51.38   & 24.15   \\
R And     & 1214                                                  &                                                         & -12.76   & 0.0878  & 7.386  & 35.45  & 38.91  & 138.95  & 156.95  & 443.30  & 790.40   & 772.19    & 717.52   & 603.70  & 680.58  \\
R And     & 1277                                                  &                                                         &          & 0.0136  & 1.344  & 8.91   & 8.91   & 44.96   & 54.42   & 136.99  & 329.49   & 378.20    & 422.51   &         &         \\
W And     & 1215                                                  & M6.5                                                    & -12.8    & 1.2990  & 21.797 & 97.65  & 93.35  & 270.92  & 272.74  & 532.97  & 754.83   & 642.28    & 474.06   & 331.76  & 225.36  \\
W And     & 1251                                                  & M7.9                                                    &          & 0.2419  & 10.195 & 51.25  & 50.13  & 191.80  & 188.69  & 336.28  &          & 559.40    & 432.35   &         &         \\
W And     & 1277                                                  & M8.3                                                    &          & 0.1056  & 5.351  & 29.49  & 31.63  & 145.49  & 124.67  & 386.10  & 572.59   & 522.06    & 412.89   &         &         \\
W And     & 1302                                                  & M8.9                                                    &          & 0.0505  & 3.079  & 17.77  & 19.96  & 112.94  & 82.37   &         &          & 559.40    & 496.41   &         &         \\
T Aqr     & 1251                                                  & M8.0                                                    &          & 0.0259  & 1.043  & 5.37   & 4.68   & 12.97   & 11.91   & 17.25   & 22.28    & 19.40     & 15.34    &         &         \\
T Ari     & 1214                                                  & M7.0                                                    & -12.9    & 1.3292  & 28.734 & 128.73 & 112.23 & 264.75  & 272.74  & 395.09  & 599.58   & 476.13    & 335.61   & 240.34  & 103.01  \\
U Ari     & 1251                                                  & M9.5                                                    &          & 0.0094  & 1.226  & 9.33   & 10.00  & 66.50   & 38.53   & 108.82  & 154.12   & 140.52    & 136.72   &         &         \\
U Ari     & 1277                                                  &                                                         &          & 0.0164  &        &        &        & 79.95   & 46.32   & 111.35  & 161.38   & 147.14    & 133.61   &         &         \\
R Boo     & 1078                                                  & MB.l                                                    &          & 0.0783  & 2.621  & 16.21  & 14.79  & 51.62   & 45.26   &         & 109.11   & 90.72     & 70.12    &         &         \\
R Cam     & 1078                                                  &                                                         &          & 0.3417  & 8.287  & 16.21  & 17.38  & 27.09   & 29.22   &         & 62.78    & 46.53     & 32.05    &         &         \\
R Cas     & 1214                                                  & M9.3                                                    & -12.22   & 0.1635  & 16.159 & 89.06  & 104.74 & 606.52  & 367.92  & 1537.09 & 2617.26  & 2739.83   & 2487.92  & 1999.03 & 2008.53 \\
R Cas     & 1251                                                  & M9.7                                                    &          & 0.0783  & 9.737  & 54.91  & 69.20  & 481.77  & 260.47  & 1015.54 & 1896.04  & 2126.79   & 1931.24  &         & 1078.65 \\
R Cas     & 1302                                                  & M9.9                                                    &          & 0.0783  & 8.480  & 45.67  & 56.25  & 449.62  & 232.14  &         &          & 2386.29   & 2545.87  &         &         \\
T Cas     & 1214                                                  & M8.9                                                    & -12.52   & 0.1131  & 10.676 & 66.02  & 70.81  & 373.98  & 260.47  & 905.10  & 1506.08  & 1405.15   & 1061.30  & 742.71  & 481.81  \\
T Cas     & 1251                                                  & M8.7                                                    &          & 0.1752  & 14.737 & 89.06  & 95.52  & 481.77  & 367.92  & 1039.20 & 1577.05  & 1471.38   & 1163.69  &         & 325.75  \\
V Cas     & 1214                                                  & M7.9                                                    & -13.13   & 0.3263  & 8.880  & 48.94  & 44.68  & 142.18  & 136.70  & 249.29  & 361.28   & 300.42    & 221.74   & 135.15  & 62.07   \\
Y Cas     & 1214                                                  & MB.l                                                    & -13.13   & 0.3496  & 11.979 & 48.94  & 47.87  & 138.95  & 121.83  & 202.63  & 307.50   & 293.58    & 254.59   & 195.35  & 170.95  \\
SS Cas    & 1277                                                  & M4.0                                                    &          & 1.2405  & 8.287  & 16.97  & 16.60  & 22.53   & 24.31   & 29.97   & 39.61    & 32.94     & 25.46    &         &         \\
T Cep     & 1214                                                  & M8.6                                                    &          & 0.8782  & 42.501 & 251.00 & 251.24 & 1210.16 & 924.17  & 2273.52 & 3371.68  & 2868.96   & 2117.56  &         &         \\
R Cyg     & 1214                                                  &                                                         &          & 1.9213  & 45.541 & 87.03  & 93.35  & 159.53  & 164.34  & 255.09  & 353.06   & 293.58    & 254.59   &         &         \\
X Cyg     & 1214                                                  & M8.2                                                    &          & 0.4504  & 49.935 & 467.38 & 398.20 & 1632.46 & 1431.38 & 2671.15 & 3696.98  & 3611.80   & 3434.29  &         &         \\
S Lac     & 1251                                                  & M8.2                                                    & -13.77   & 0.0517  & 1.653  & 8.12   & 8.13   & 32.57   & 28.56   & 57.11   & 79.04    & 72.06     & 53.19    &         & 12.67   \\
R Leo     & 1109                                                  &                                                         &          &         &        &        &        &         &         &         & 15770.54 & 122383.85 & 10135.30 & 5634.02 & 3333.33 \\
W Lyr     & 1214                                                  & M6.5                                                    &          & 0.1835  & 2.682  & 10.23  & 9.12   & 18.74   & 18.87   & 26.10   & 36.13    & 30.04     & 23.76    &         &         \\
V Mon     & 1277                                                  & M5.0                                                    &          & 2.9081  & 39.665 & 114.73 & 109.67 & 200.84  & 211.72  & 313.83  & 454.83   & 361.18    & 266.59   &         &         \\
R Tri     & 1214                                                  & M5.0                                                    & -13.03   & 10.3182 & 68.929 & 154.76 & 147.94 & 220.21  & 226.86  & 273.34  & 378.31   & 280.37    & 211.76   & 178.16  & 78.14   \\
R Tri     & 1251                                                  & M5.5                                                    &          & 5.0536  & 51.098 & 151.24 & 141.28 & 241.46  & 248.74  & 321.14  & 454.83   & 344.92    & 248.79   &         &         \\
R Tri     & 1277                                                  & M7.0                                                    &          & 1.3602  & 21.797 & 99.92  & 89.14  & 210.30  & 216.65  & 321.14  & 434.36   & 352.96    & 254.59   &         &         \\
W Peg     & 1214                                                  & M8.9                                                    & -12.84   & 0.1185  & 7.218  & 34.65  & 38.91  & 187.43  & 133.58  & 433.21  & 688.41   & 657.24    & 519.80   & 363.76  & 241.48  \\
Z Peg     & 1214                                                  & M8.5                                                    & -13.3    & 0.0529  & 2.940  & 16.21  & 16.99  & 71.26   & 58.31   & 172.46  & 244.26   & 227.89    & 176.13   & 91.37   & 52.83   \\
          &                                                       &                                                         &          &         &        &        &        &         &         &         &          &           &          &         &         \\
          &                                                       &                                                         &          &         &        &        &        &         &         &         &          &           &          &         &
\end{tabular}
\end{adjustbox}
\end{table}

% Please add the following required packages to your document preamble:
% \usepackage{multirow}
\begin{table}[h]
\caption{\small Spectral energy distribution fitting results, including fit spectral type, $\chi^2_{\rm DOF}$ for both DLC74 and our spectral type, estimates for reddening $(A_{\rm V}$, in mag), angular size $(\theta_{\rm EST}$, in mas), and fit bolometric flux as compared to DLC74 flux$(F_{\rm BOL}$, in erg s$^{-1}$ cm$^{-2}$).}\label{table-FBOLresults}
\begin{adjustbox}{width=0.75\textwidth}
\small
\begin{tabular}{lcccccrrrr}
 &                                                       &                                                        & \multicolumn{6}{c}{SED Fitting Results}                                                                                               &                                                      \\
Star                       & \begin{tabular}[c]{@{}l@{}}JD\\ 2440000+\end{tabular} & \begin{tabular}[c]{@{}l@{}}DLC74\\ SpType\end{tabular} & \begin{tabular}[c]{@{}l@{}}Fit\\ SpType\end{tabular} & \begin{tabular}[c]{@{}l@{}}DLC74\\ $\chi^2/$DOF\end{tabular} & \begin{tabular}[c]{@{}l@{}}Fit\\ $\chi^2/$DOF\end{tabular}& $A_V$ (mag)              & $\theta_{\rm EST}$         & $F_{\rm BOL}$                            & \begin{tabular}[c]{@{}l@{}}DLC74\\ $F_{\rm BOL}$\end{tabular} \\
\multirow{2}{*}{IRC-30217$\Big\{$} & 1050 & M9.8 & M8III & 428.16 & 0.82 & $3.01 \pm 0.35$ & $1.81 \pm 0.31$ & $(9.41\pm 2.89) \times 10^{-8}$ & $1.00 \times 10^{-7}$ \\
  & 1107 & M9.3 & M8III & 549.87 & 0.13 & $0.56 \pm 0.38$ & $1.69 \pm 0.18$ & $(8.17\pm 1.30) \times 10^{-8}$ &  \\
IRC-20293 & 1051 & M7.9 & M7III & 33.41 & 0.58 & $1.90 \pm 0.30$ & $2.52 \pm 0.32$ & $(2.08\pm 0.42) \times 10^{-7}$ & $1.82 \times 10^{-7}$ \\
IRC-10236 & 1049 & c & M2III & 355.15 & 18.15 & $8.81 \pm 0.40$ & $2.46 \pm 6.92$ & $(4.73\pm 0.27) \times 10^{-7}$ &  \\
IRC+00028 & 1215 & M7.9 & M7III & 31.44 & 0.58 & $0.98 \pm 0.29$ & $2.85 \pm 0.29$ & $(2.67\pm 3.63) \times 10^{-7}$ & $2.75 \times 10^{-7}$ \\
\multirow{2}{*}{IRC+00266$\Big\{$} & 1050 & M9.9 & M8III & 421.87 & 0.48 & $5.47 \pm 0.35$ & $1.69 \pm 0.69$ & $(8.19\pm 0.66) \times 10^{-8}$ &  \\
  & 1107 & M9.6 & M8III & 298.95 & 1.10 & $4.51 \pm 0.41$ & $1.26 \pm 0.46$ & $(4.50\pm 0.32) \times 10^{-8}$ & $5.62 \times 10^{-8}$ \\
IRC+00281 & 1049 & M7.8 & M7III & 41.11 & 0.36 & $1.26 \pm 0.30$ & $2.36 \pm 0.26$ & $(1.84\pm 0.29) \times 10^{-7}$ &  \\
IRC+10011 & 1251 & M8.5 & M9.5III & 319.69 & 0.45 & $5.92 \pm 0.38$ & $2.43 \pm 1.07$ & $(1.36\pm 0.12) \times 10^{-7}$ &  \\
IRC+10050 & 1277 & M9.5 & M8III & 405.86 & 0.50 & $3.39 \pm 0.34$ & $10.76 \pm 1.95$ & $(3.31\pm 1.10) \times 10^{-6}$ &  \\
IRC+10313 & 1215 & M8.4 & M7.5III & 11.12 & 0.87 & $2.06 \pm 0.31$ & $1.88 \pm 0.23$ & $(1.09\pm 0.22) \times 10^{-7}$ &  \\
\multirow{2}{*}{IRC+10523$\Big\{$} & 1215 & M9.2 & M8III & 11.87 & 0.69 & $1.95 \pm 0.33$ & $3.31 \pm 0.40$ & $(3.14\pm 0.62) \times 10^{-7}$ &  \\
  & 1251 & M9.1 & M8III & 9.46 & 0.71 & $2.04 \pm 0.33$ & $3.24 \pm 0.40$ & $(3.00\pm 0.60) \times 10^{-7}$ & $3.98 \times 10^{-7}$ \\
IRC+10525 & 1251 & M8.5 & M9III & 2.70 & 0.13 & $0.00 \pm 0.32$ & $2.07 \pm 0.18$ & $(1.06\pm 0.11) \times 10^{-7}$ &  \\
IRC+20052 & 1215 & M8.1 & M7III & 25.20 & 1.19 & $2.55 \pm 0.30$ & $3.79 \pm 0.54$ & $(4.72\pm 1.15) \times 10^{-7}$ & $3.80 \times 10^{-7}$ \\
IRC+20281 & 1109 & M9.3 & M8.5III & 157.80 & 0.52 & $6.74 \pm 0.43$ & $2.03 \pm 1.77$ & $(1.10\pm 0.19) \times 10^{-7}$ &  \\
IRC+20328 & 1106 & M9.8 & M8III & 450.70 & 0.73 & $3.95 \pm 0.40$ & $2.70 \pm 0.80$ & $(2.09\pm 0.12) \times 10^{-7}$ &  \\
\multirow{4}{*}{IRC+30021$\Bigg\{$} & 1106 & M9.6 & M8.5III & 272.57 & 0.11 & $2.48 \pm 0.39$ & $2.01 \pm 0.36$ & $(1.08\pm 0.35) \times 10^{-7}$ &  \\
  & 1214 & M8.9 & M8III & 2.18 & 0.81 & $5.95 \pm 0.34$ & $2.25 \pm 1.02$ & $(1.45\pm 0.13) \times 10^{-7}$ &  \\
  & 1251 & M9.4 & M8III & 278.96 & 0.75 & $6.01 \pm 0.35$ & $2.08 \pm 0.97$ & $(1.24\pm 0.11) \times 10^{-7}$ & $1.29 \times 10^{-7}$ \\
  & 1277 & M9.7 & M8III & 241.15 & 1.15 & $6.55 \pm 0.34$ & $2.06 \pm 1.17$ & $(1.22\pm 0.14) \times 10^{-7}$ &  \\
IRC+30055 & 1215 & M7.9 & M7III & 31.15 & 0.58 & $2.19 \pm 0.30$ & $2.41 \pm 0.31$ & $(1.91\pm 4.07) \times 10^{-7}$ & $1.66 \times 10^{-7}$ \\
\multirow{2}{*}{IRC+30292$\Big\{$} & 1074 & M8.6 & M9III & 1.88 & 0.08 & $3.39 \pm 0.34$ & $2.12 \pm 0.41$ & $(1.11\pm 0.40) \times 10^{-7}$ & $1.35 \times 10^{-7}$ \\
  & 1215 & M9.6 & M8III & 156.50 & 0.50 & $7.78 \pm 0.36$ & $1.99 \pm 1.83$ & $(1.13\pm 0.21) \times 10^{-7}$ & $8.71 \times 10^{-8}$ \\
IRC+30515 & 1215 & M7.9 & M7III & 32.94 & 0.70 & $1.43 \pm 0.29$ & $4.18 \pm 0.46$ & $(5.76\pm 0.93) \times 10^{-7}$ & $5.62 \times 10^{-7}$ \\
\multirow{2}{*}{IRC+40004$\Big\{$} & 1215 & M9.1 & M9.5III & 8.38 & 12.28 & $1.05 \pm 0.33$ & $2.41 \pm 0.25$ & $(1.33\pm 0.19) \times 10^{-7}$ &  \\
  & 1251 & M9.0 & M9.5III & 10.62 & 8.52 & $1.33 \pm 0.33$ & $2.25 \pm 0.24$ & $(1.16\pm 0.18) \times 10^{-7}$ & $3.09 \times 10^{-7}$ \\
IRC+40442 & 1251 & M9.3 & M8III & 355.87 & 1.49 & $4.84 \pm 0.33$ & $3.89 \pm 1.16$ & $(4.33\pm 0.25) \times 10^{-7}$ &  \\
IRC+40448 & 1215 & M7.0 & M9.5III & 204.46 & 14.00 & $2.96 \pm 0.34$ & $4.43 \pm 0.73$ & $(4.51\pm 0.13) \times 10^{-7}$ & $2.34 \times 10^{-6}$ \\
\multirow{2}{*}{IRC+40485$\Big\{$} & 1215 & c & M2III & 350.92 & 6.39 & $8.48 \pm 0.34$ & $2.97 \pm 5.78$ & $(6.89\pm 2.69) \times 10^{-7}$ & $3.39 \times 10^{-7}$ \\
  & 1251 & c & M2III & 388.18 & 6.13 & $8.12 \pm 0.33$ & $3.13 \pm 5.23$ & $(7.68\pm 2.56) \times 10^{-7}$ &  \\
\multirow{2}{*}{IRC+50096$\Big\{$} & 1215 & c & M2III & 463.34 & 8.76 & $6.90 \pm 0.33$ & $3.30 \pm 3.50$ & $(8.53\pm 1.80) \times 10^{-7}$ & $7.41 \times 10^{-7}$ \\
  & 1251 & c & M2III & 425.89 & 8.06 & $7.23 \pm 0.33$ & $3.70 \pm 4.42$ & $(1.07\pm 0.26) \times 10^{-6}$ &  \\
IRC+50260 & 1054 & M7.0 & M6.5III & 2.45 & 0.30 & $0.24 \pm 0.29$ & $1.54 \pm 0.14$ & $(8.40\pm 9.29) \times 10^{-8}$ &  \\
IRC+50261 & 1078 & M6.0 & M5.5III & 2.35 & 0.28 & $0.00 \pm 0.26$ & $1.23 \pm 0.11$ & $(6.12\pm 0.06) \times 10^{-8}$ &  \\
IRC+50357 & 1215 & c & M9.5III & 89.99 & 14.10 & $4.72 \pm 0.37$ & $1.25 \pm 0.35$ & $(3.61\pm 0.20) \times 10^{-8}$ & $1.48 \times 10^{-7}$ \\
IRC+60015 & 1251 & M7.8 & M7III & 40.14 & 0.35 & $1.33 \pm 0.29$ & $2.85 \pm 0.31$ & $(2.67\pm 4.16) \times 10^{-7}$ &  \\
IRC+60052 & 1251 & M6.0 & M5.5III & 1.63 & 0.17 & $2.32 \pm 0.27$ & $2.44 \pm 0.33$ & $(2.40\pm 0.55) \times 10^{-7}$ &  \\
IRC+60092 & 1215 & M8.6 & M9III & 2.00 & 0.13 & $3.24 \pm 0.33$ & $3.94 \pm 0.69$ & $(3.83\pm 0.12) \times 10^{-7}$ & $5.75 \times 10^{-7}$ \\
IRC+60169 & 1054 & M9.3 & M8III & 347.40 & 0.26 & $4.81 \pm 0.35$ & $5.35 \pm 1.71$ & $(8.19\pm 0.51) \times 10^{-7}$ &  \\
\multirow{2}{*}{IRC+60184$\Big\{$} & 1054 & M8.8 & M8.5III & 18.37 & 0.27 & $0.00 \pm 0.32$ & $1.75 \pm 0.16$ & $(8.14\pm 0.85) \times 10^{-8}$ &  \\
  & 1106 & M9.3 & M8III & 496.80 & 0.49 & $2.13 \pm 0.39$ & $2.12 \pm 0.33$ & $(1.28\pm 0.36) \times 10^{-7}$ &  \\
IRC+60288 & 1214 & M8.2 & M7III & 17.24 & 0.53 & $0.58 \pm 0.30$ & $2.33 \pm 0.23$ & $(1.79\pm 0.22) \times 10^{-7}$ &  \\
  & 1251 & M8.2 & M7.5III & 20.50 & 0.48 & $0.91 \pm 0.31$ & $2.50 \pm 0.25$ & $(1.92\pm 0.27) \times 10^{-7}$ & $2.29 \times 10^{-7}$ \\
IRC+60289 & 1215 & M7.9 & M7III & 32.03 & 0.78 & $1.95 \pm 0.29$ & $1.75 \pm 0.22$ & $(1.01\pm 1.96) \times 10^{-7}$ &  \\
IRC+60316 & 1215 & M6.0 & M6III & 2.52 & 0.28 & $2.75 \pm 0.28$ & $2.68 \pm 0.41$ & $(2.70\pm 0.73) \times 10^{-7}$ & $1.95 \times 10^{-7}$ \\
IRC+60334 & 1251 & M8.9 & M8.5III & 1.96 & 0.14 & $2.37 \pm 0.33$ & $2.58 \pm 0.35$ & $(1.78\pm 0.41) \times 10^{-7}$ &  \\
IRC+70102 & 1052 & M9.0 & M8III & 8.50 & 0.29 & $1.82 \pm 0.34$ & $1.86 \pm 0.23$ & $(9.86\pm 0.20) \times 10^{-8}$ & $1.20 \times 10^{-7}$ \\
\multirow{2}{*}{IRC+70171$\Big\{$} & 1215 & M9.3 & M8III & 308.81 & 0.88 & $5.24 \pm 0.34$ & $2.95 \pm 1.03$ & $(2.49\pm 0.17) \times 10^{-7}$ & $1.45 \times 10^{-7}$ \\
  & 1251 & M9.2 & M8III & 11.22 & 0.34 & $5.23 \pm 0.34$ & $2.60 \pm 0.90$ & $(1.93\pm 1.31) \times 10^{-7}$ & $1.86 \times 10^{-7}$ \\
\multirow{2}{*}{IRC+80005$\Big\{$} & 1050 & M9.2 & M8III & 28.41 & 2.07 & $4.43 \pm 0.33$ & $2.42 \pm 0.62$ & $(1.67\pm 0.83) \times 10^{-7}$ & $1.70 \times 10^{-7}$ \\
  & 1215 & M9.5 & M8III & 388.53 & 2.40 & $4.42 \pm 0.33$ & $2.74 \pm 0.70$ & $(2.15\pm 0.11) \times 10^{-7}$ &  \\
\multirow{2}{*}{R And-A$\Big\{$} & 1214 & S-type & M10III & 476.42 & 1.71 & $4.23 \pm 0.30$ & $8.80 \pm 2.19$ & $(2.55\pm 0.12) \times 10^{-6}$ &  \\
  & 1277 & S-type & M10III & 629.77 & 1.29 & $5.67 \pm 0.30$ & $6.53 \pm 2.75$ & $(1.40\pm 0.12) \times 10^{-6}$ & $1.74 \times 10^{-6}$ \\
\multirow{4}{*}{W And-A$\Bigg\{$} & 1215 & M8.9 & M9III & 2.16 & 0.66 & $1.09 \pm 0.29$ & $6.70 \pm 0.69$ & $(1.48\pm 0.21) \times 10^{-6}$ &  \\
  & 1251 & M8.3 & M8.5III & 24.07 & 0.73 & $2.04 \pm 0.34$ & $6.52 \pm 0.93$ & $(1.40\pm 0.34) \times 10^{-6}$ &  \\
  & 1277 & M7.9 & M8III & 29.15 & 2.30 & $2.89 \pm 0.31$ & $6.58 \pm 1.03$ & $(1.33\pm 0.37) \times 10^{-6}$ &  \\
  & 1302 & M6.5 & M6.5III & 5.25 & 0.12 & $2.15 \pm 0.39$ & $5.91 \pm 0.95$ & $(9.30\pm 0.27) \times 10^{-7}$ & $1.58 \times 10^{-6}$ \\
T Aqr-A & 1251 & M8.0 & M8III & 40.41 & 0.88 & $0.00 \pm 0.27$ & $1.10 \pm 0.09$ & $(4.26\pm 0.39) \times 10^{-8}$ &  \\
T Ari-A & 1214 & M7.0 & M7III & 3.32 & 0.22 & $0.60 \pm 0.27$ & $5.54 \pm 0.50$ & $(1.16\pm 0.13) \times 10^{-6}$ & $1.26 \times 10^{-6}$ \\
\multirow{2}{*}{U Ari-A$\Big\{$} & 1251 & M9.5 & M9.5III & 478.73 & 1.48 & $1.24 \pm 0.33$ & $3.05 \pm 0.31$ & $(2.66\pm 0.40) \times 10^{-7}$ &  \\
  & 1277 & mid M? & M9.5III & 948.13 & 0.80 & $0.43 \pm 0.36$ & $2.92 \pm 0.27$ & $(2.44\pm 0.30) \times 10^{-7}$ &  \\
R Boo-A & 1078 & M8.1 & M8III & 17.42 & 0.54 & $0.00 \pm 0.30$ & $2.29 \pm 0.20$ & $(1.61\pm 0.17) \times 10^{-7}$ &  \\
R Cam-A & 1078 & S-type & M10III & 1169.95 & 0.49 & $0.71 \pm 0.23$ & $1.66 \pm 0.15$ & $(1.34\pm 0.16) \times 10^{-7}$ &  \\
\multirow{3}{*}{R Cas-A$\Bigg\{$} & 1214 & M9.7 & M9.5III & 394.35 & 1.77 & $3.86 \pm 0.33$ & $15.35 \pm 3.24$ & $(6.74\pm 0.27) \times 10^{-6}$ &  \\
  & 1251 & M9.9 & M10III & 423.20 & 1.18 & $3.86 \pm 0.34$ & $13.35 \pm 2.81$ & $(5.10\pm 0.20) \times 10^{-6}$ &  \\
  & 1302 & M9.3 & M9.5III & 428.99 & 1.09 & $4.44 \pm 0.41$ & $14.06 \pm 4.95$ & $(5.65\pm 0.39) \times 10^{-6}$ & $6.03 \times 10^{-6}$ \\
\multirow{2}{*}{T Cas-A$\Big\{$} & 1214 & M8.9 & M9III & 18.51 & 1.96 & $3.09 \pm 0.32$ & $10.68 \pm 1.76$ & $(3.26\pm 0.97) \times 10^{-6}$ & $3.02 \times 10^{-6}$ \\
  & 1251 & M8.7 & M8III & 13.98 & 1.54 & $2.44 \pm 0.31$ & $10.70 \pm 1.46$ & $(3.52\pm 0.81) \times 10^{-6}$ &  \\
V Cas-A & 1214 & M7.9 & M8III & 31.07 & 0.61 & $0.86 \pm 0.29$ & $4.55 \pm 0.45$ & $(6.83\pm 0.89) \times 10^{-7}$ & $7.41 \times 10^{-7}$ \\
Y Cas-A & 1214 & M8.1 & M8III & 21.31 & 0.47 & $0.59 \pm 0.30$ & $4.22 \pm 0.41$ & $(5.85\pm 0.72) \times 10^{-7}$ & $7.41 \times 10^{-7}$ \\
SS Cas-A & 1277 & M4.0 & M4III & 3.44 & 0.68 & $0.00 \pm 0.20$ & $1.35 \pm 0.11$ & $(8.93\pm 0.71) \times 10^{-8}$ &  \\
T Cep-A & 1214 & M8.6 & M8III & 11.20 & 1.42 & $1.55 \pm 0.31$ & $14.37 \pm 1.58$ & $(6.35\pm 0.11) \times 10^{-6}$ &  \\
R Cyg-A & 1214 & S-type & M10III & 1056.44 & 0.56 & $0.76 \pm 0.23$ & $4.14 \pm 0.38$ & $(7.88\pm 0.93) \times 10^{-7}$ &  \\
X Cyg-A & 1214 & M8.2 & M8III & 15.39 & 0.71 & $0.77 \pm 0.31$ & $14.84 \pm 1.44$ & $(6.77\pm 0.87) \times 10^{-6}$ &  \\
S Lac-A & 1251 & M8.2 & M8III & 16.18 & 0.95 & $0.93 \pm 0.30$ & $2.16 \pm 0.21$ & $(1.43\pm 0.19) \times 10^{-7}$ & $1.70 \times 10^{-7}$ \\
W Lyr-A & 1214 & M6.5 & M6.5III & 13.15 & 0.21 & $0.00 \pm 0.26$ & $1.33 \pm 0.11$ & $(6.66\pm 0.62) \times 10^{-8}$ &  \\
V Mon-A & 1277 & M5.0 & M5III & 2.24 & 0.42 & $0.78 \pm 0.24$ & $4.78 \pm 0.44$ & $(9.84\pm 0.12) \times 10^{-7}$ &  \\
\multirow{3}{*}{R Tri-A$\Bigg\{$} & 1214 & M7.0 & M7III & 3.52 & 0.85 & $0.00 \pm 0.20$ & $4.09 \pm 0.33$ & $(8.16\pm 0.64) \times 10^{-7}$ &  \\
  & 1251 & M5.5 & M5.5III & 8.62 & 0.79 & $0.00 \pm 0.23$ & $4.51 \pm 0.37$ & $(8.73\pm 0.75) \times 10^{-7}$ &  \\
  & 1277 & M5.0 & M5III & 13.95 & 0.43 & $0.00 \pm 0.27$ & $4.67 \pm 0.40$ & $(7.68\pm 0.71) \times 10^{-7}$ & $9.33 \times 10^{-7}$ \\
W Peg-A & 1214 & M8.9 & M9III & 15.10 & 1.69 & $2.96 \pm 0.32$ & $7.32 \pm 1.17$ & $(1.65\pm 0.47) \times 10^{-6}$ & $1.45 \times 10^{-6}$ \\
Z Peg-A & 1214 & M8.5 & M8.5III & 19.11 & 1.75 & $2.39 \pm 0.31$ & $4.18 \pm 0.57$ & $(5.37\pm 0.12) \times 10^{-7}$ & $5.01 \times 10^{-7}$ \\
\end{tabular}
\end{adjustbox}
\end{table}

%\section{Revised zero points for the Lockwood and Wing narrowband filters}

\section{{\tt sedFit} Plots}

The summary spectral energy distribution fitting plots from {\tt sedFit} for each of the stars discussed in \S \ref{sec-sedFitting} are presented in Figures A1-A14.

\makeatletter
\renewcommand{\thefigure}{A\@arabic\c@figure}
\makeatother

\setcounter{figure}{0}

\begin{figure}
\subfigure[IRC+00028-A (M8III)]{\includegraphics[width = 2.35in,angle=270]{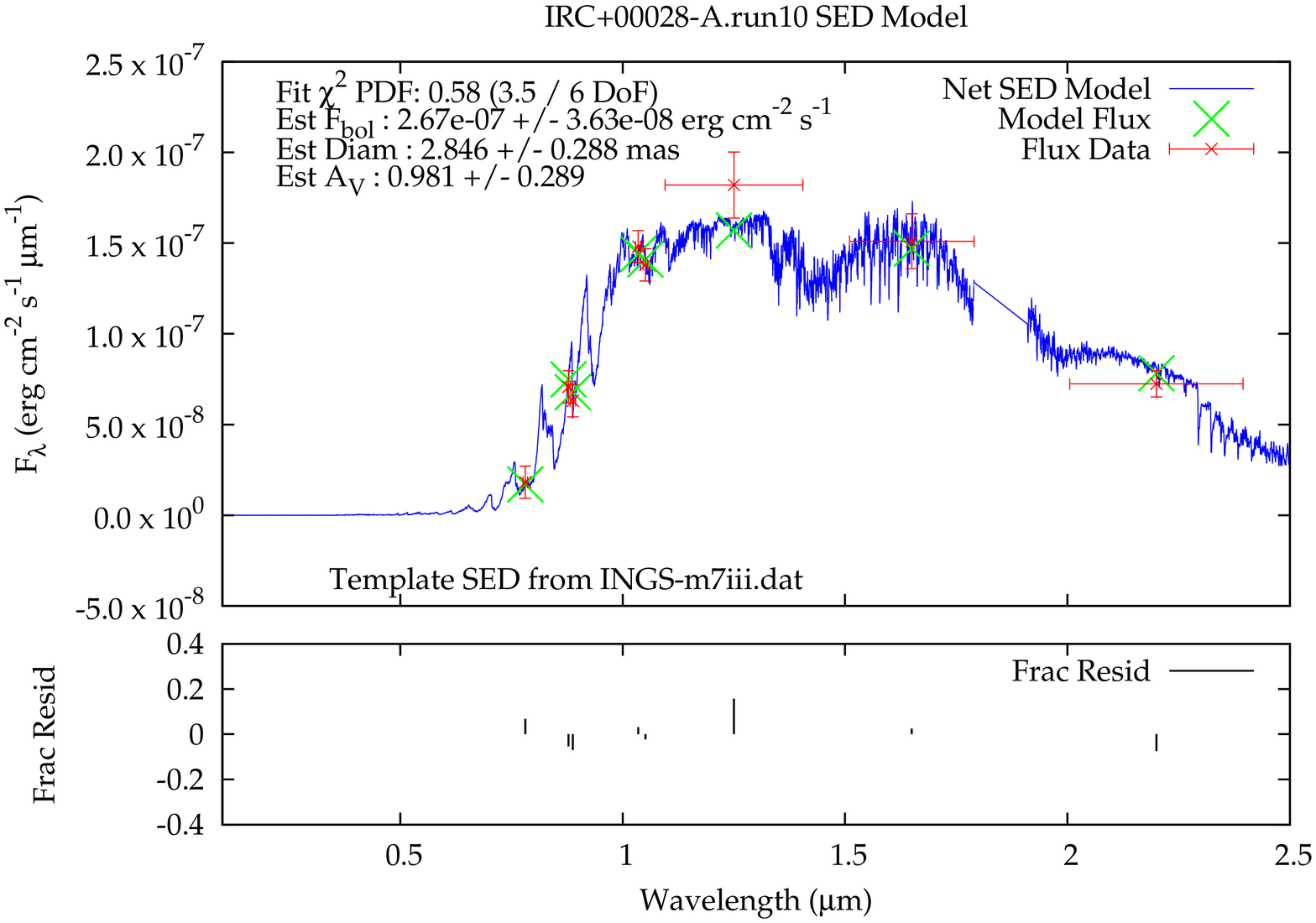}}
\subfigure[IRC+00266-A (M10III)]{\includegraphics[width = 2.35in,angle=270]{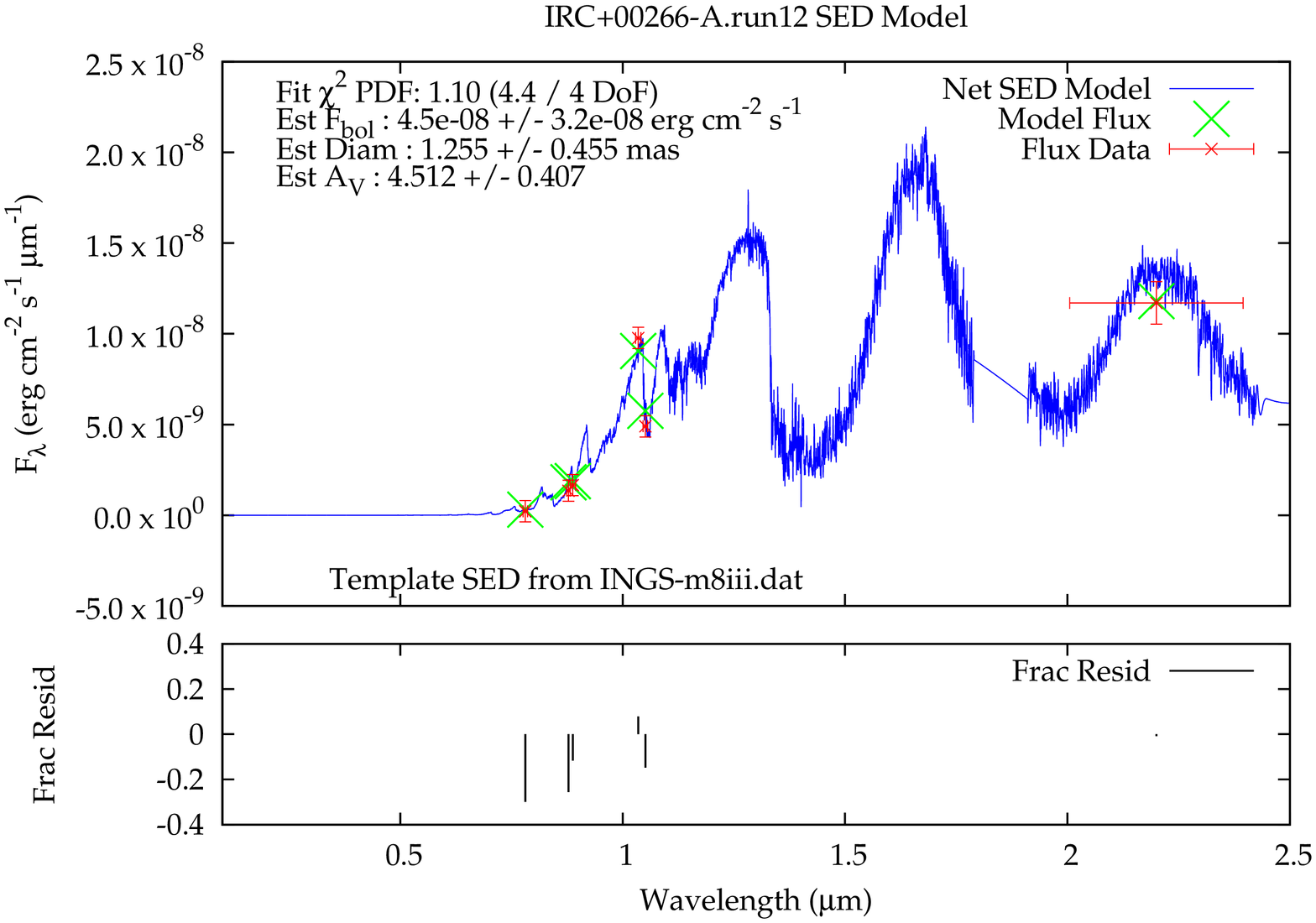}} \\
\subfigure[IRC+00266-B (M9.5III)]{\includegraphics[width = 2.35in,angle=270]{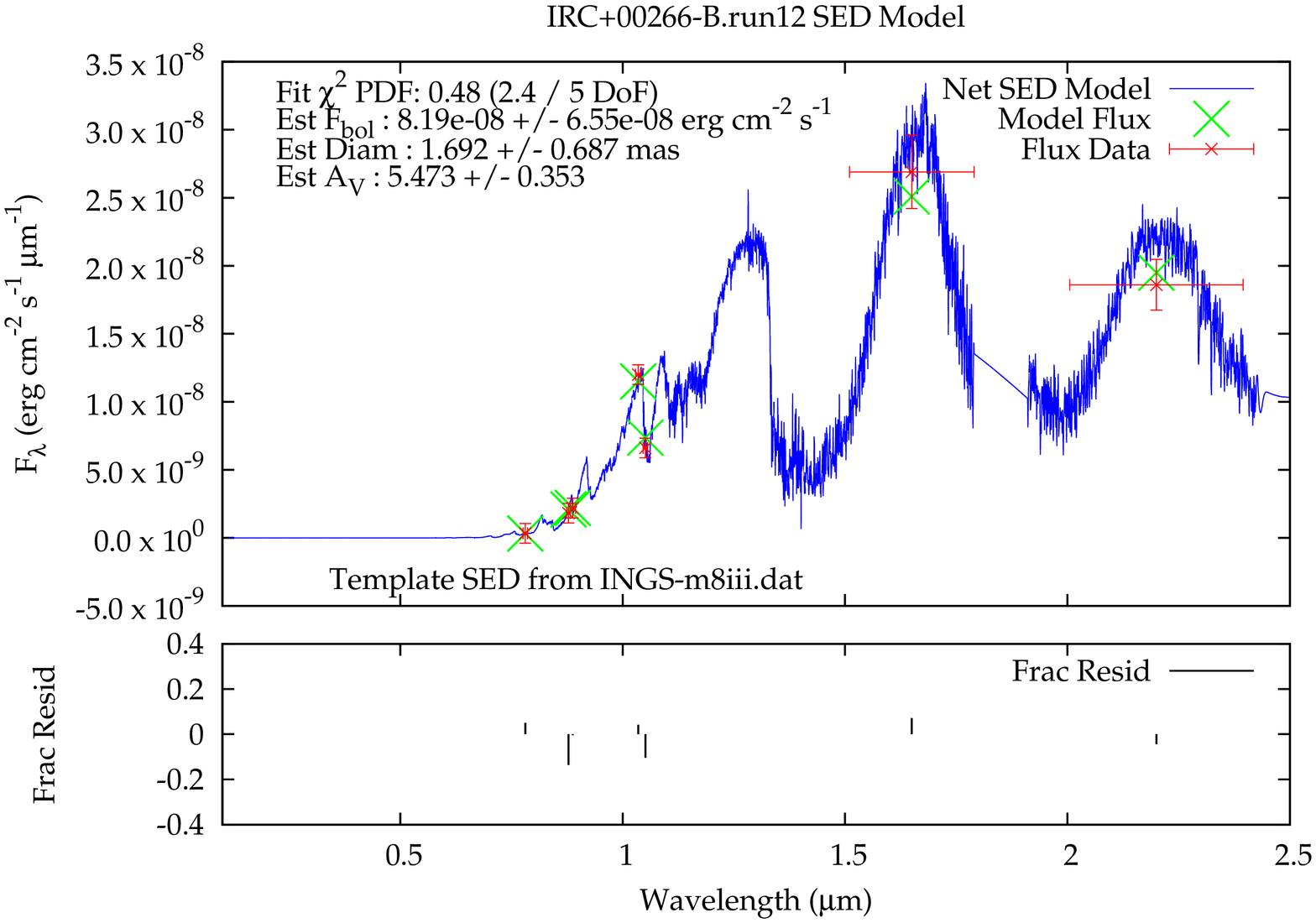}}
\subfigure[IRC+00281-A (M8III)]{\includegraphics[width = 2.35in,angle=270]{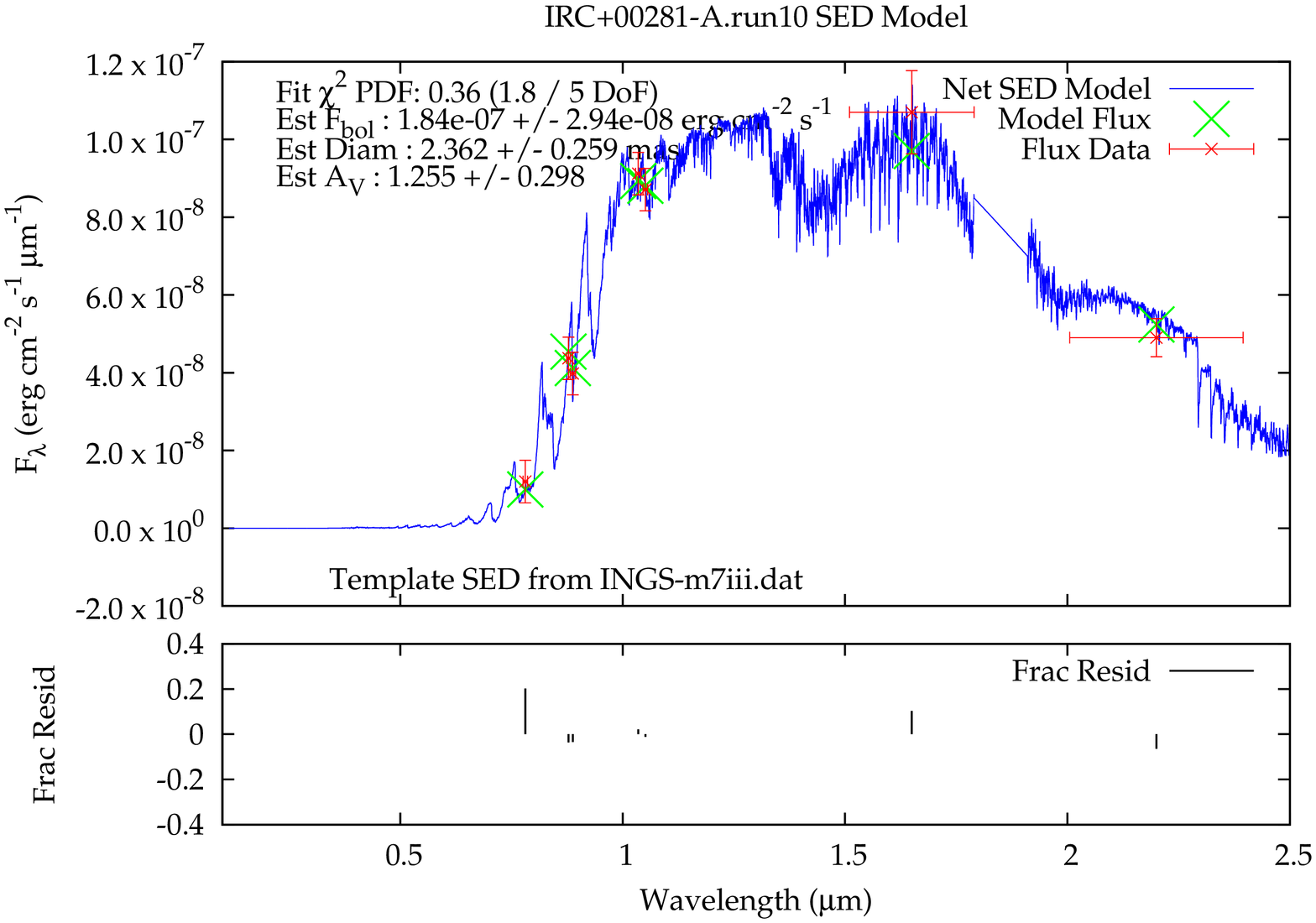}} \\
\subfigure[IRC+10011-A (M8.5III)]{\includegraphics[width = 2.35in,angle=270]{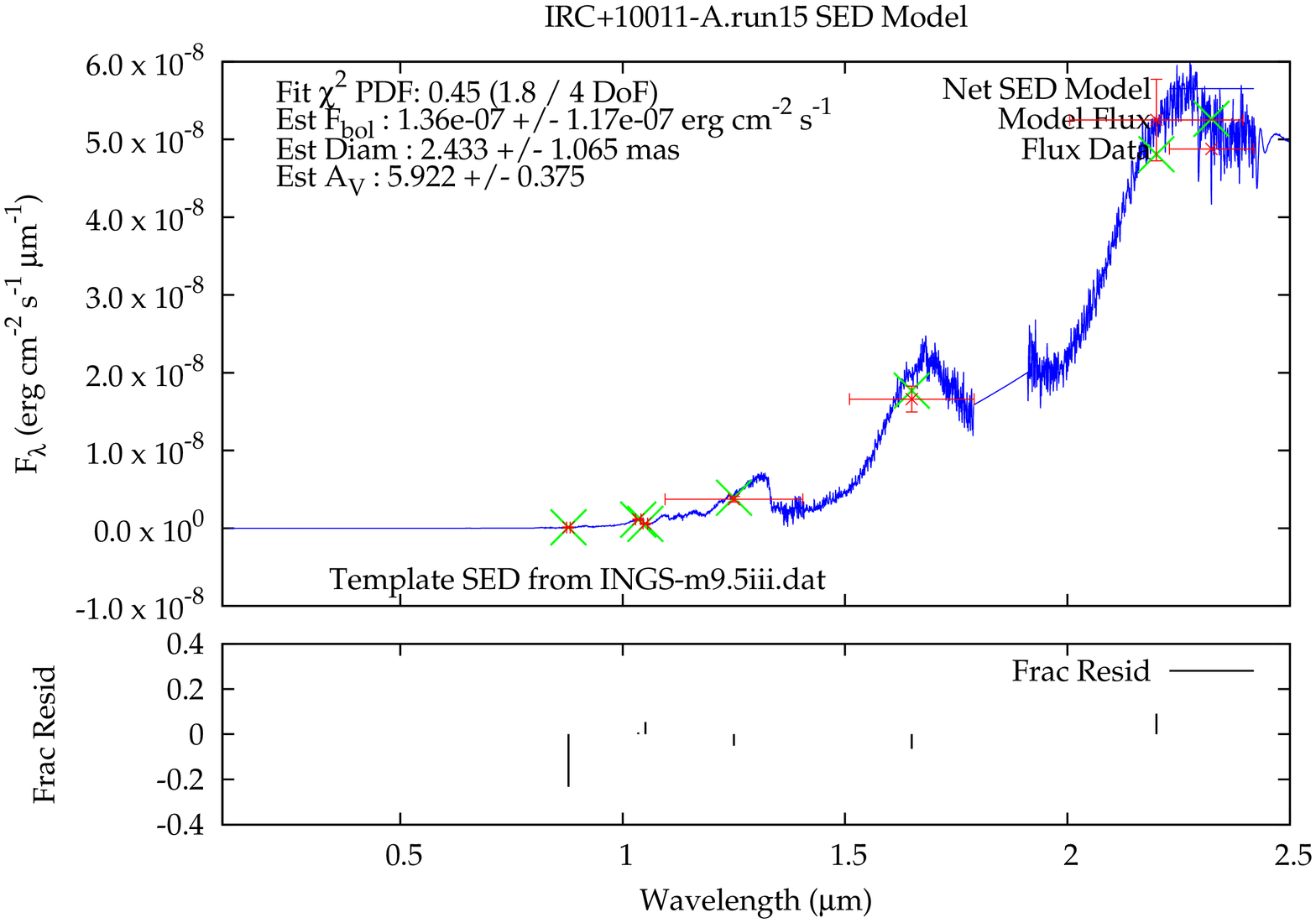}}
\subfigure[IRC+10050-A (M9.5III)]{\includegraphics[width = 2.35in,angle=270]{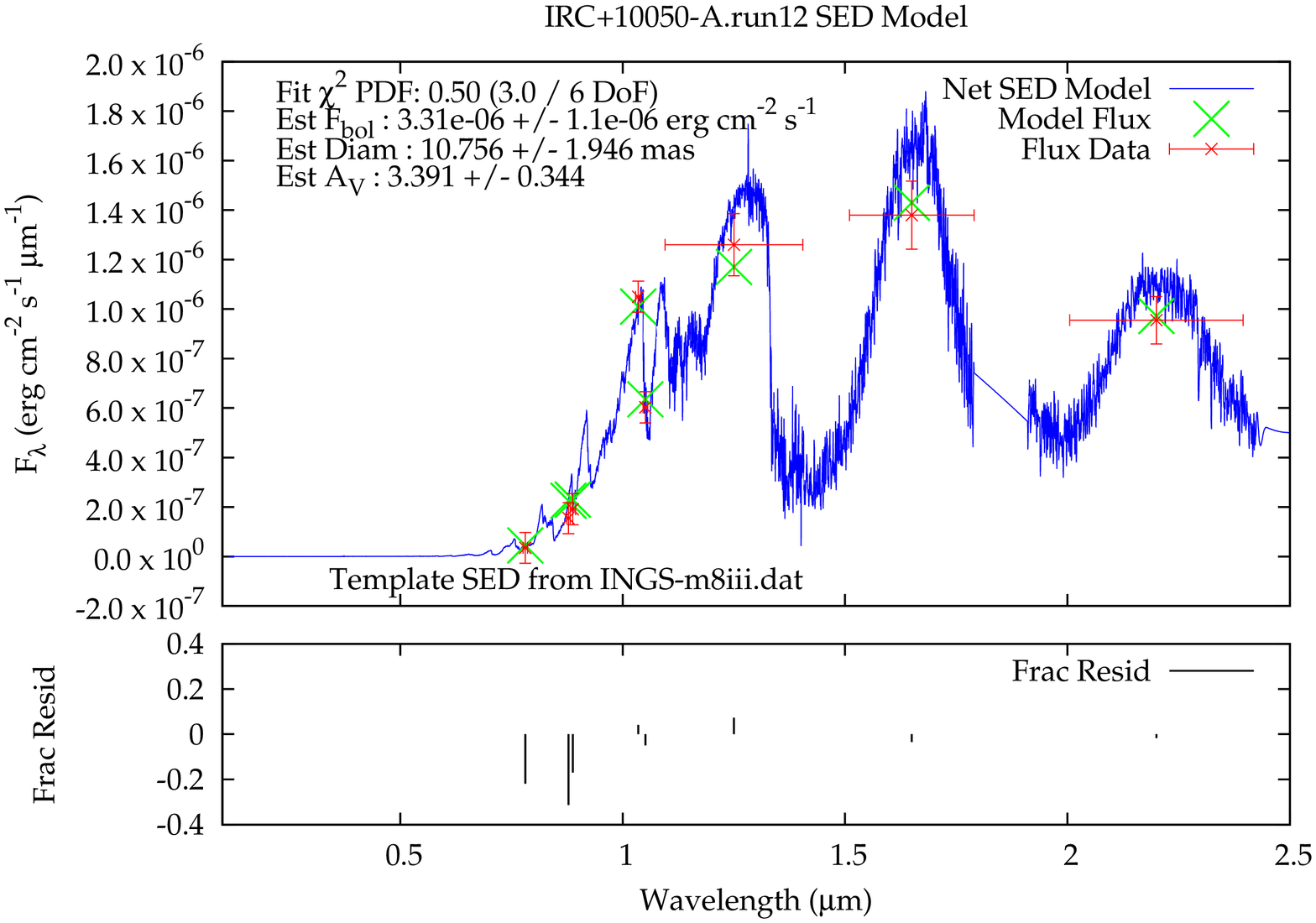}} \\
\caption{SED fits as described in \S 2.2.}
\end{figure}

\begin{figure}
\subfigure[IRC+10313-A (M8.5III)]{\includegraphics[width = 2.35in,angle=270]{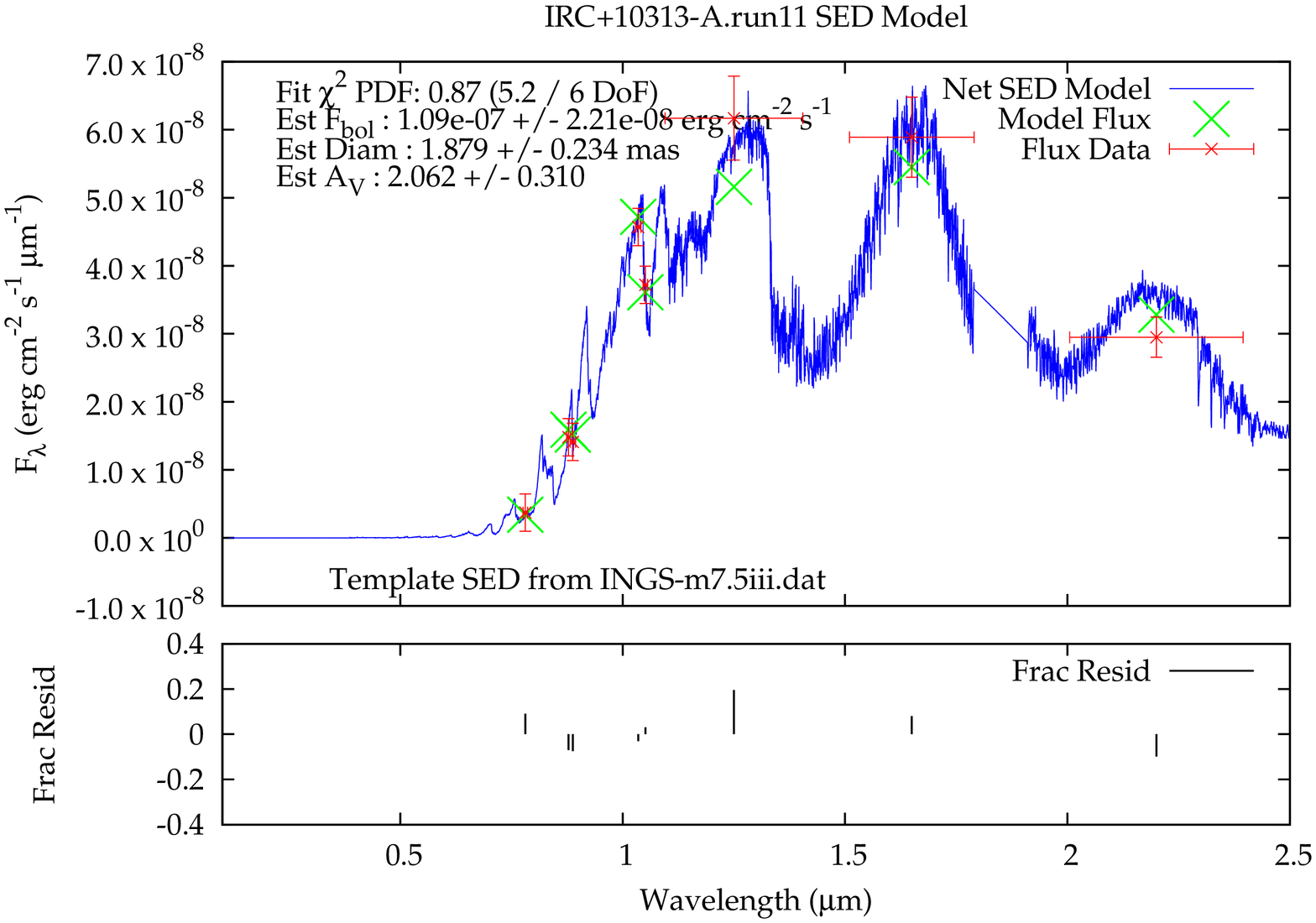}}
\subfigure[IRC+10523-A (M9III)]{\includegraphics[width = 2.35in,angle=270]{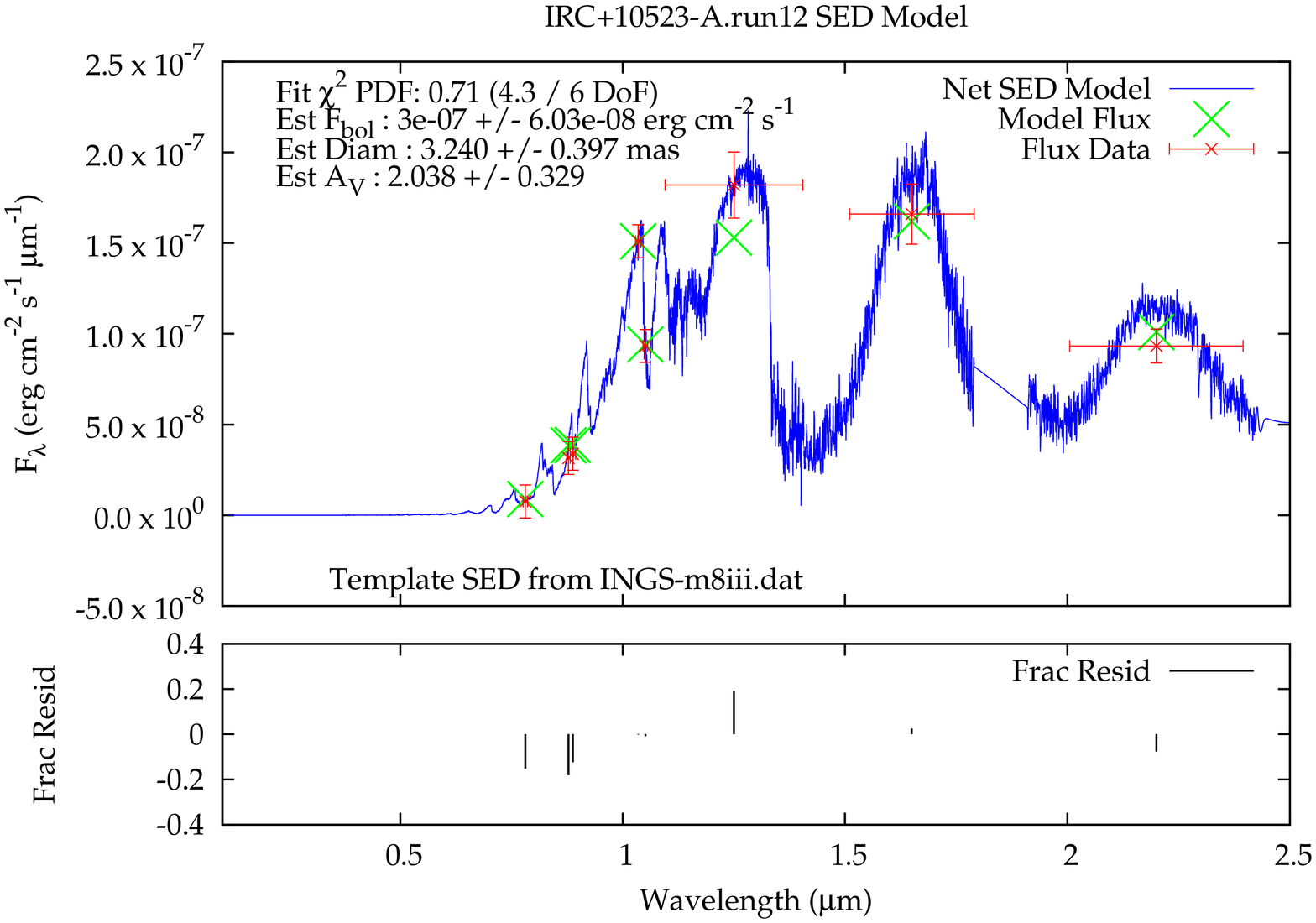}} \\
\subfigure[IRC+10523-B (M9III)]{\includegraphics[width = 2.35in,angle=270]{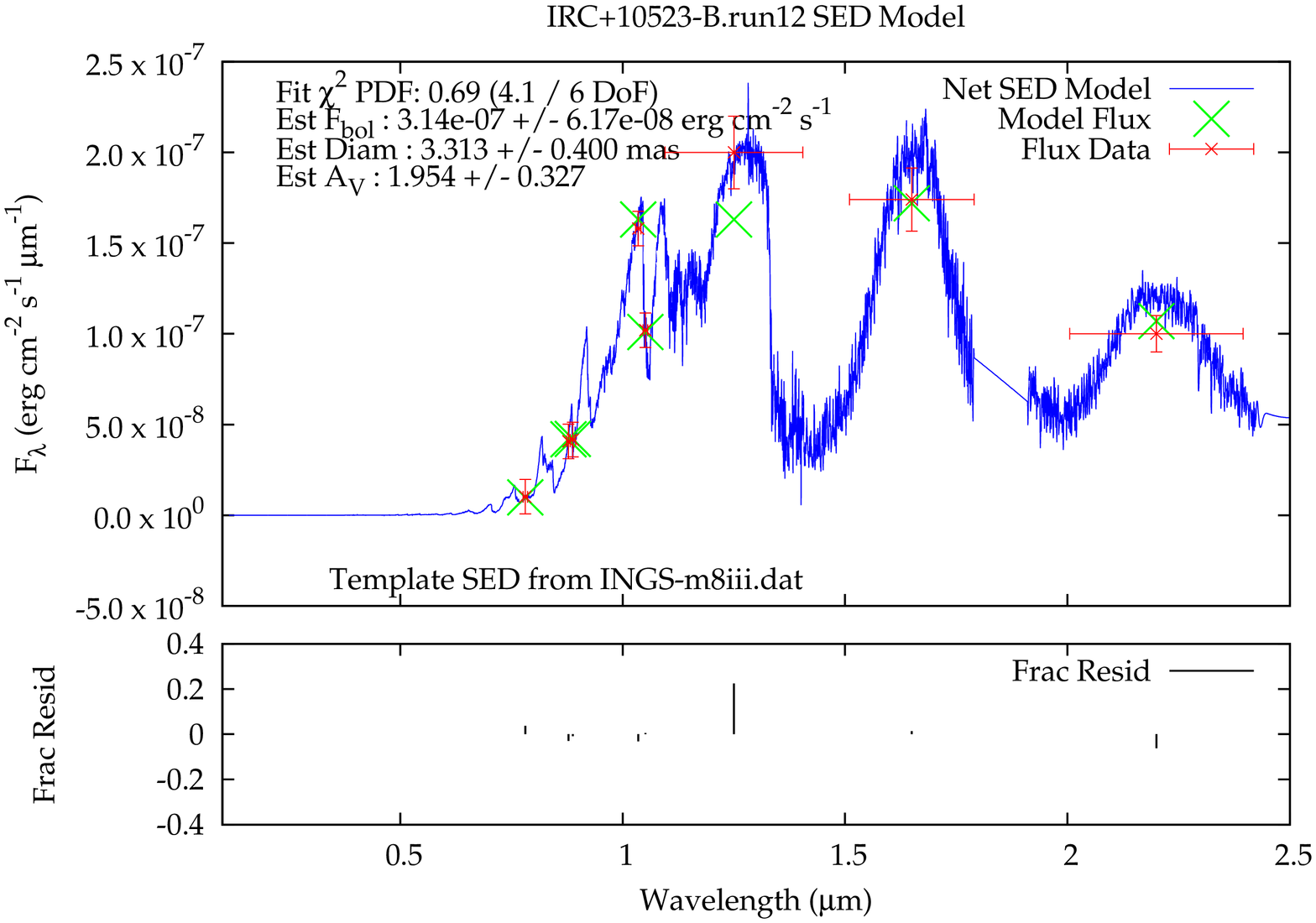}}
\subfigure[IRC+10525-A (M8.5III)]{\includegraphics[width = 2.35in,angle=270]{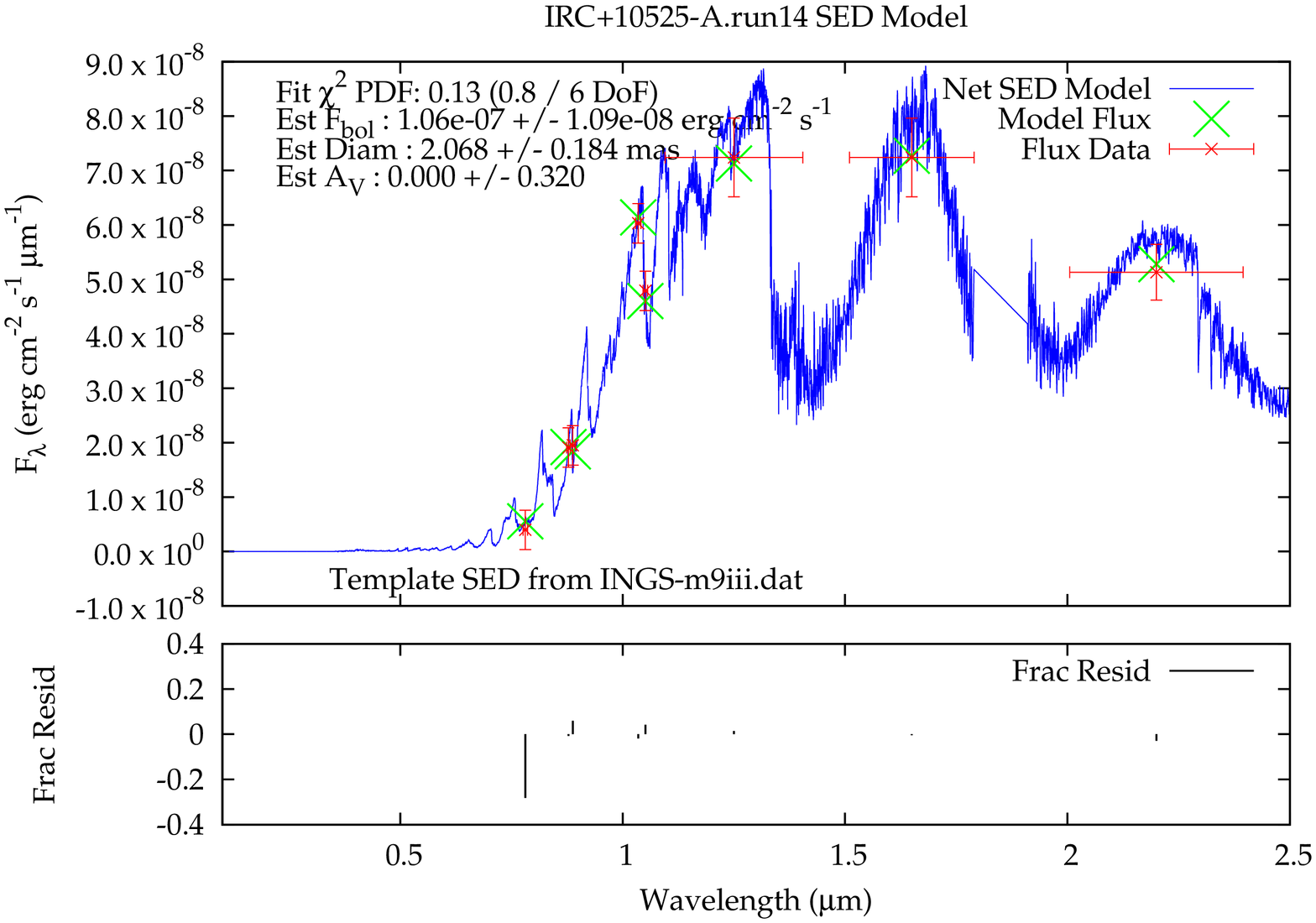}} \\
\subfigure[IRC+20052-A (M8III)]{\includegraphics[width = 2.35in,angle=270]{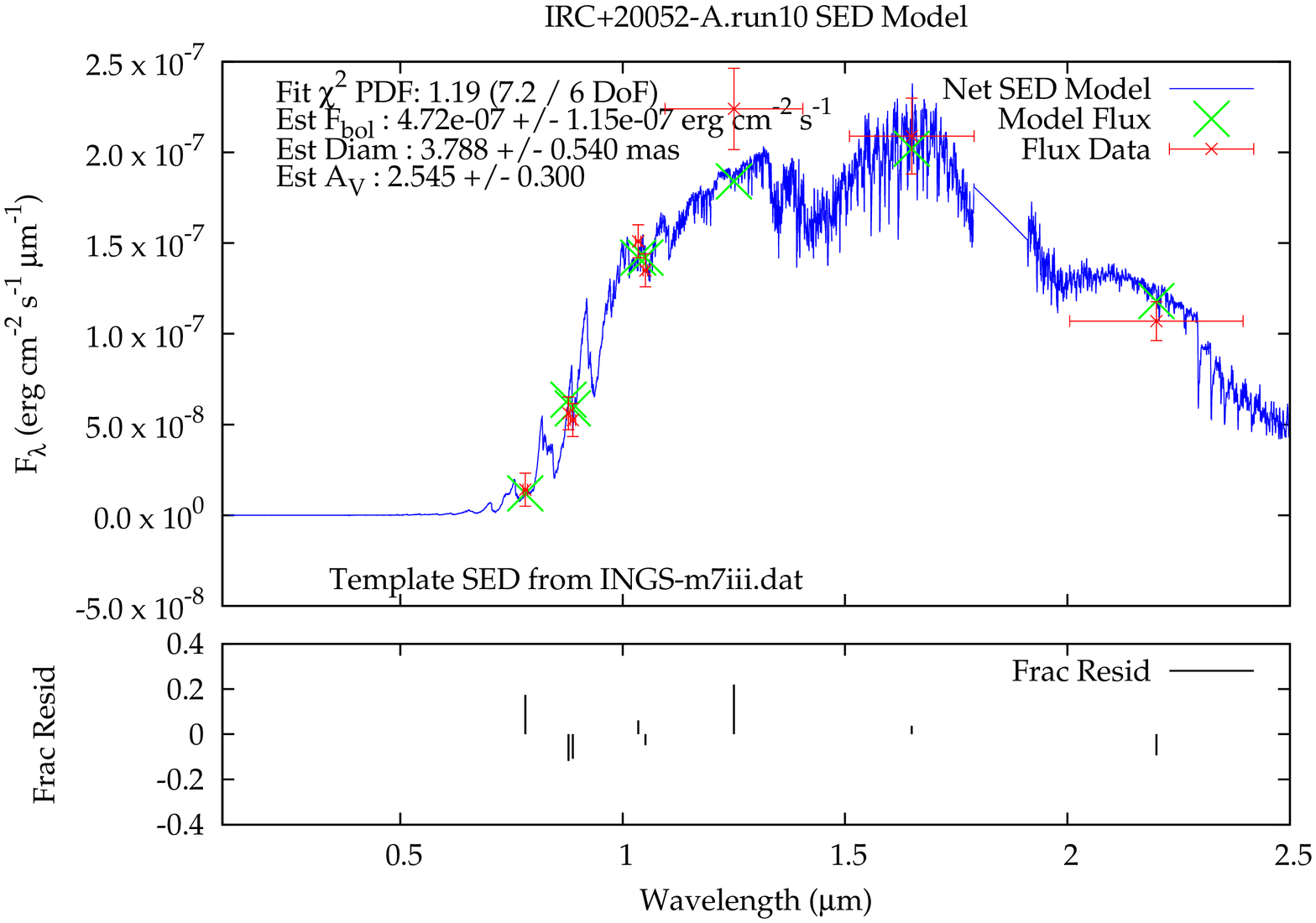}}
\subfigure[IRC+20281-A (M9.5III)]{\includegraphics[width = 2.35in,angle=270]{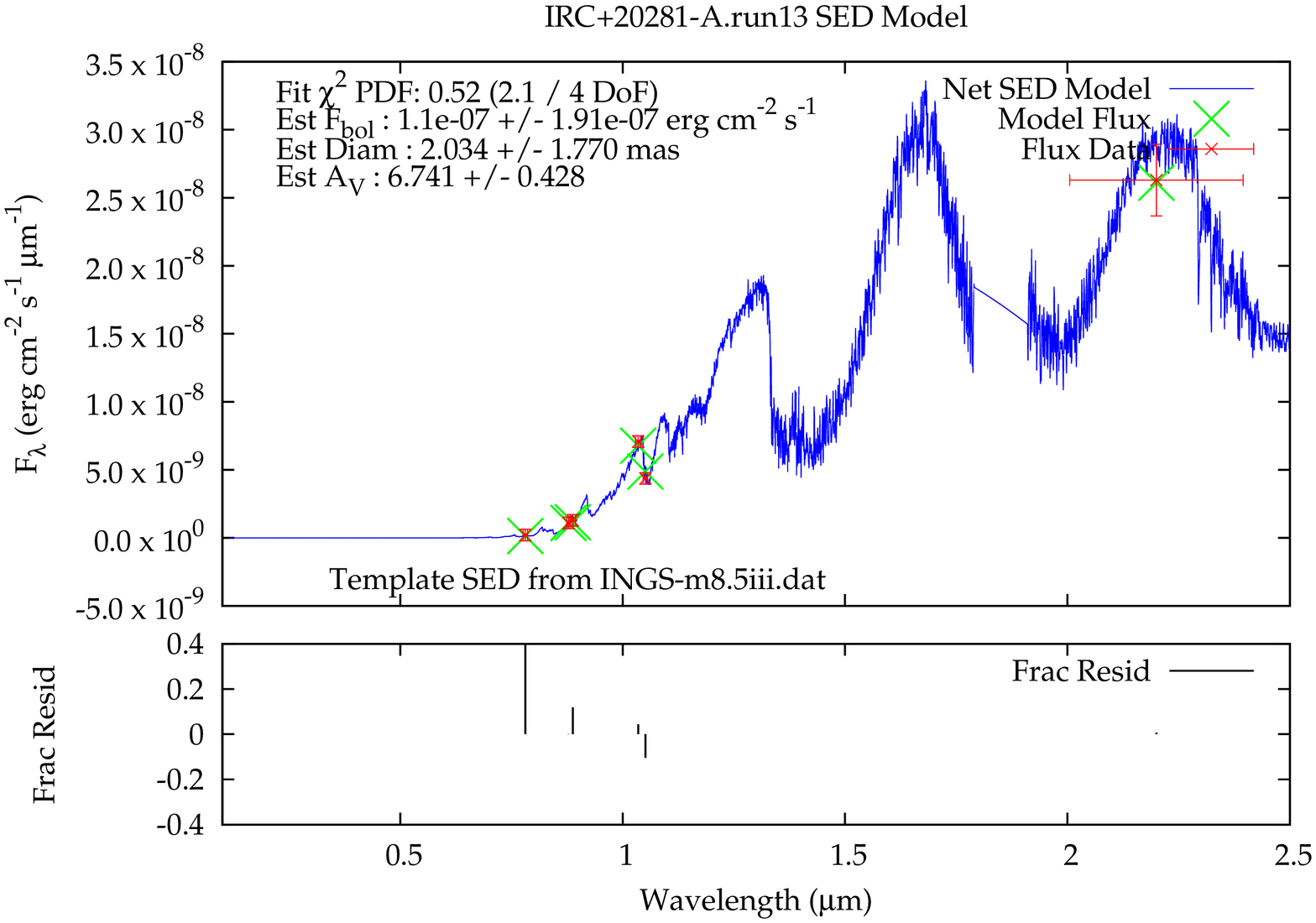}} \\
\caption{SED fits as described in \S 2.2.}
\end{figure}

\begin{figure}
\subfigure[IRC+20328-A (M10III)]{\includegraphics[width = 2.35in,angle=270]{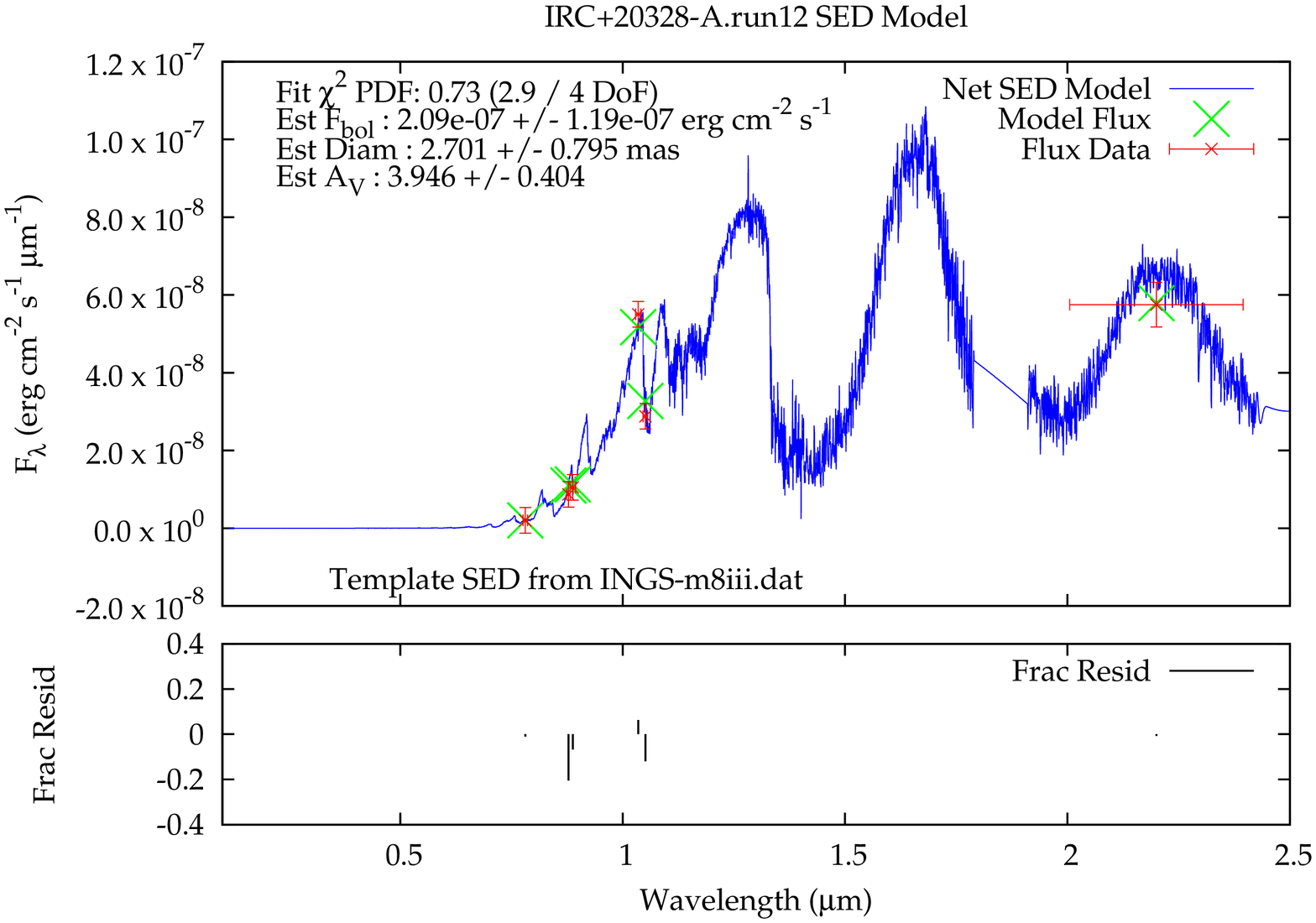}}
\subfigure[IRC+30021-A (M9.5III)]{\includegraphics[width = 2.35in,angle=270]{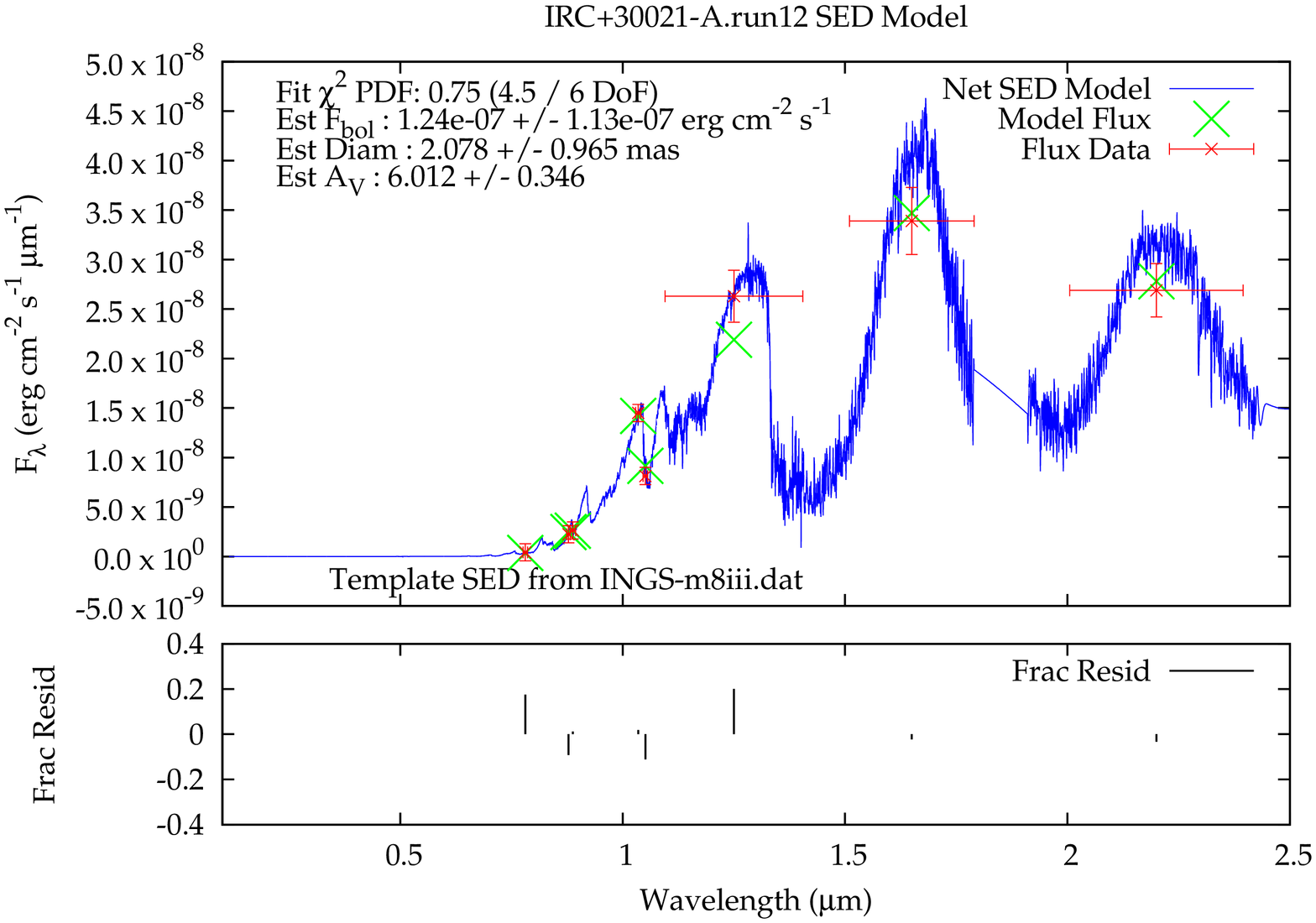}} \\
\subfigure[IRC+30021-B (M9III)]{\includegraphics[width = 2.35in,angle=270]{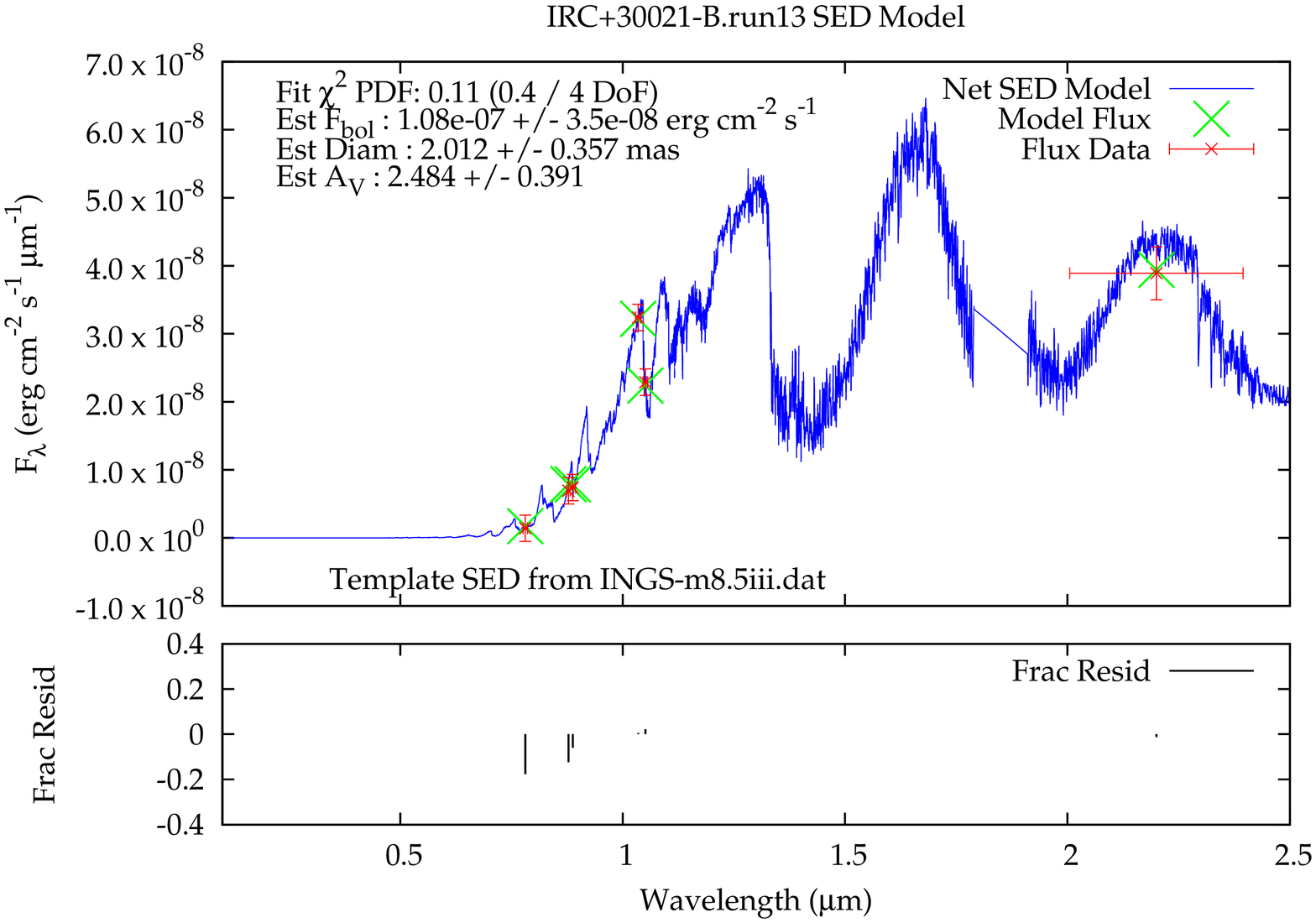}}
\subfigure[IRC+30021-C (M9.5III)]{\includegraphics[width = 2.35in,angle=270]{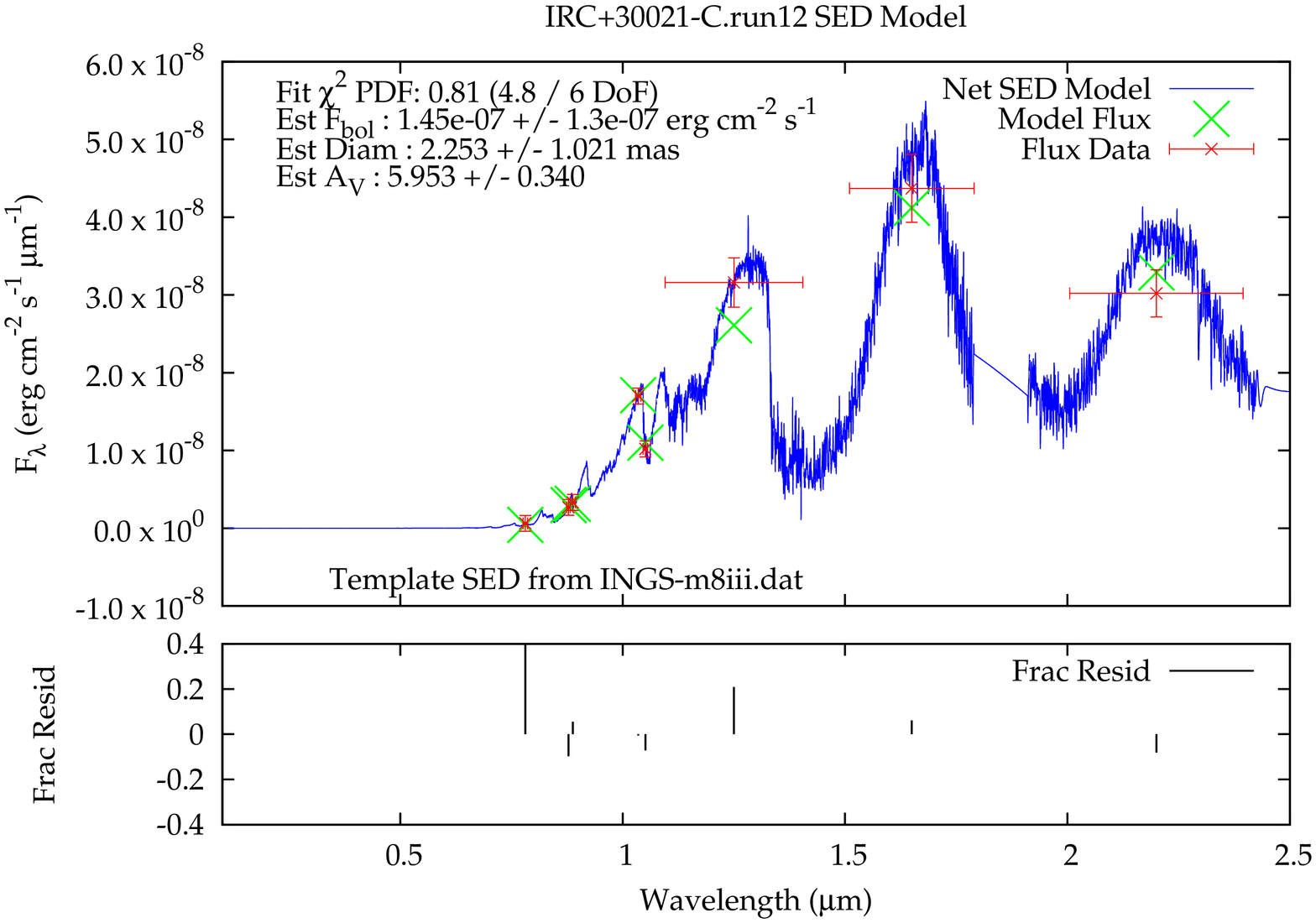}} \\
\subfigure[IRC+30021-D (M9.5III)]{\includegraphics[width = 2.35in,angle=270]{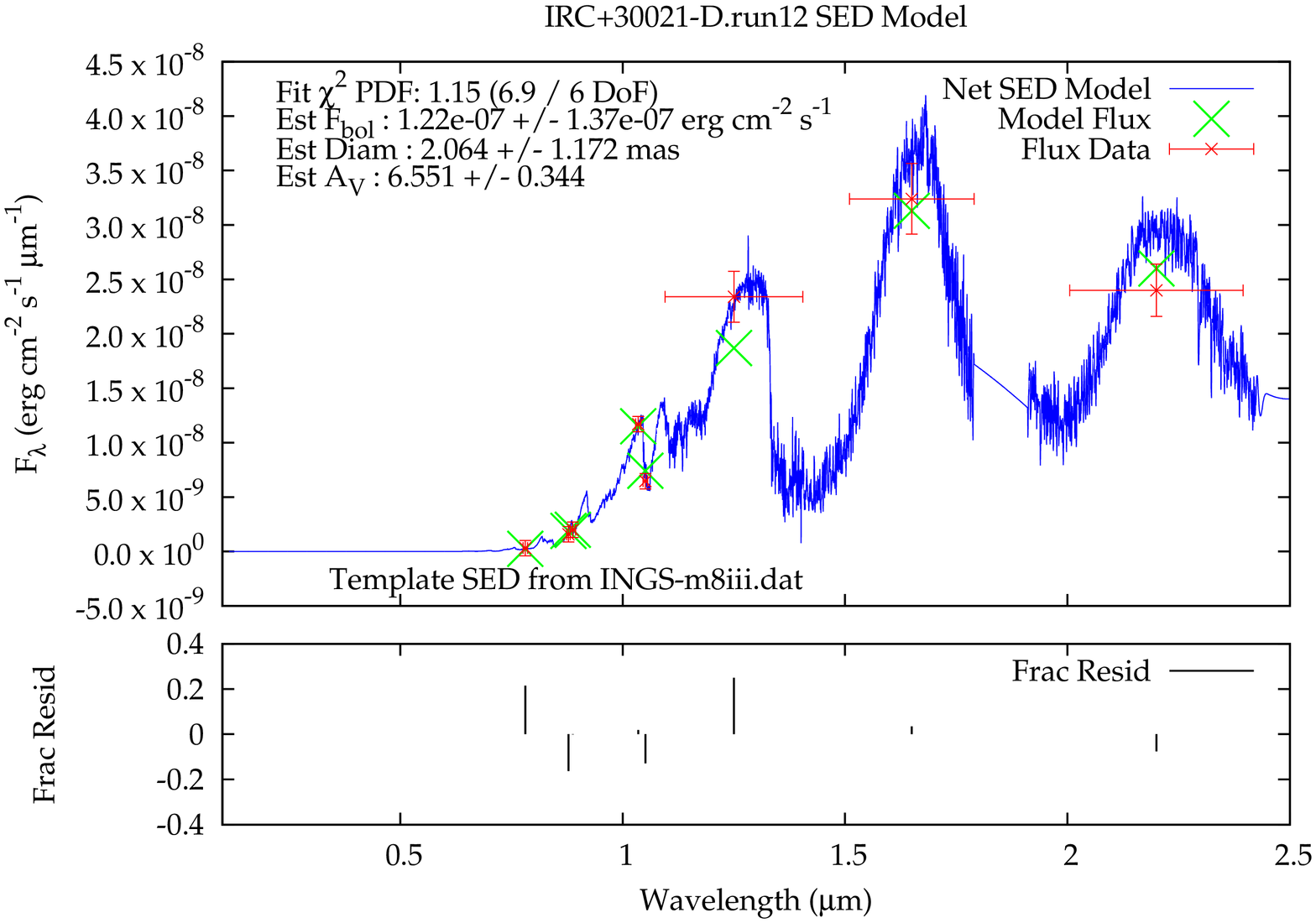}}
\subfigure[IRC+30055-A (M8III)]{\includegraphics[width = 2.35in,angle=270]{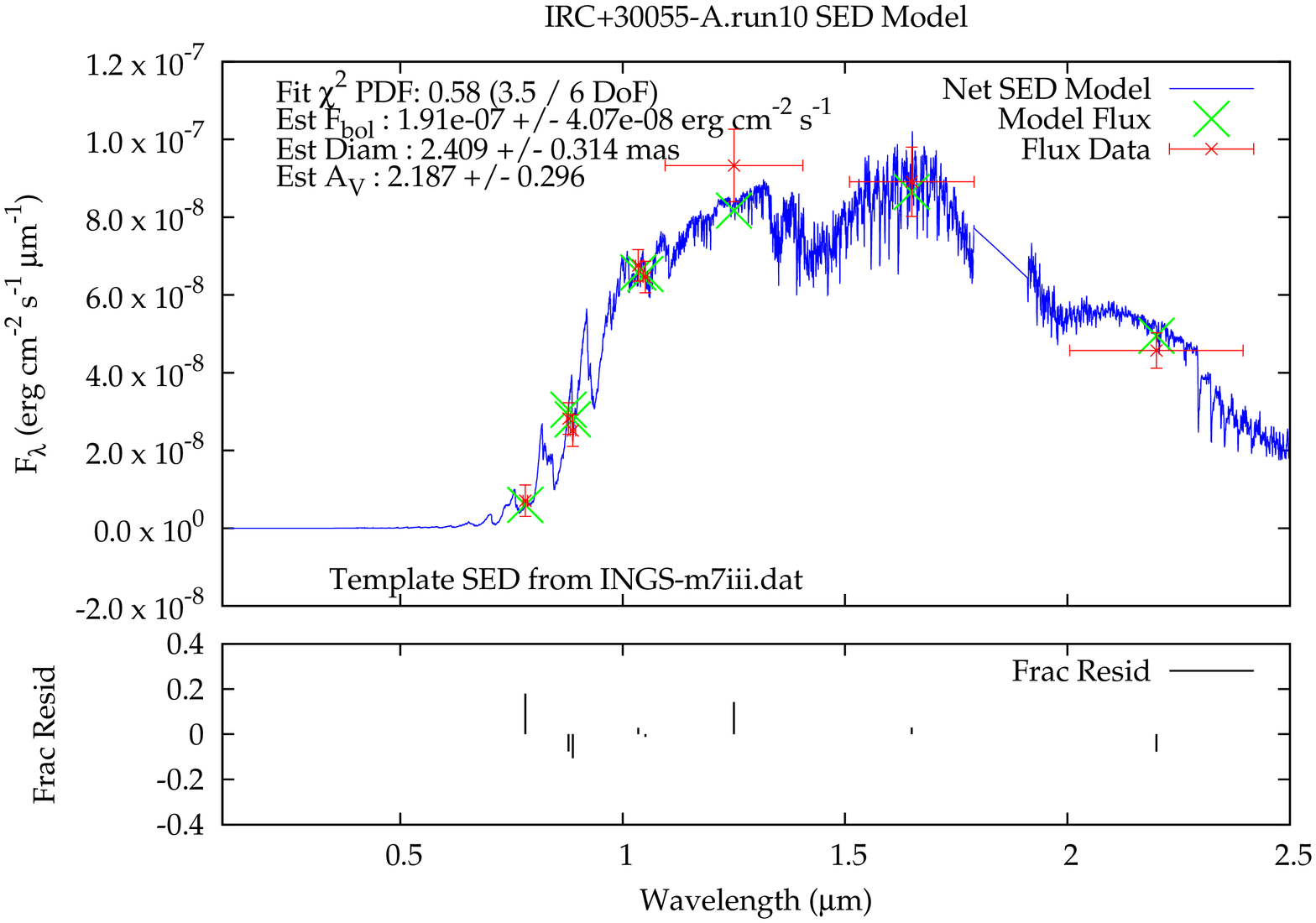}} \\
\caption{SED fits as described in \S 2.2.}
\end{figure}

\begin{figure}
\subfigure[IRC+30292-A (M8.5III)]{\includegraphics[width = 2.35in,angle=270]{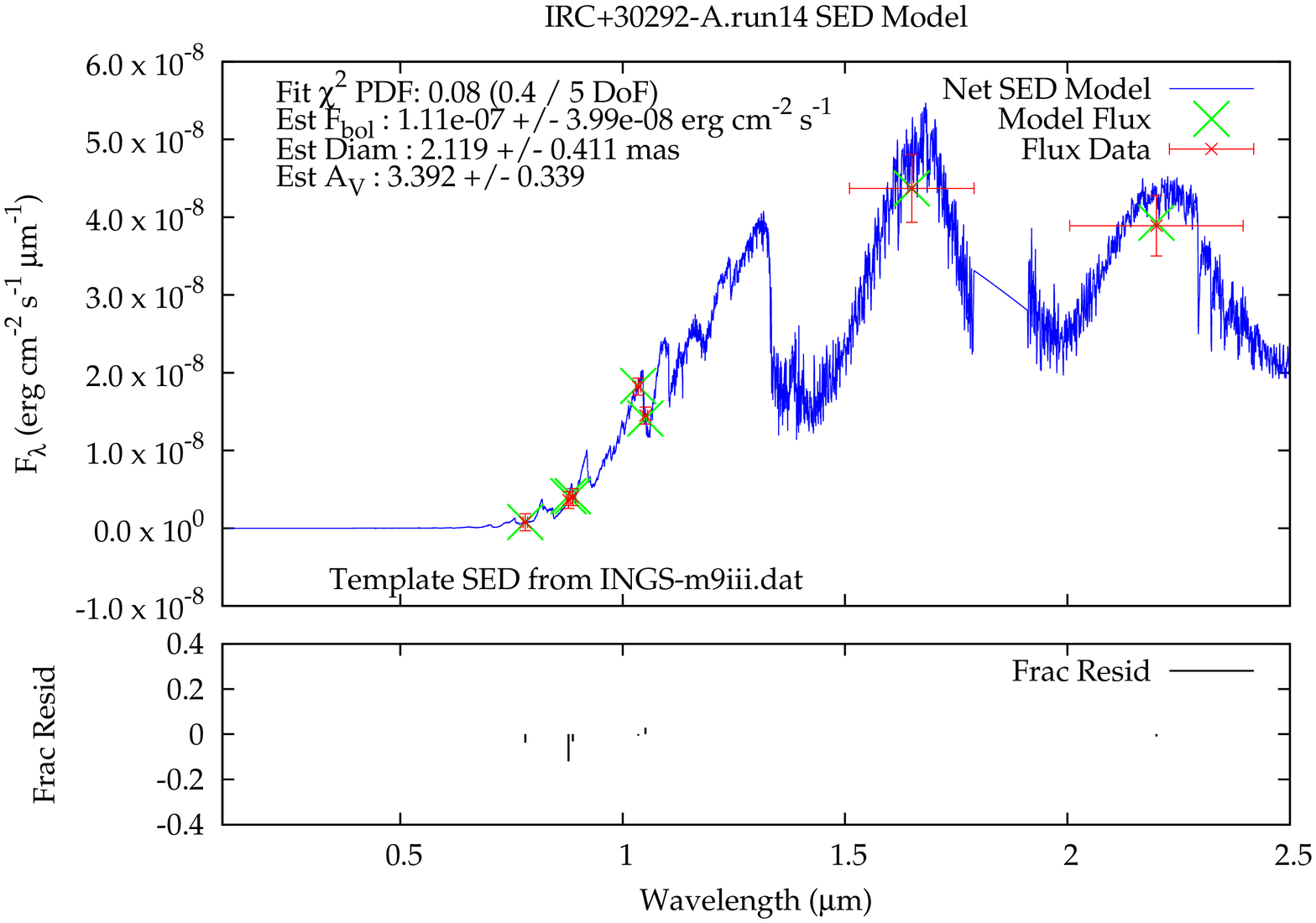}}
\subfigure[IRC+30292-B (M9.5III)]{\includegraphics[width = 2.35in,angle=270]{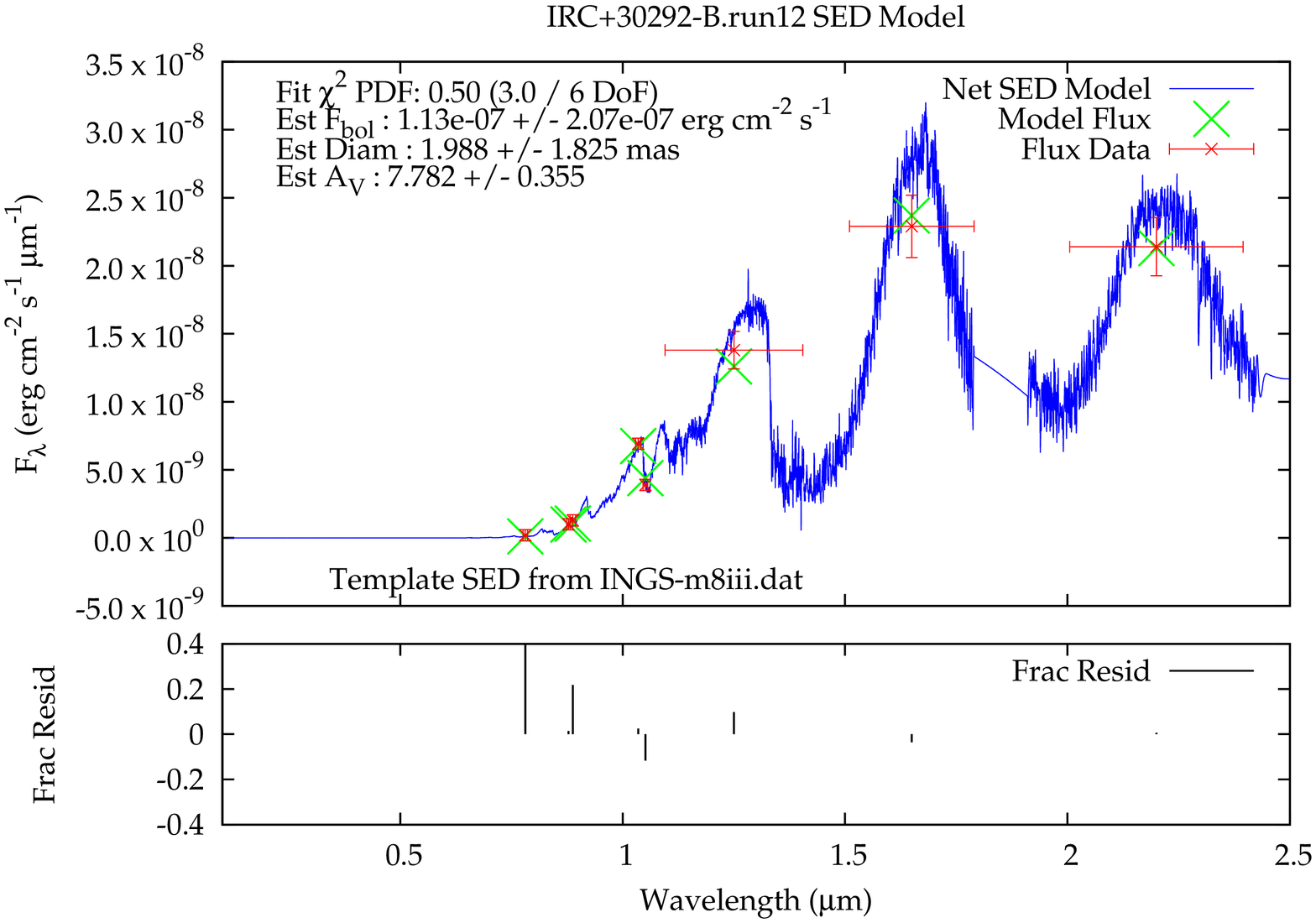}} \\
\subfigure[IRC+30515-A (M8III)]{\includegraphics[width = 2.35in,angle=270]{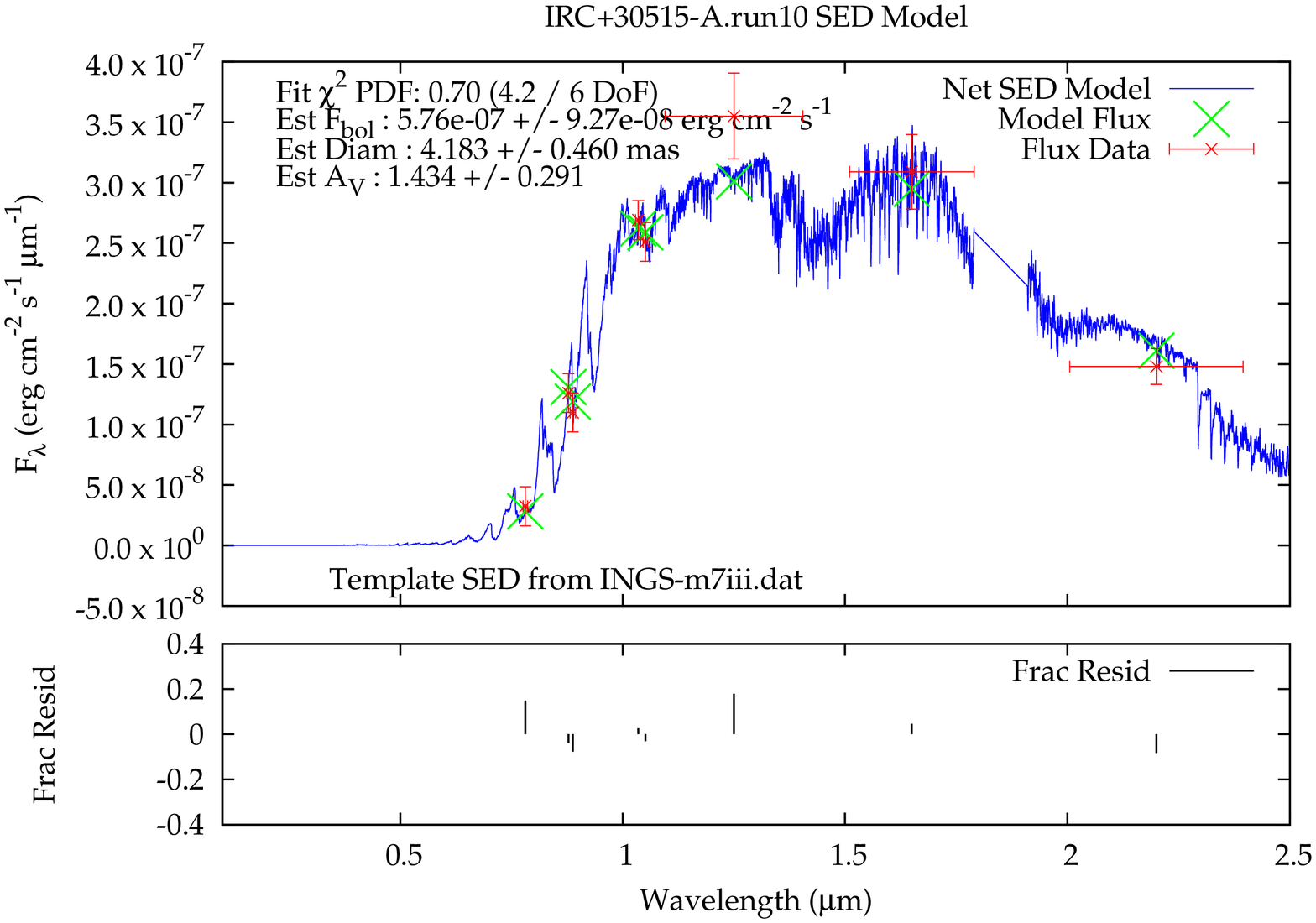}}
\subfigure[IRC+40004-A (M9III)]{\includegraphics[width = 2.35in,angle=270]{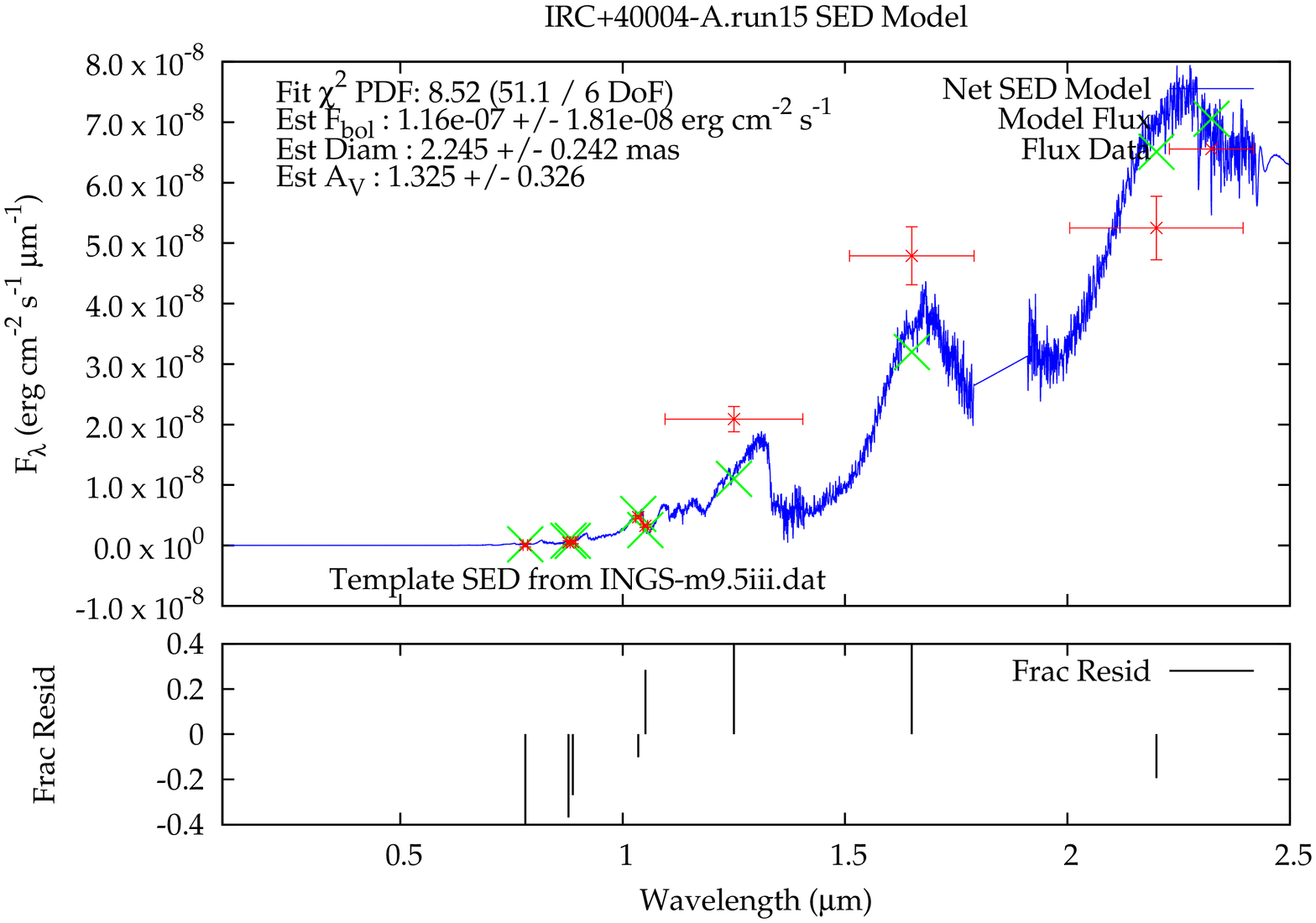}} \\
\subfigure[IRC+40004-B (M9III)]{\includegraphics[width = 2.35in,angle=270]{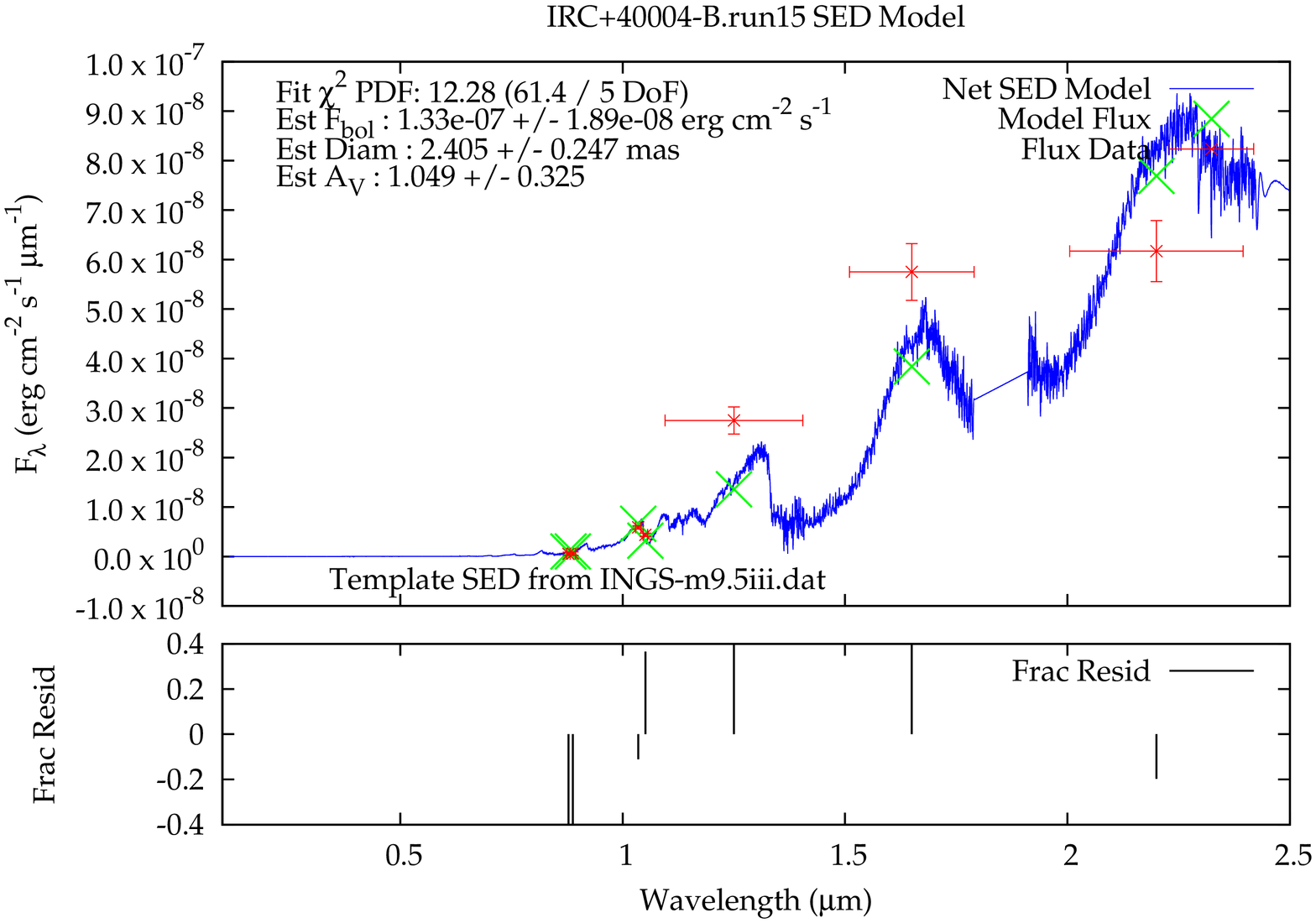}}
\subfigure[IRC+40442-A (M9.5III)]{\includegraphics[width = 2.35in,angle=270]{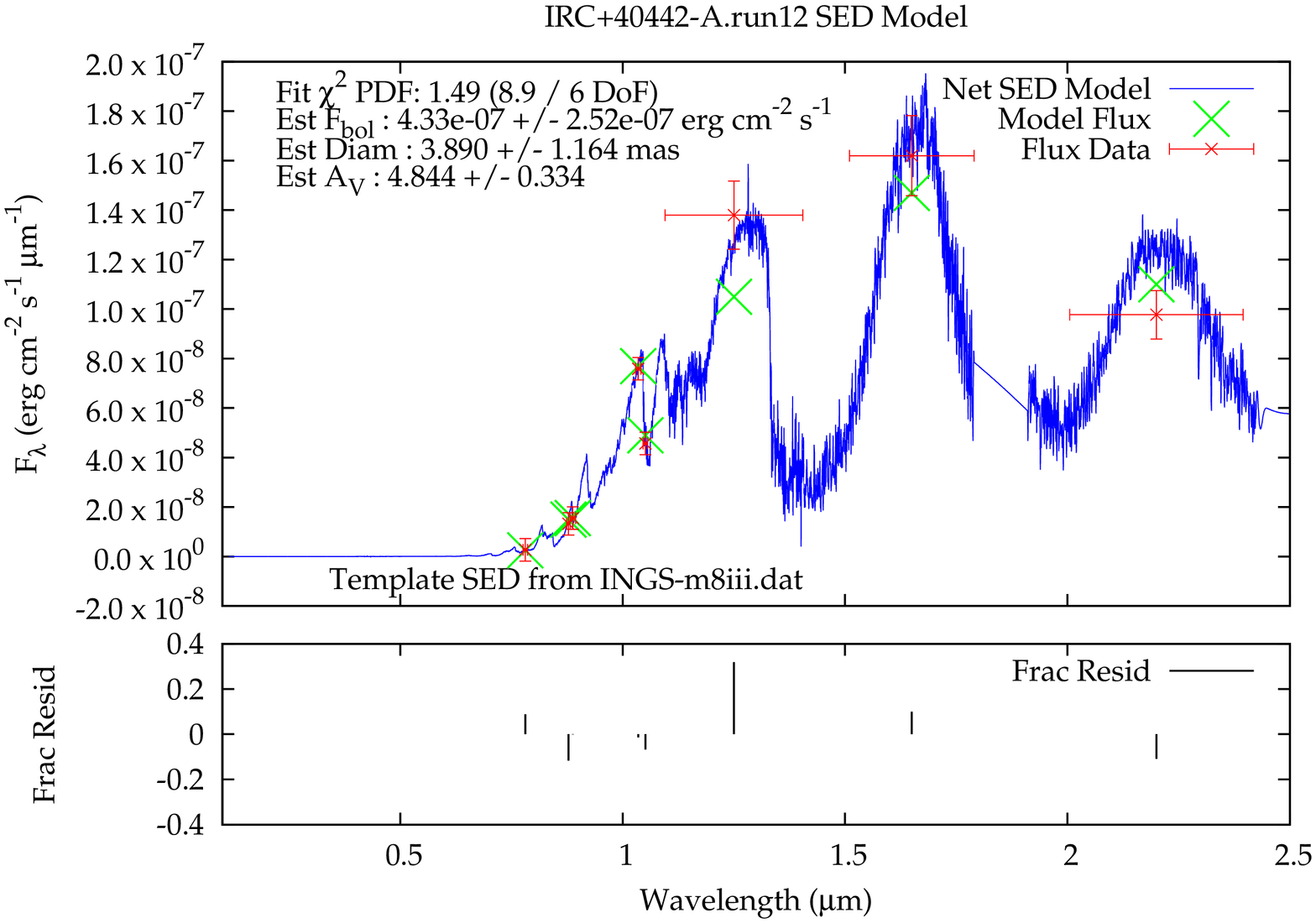}} \\
\caption{SED fits as described in \S 2.2.}
\end{figure}

\begin{figure}
\subfigure[IRC+40448-A (M7III)]{\includegraphics[width = 2.35in,angle=270]{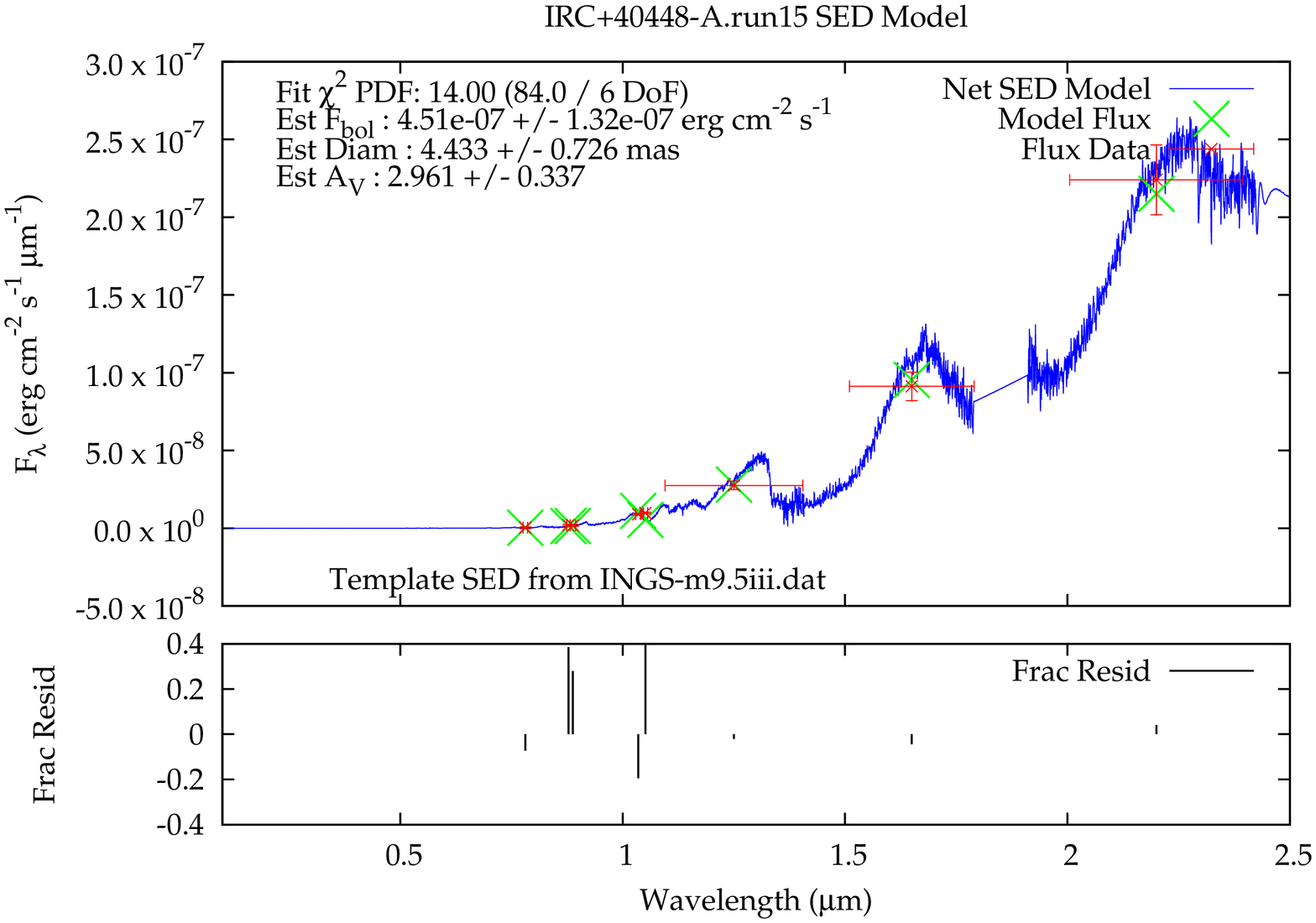}}
\subfigure[IRC+40485-A (M10III)]{\includegraphics[width = 2.35in,angle=270]{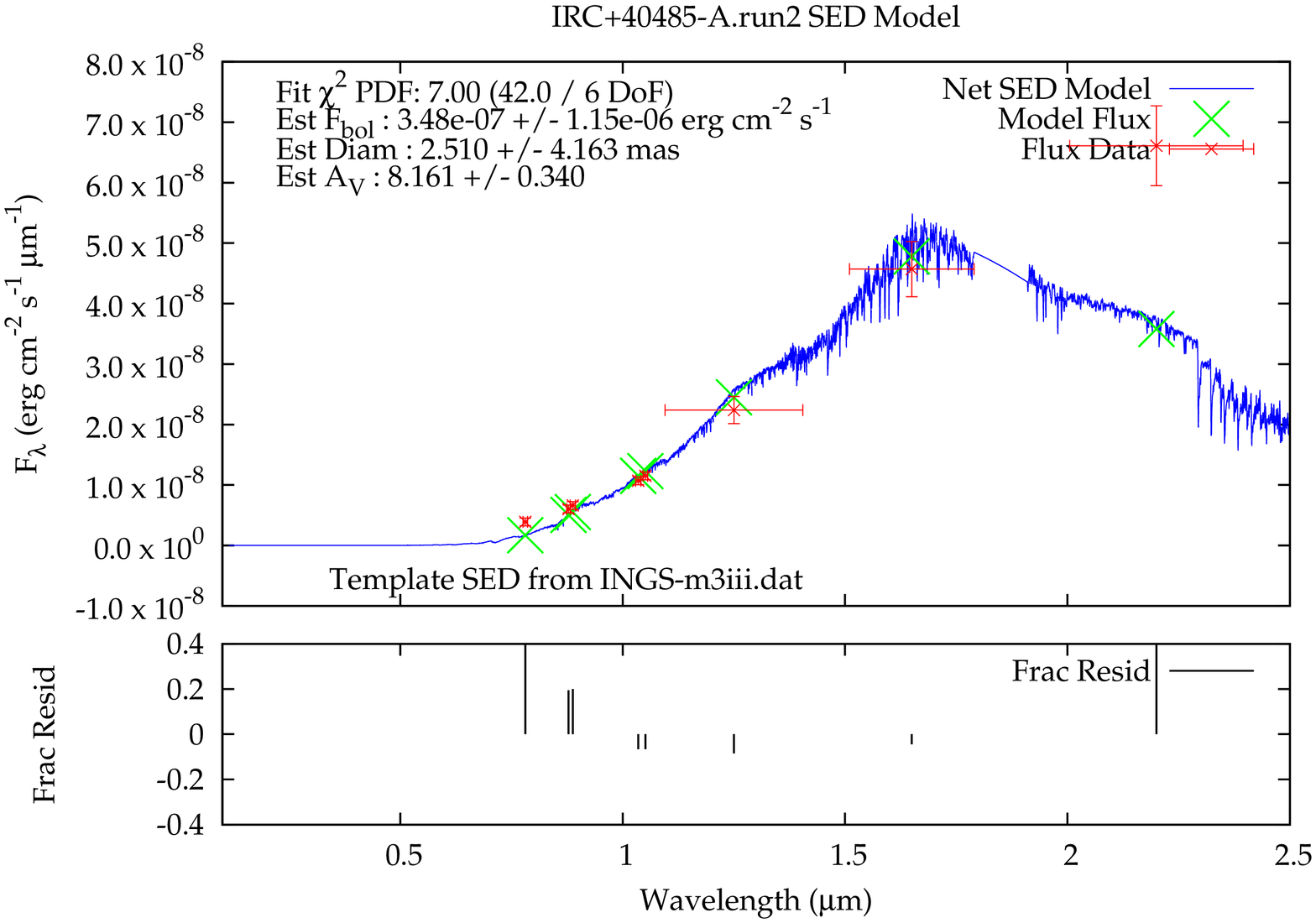}} \\
\subfigure[IRC+40485-B (M10III)]{\includegraphics[width = 2.35in,angle=270]{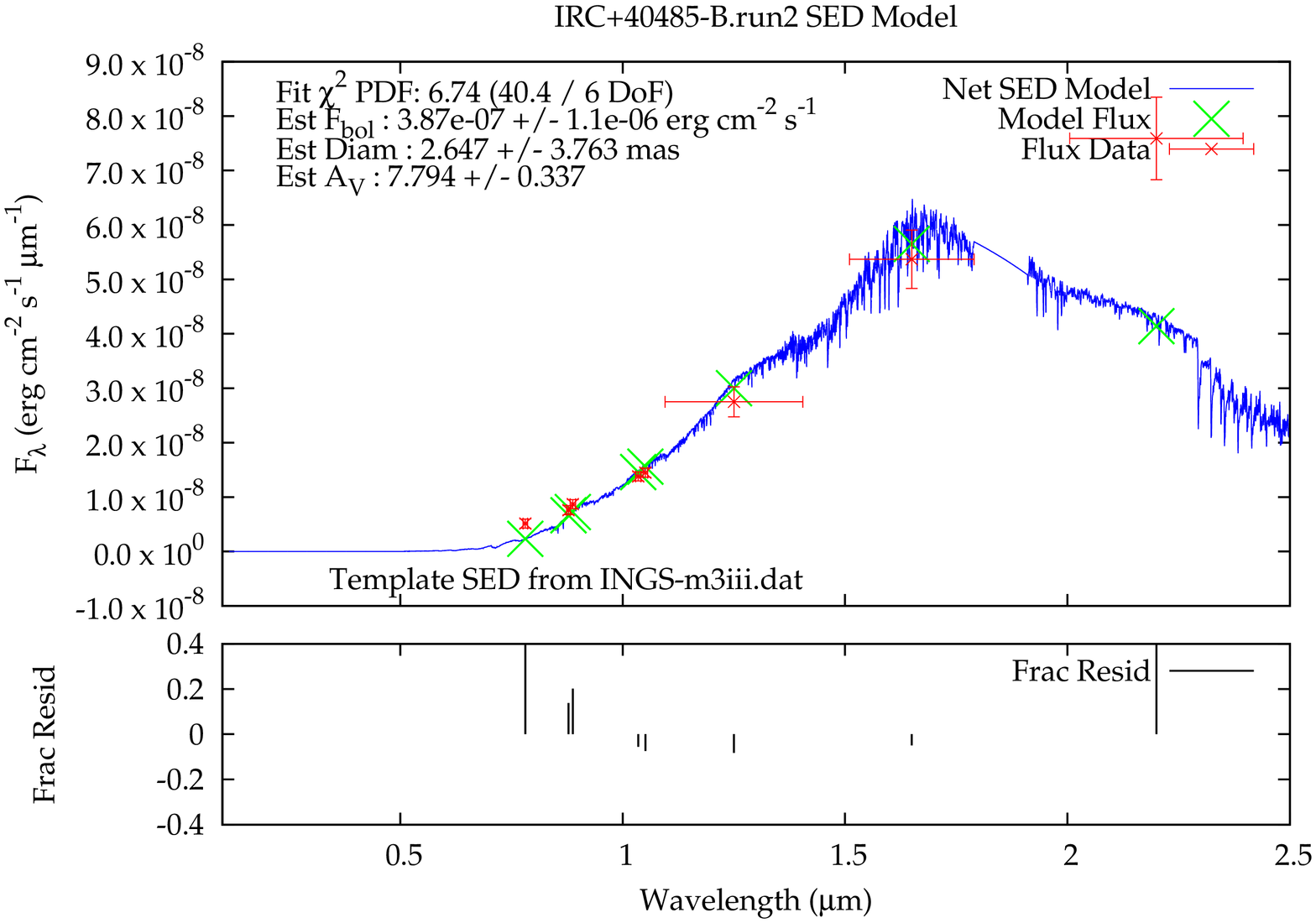}}
\subfigure[IRC+50096-A (M10III)]{\includegraphics[width = 2.35in,angle=270]{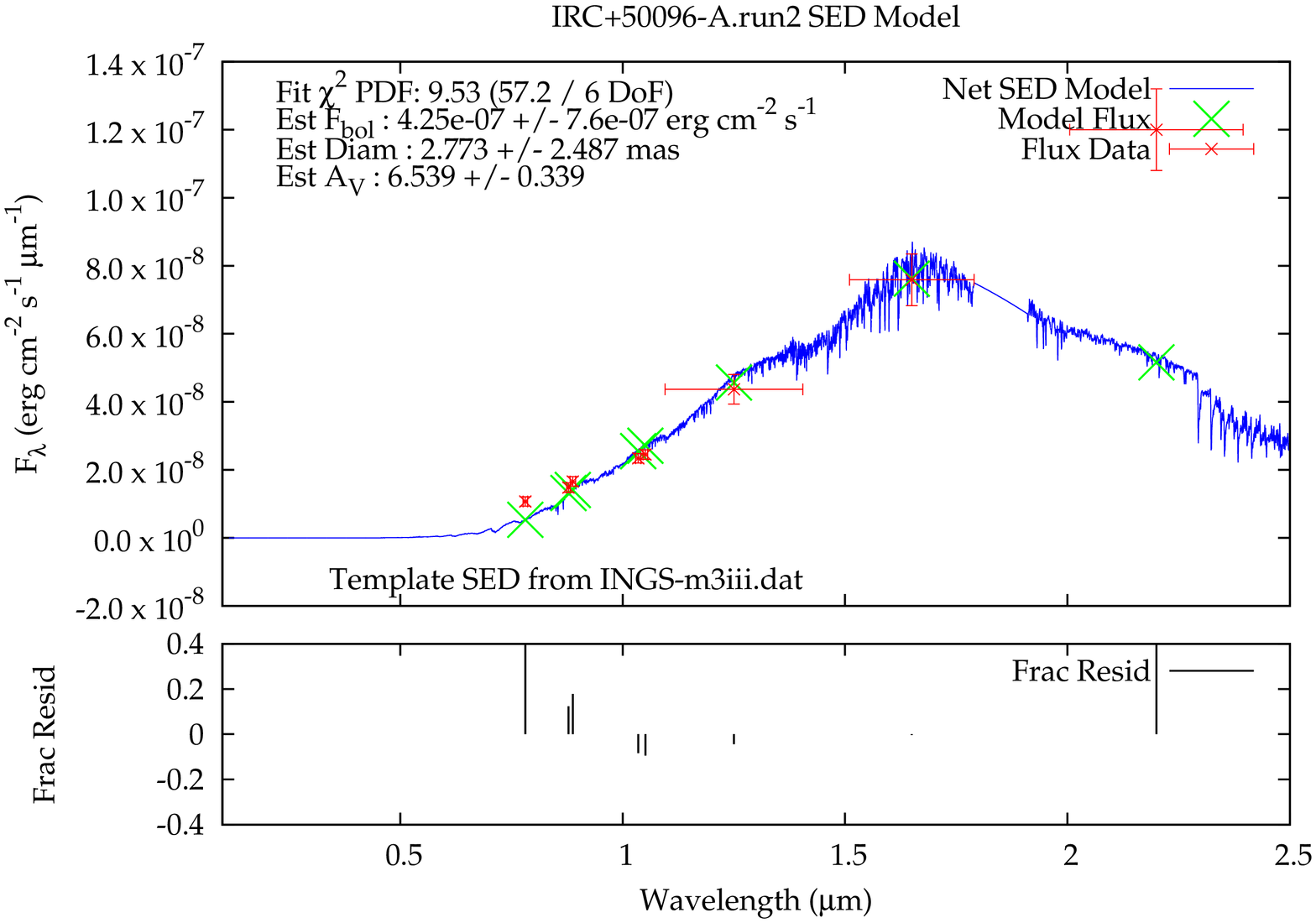}} \\
\subfigure[IRC+50096-B (M10III)]{\includegraphics[width = 2.35in,angle=270]{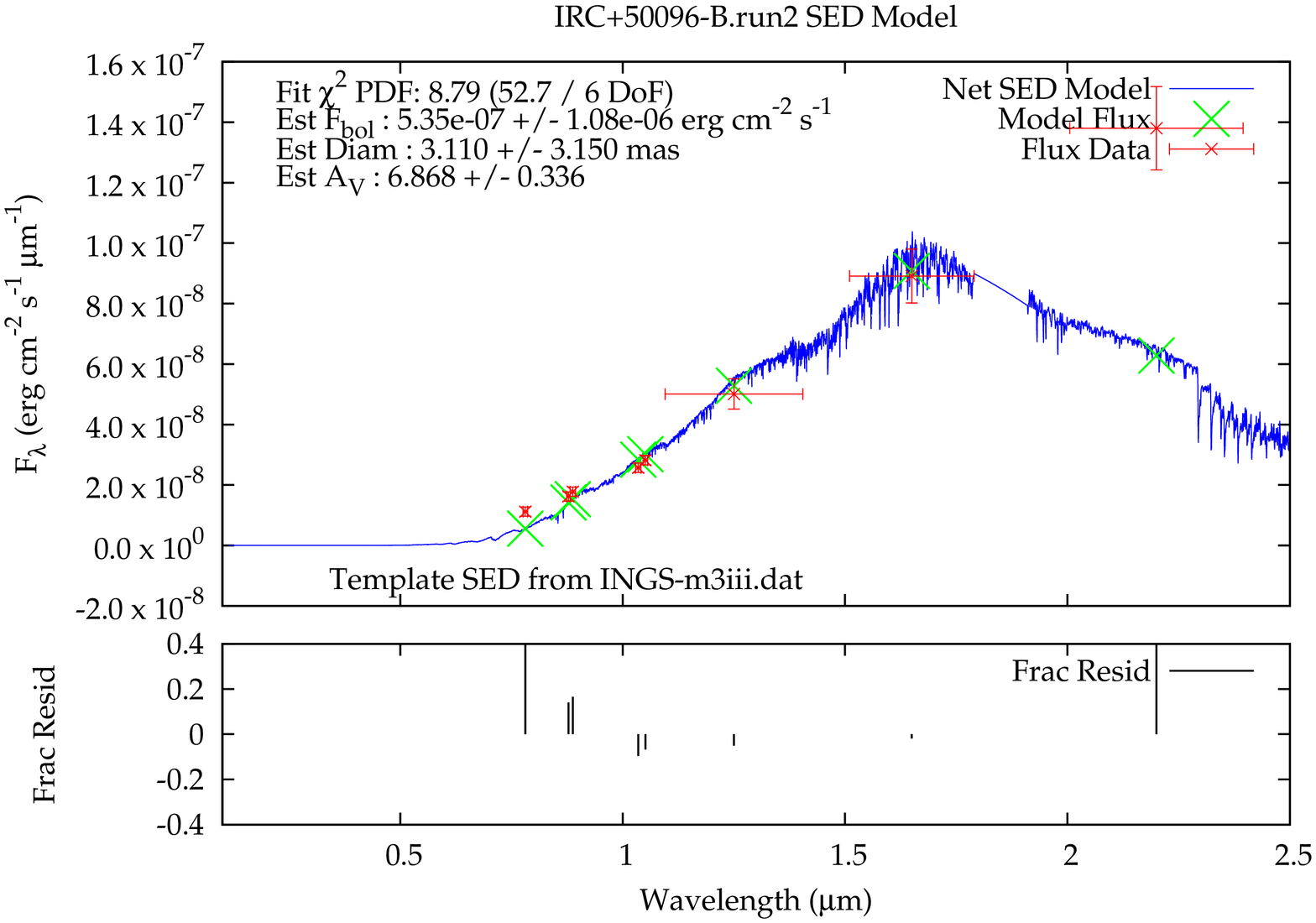}}
\subfigure[IRC+50260-A (M7III)]{\includegraphics[width = 2.35in,angle=270]{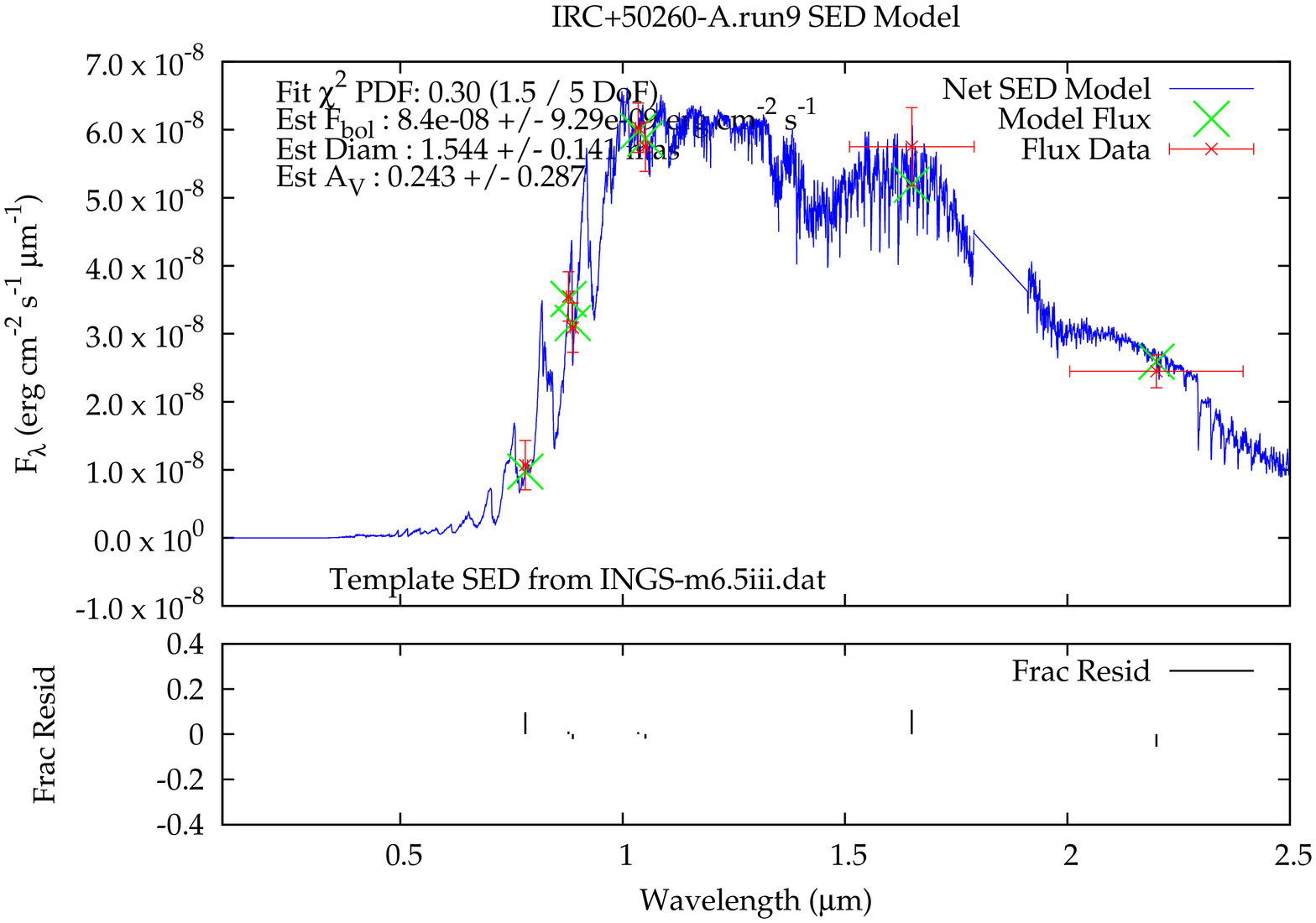}} \\
\caption{SED fits as described in \S 2.2.}
\end{figure}

\begin{figure}
\subfigure[IRC+50261-A (M6III)]{\includegraphics[width = 2.35in,angle=270]{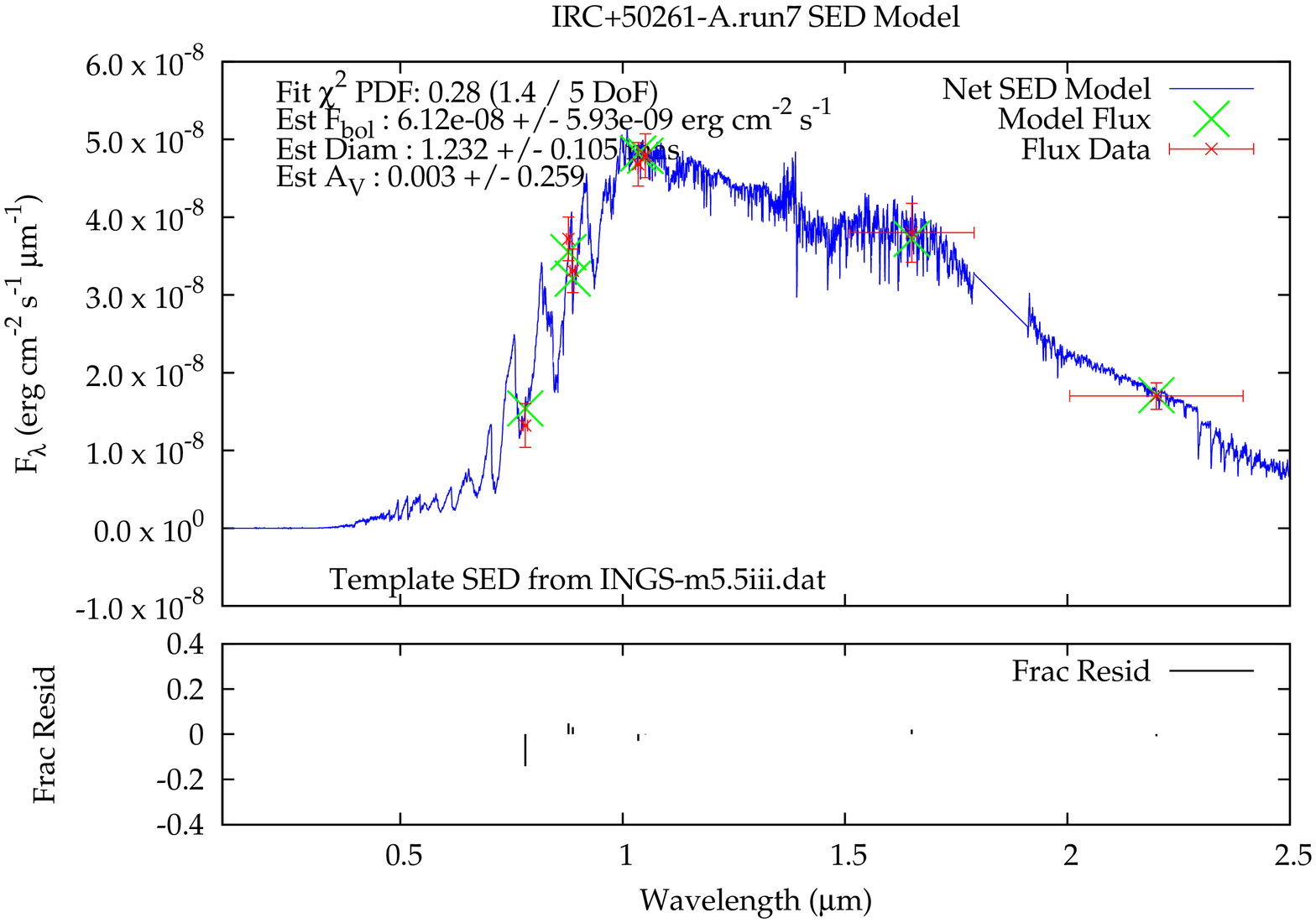}}
\subfigure[IRC+50357-A (M10III)]{\includegraphics[width = 2.35in,angle=270]{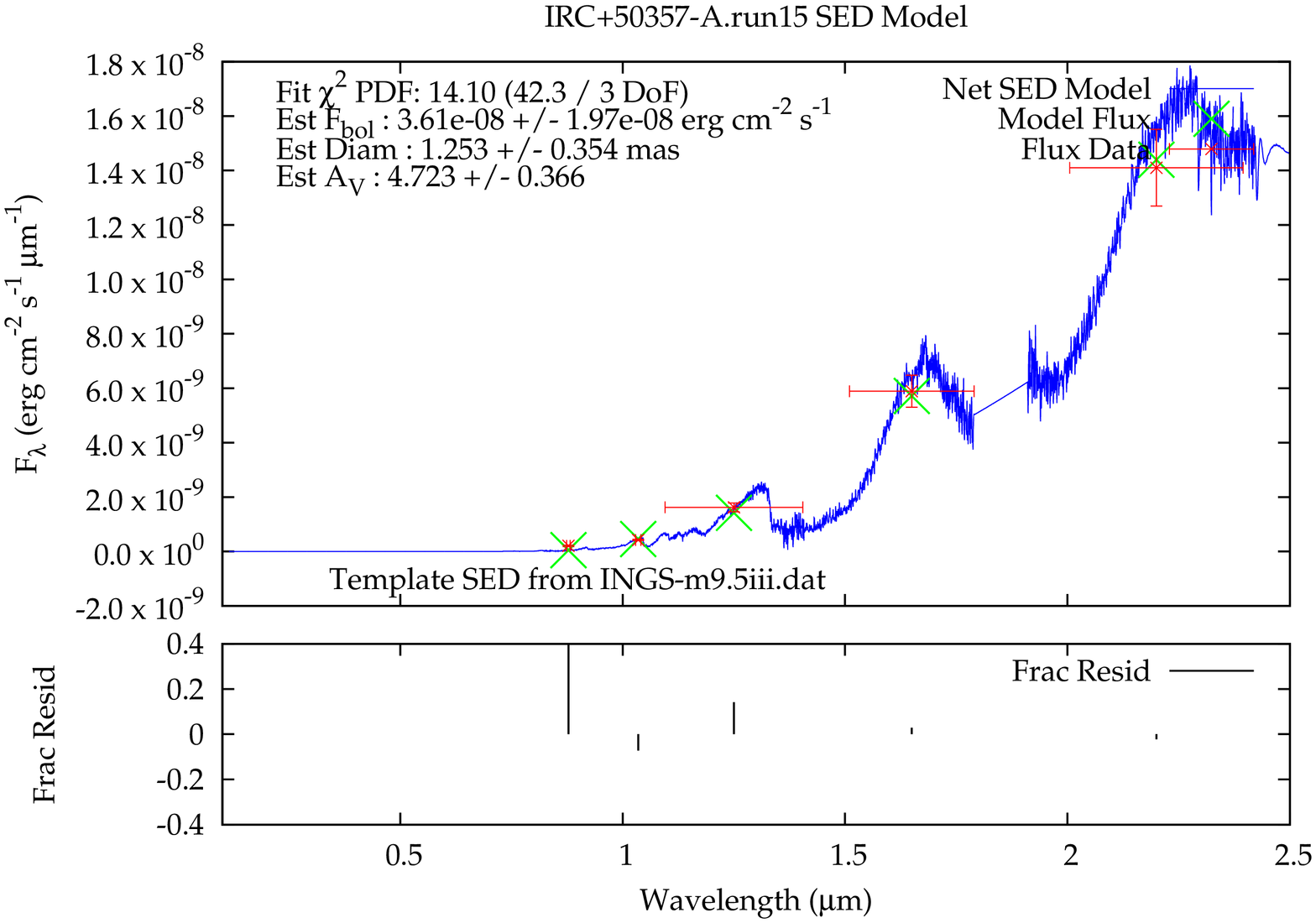}} \\
\subfigure[IRC+60015-A (M8III)]{\includegraphics[width = 2.35in,angle=270]{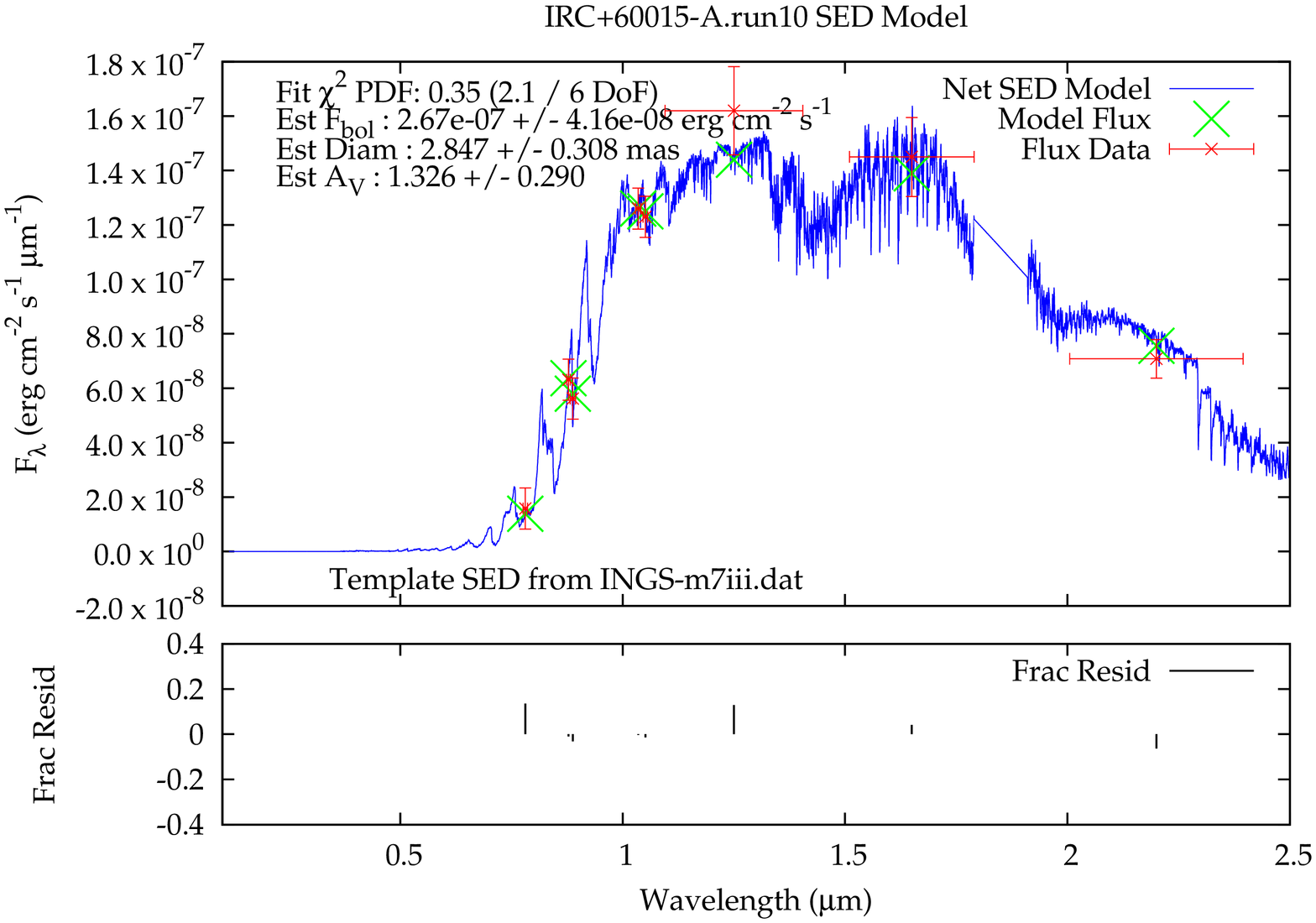}}
\subfigure[IRC+60052-A (M6III)]{\includegraphics[width = 2.35in,angle=270]{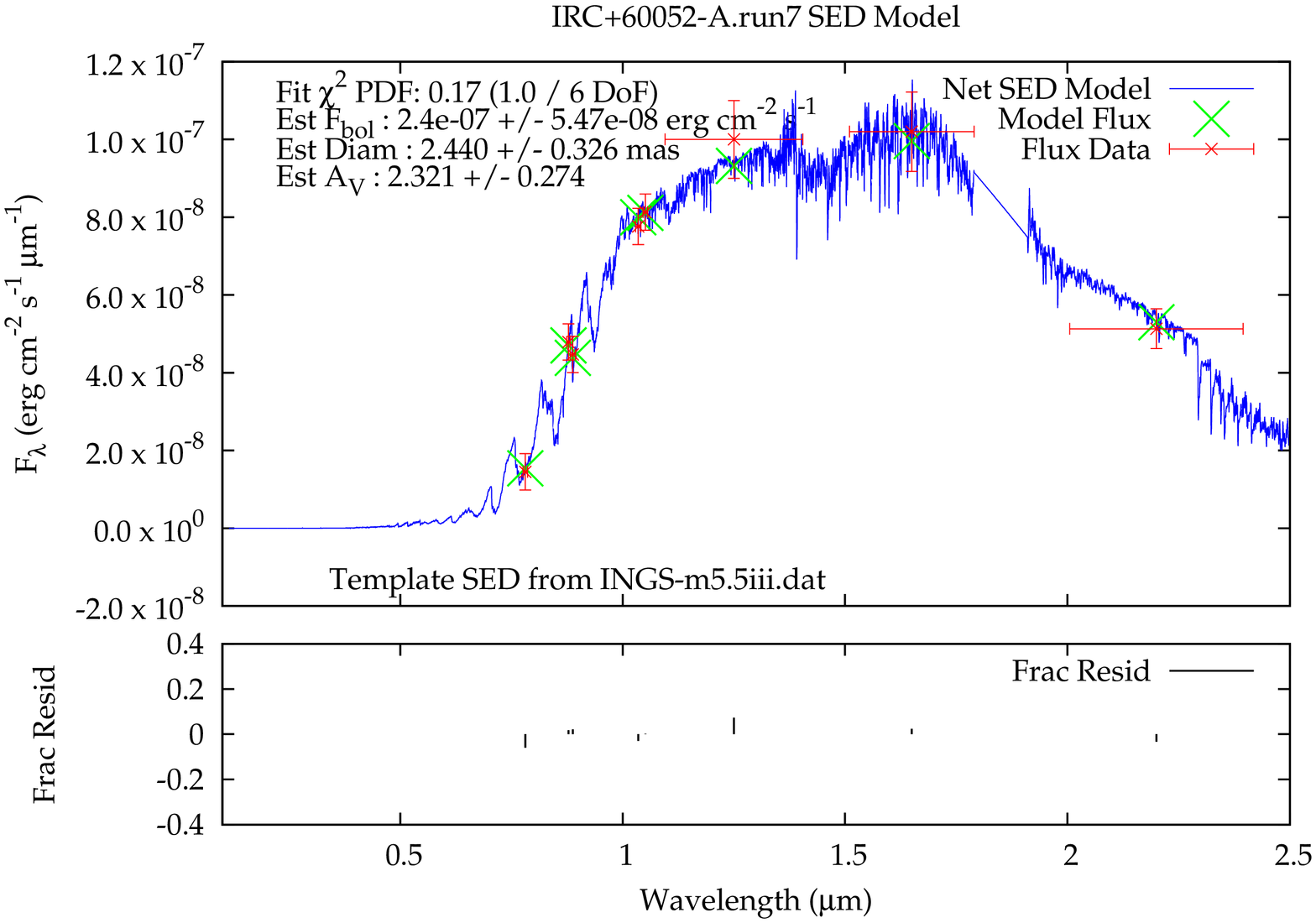}} \\
\subfigure[IRC+60092-A (M8.5III)]{\includegraphics[width = 2.35in,angle=270]{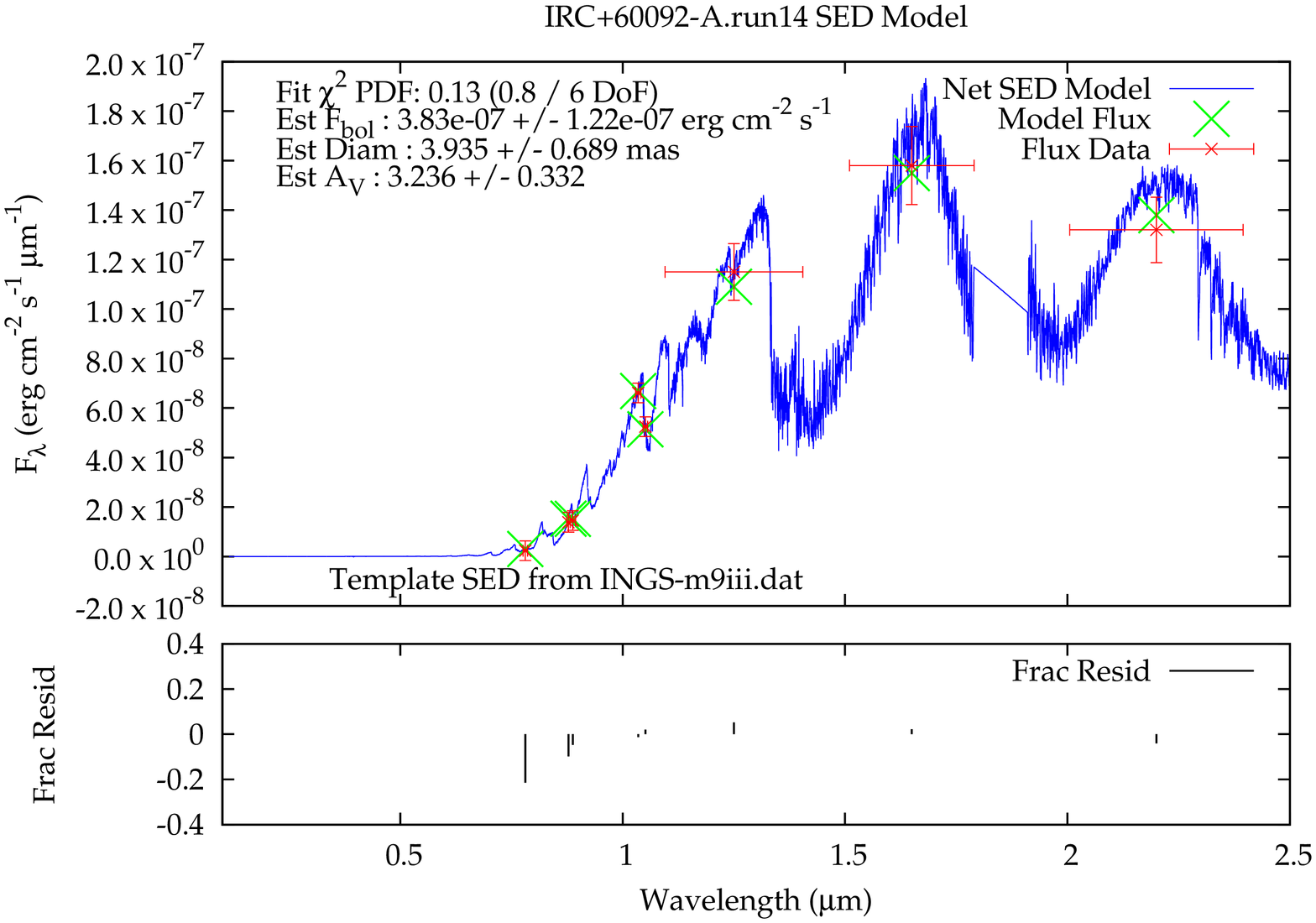}}
\subfigure[IRC+60169-A (M9.5III)]{\includegraphics[width = 2.35in,angle=270]{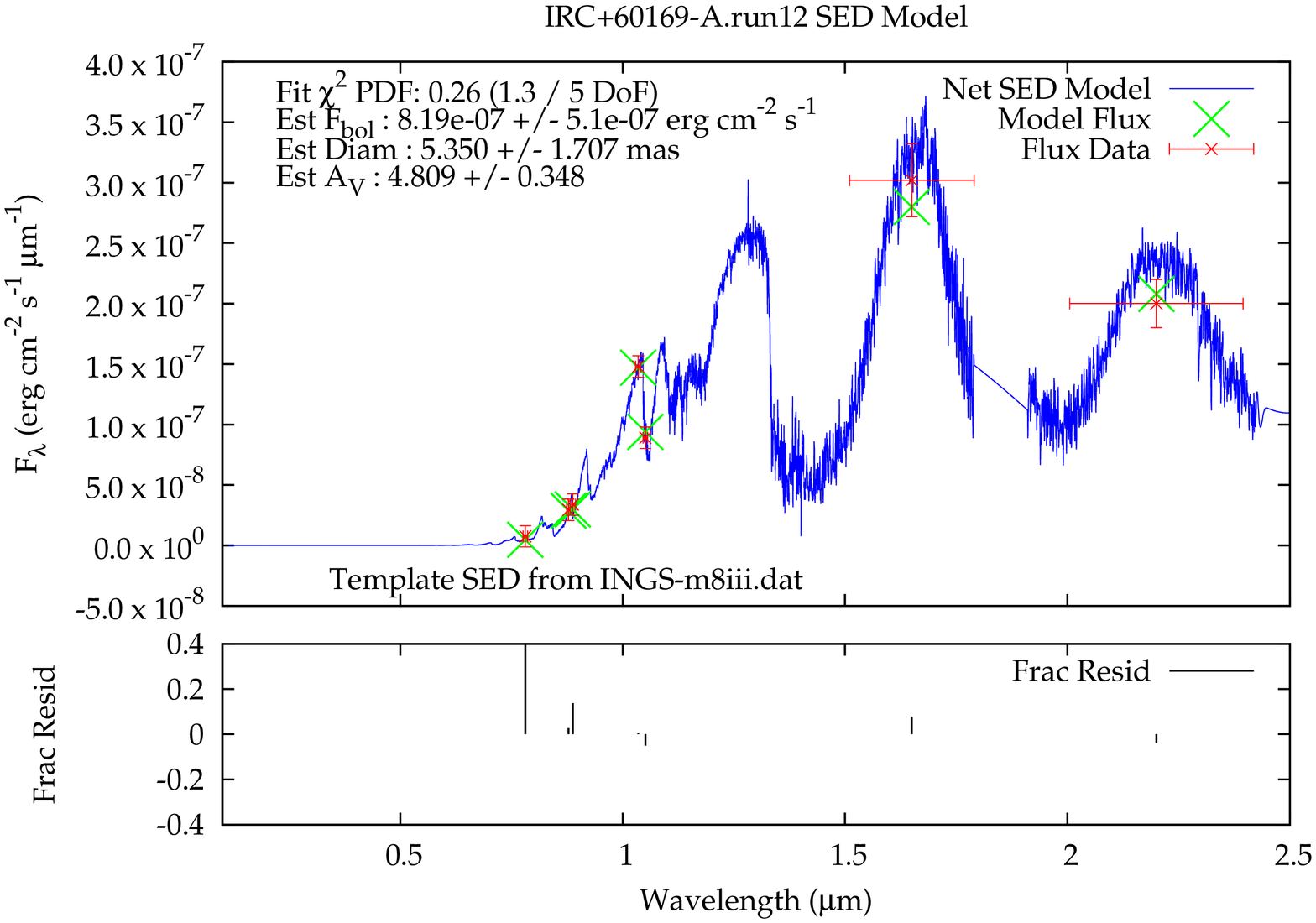}} \\
\caption{SED fits as described in \S 2.2.}
\end{figure}

\begin{figure}
\subfigure[IRC+60184-A (M9III)]{\includegraphics[width = 2.35in,angle=270]{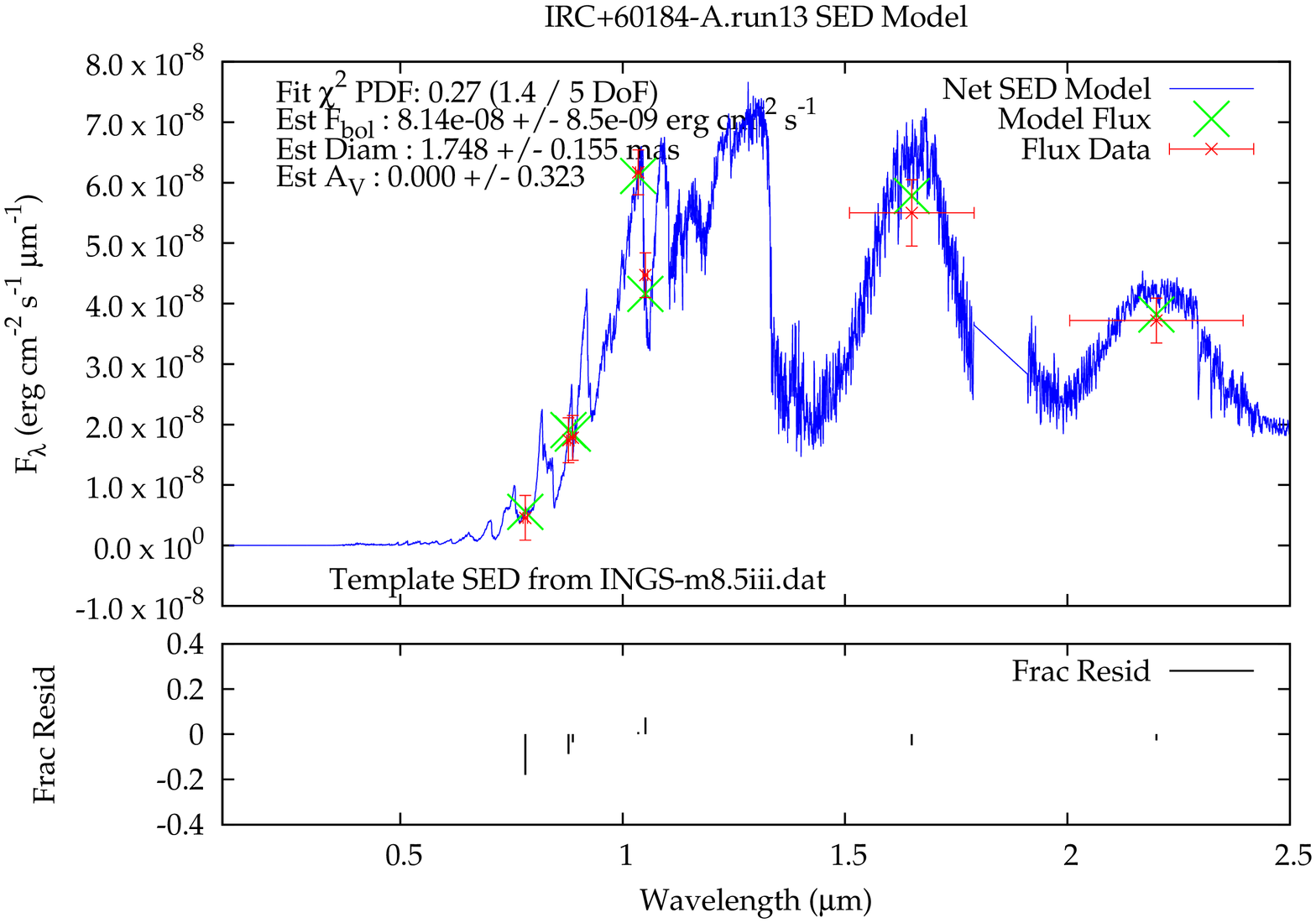}}
\subfigure[IRC+60288-A (M9.5III)]{\includegraphics[width = 2.35in,angle=270]{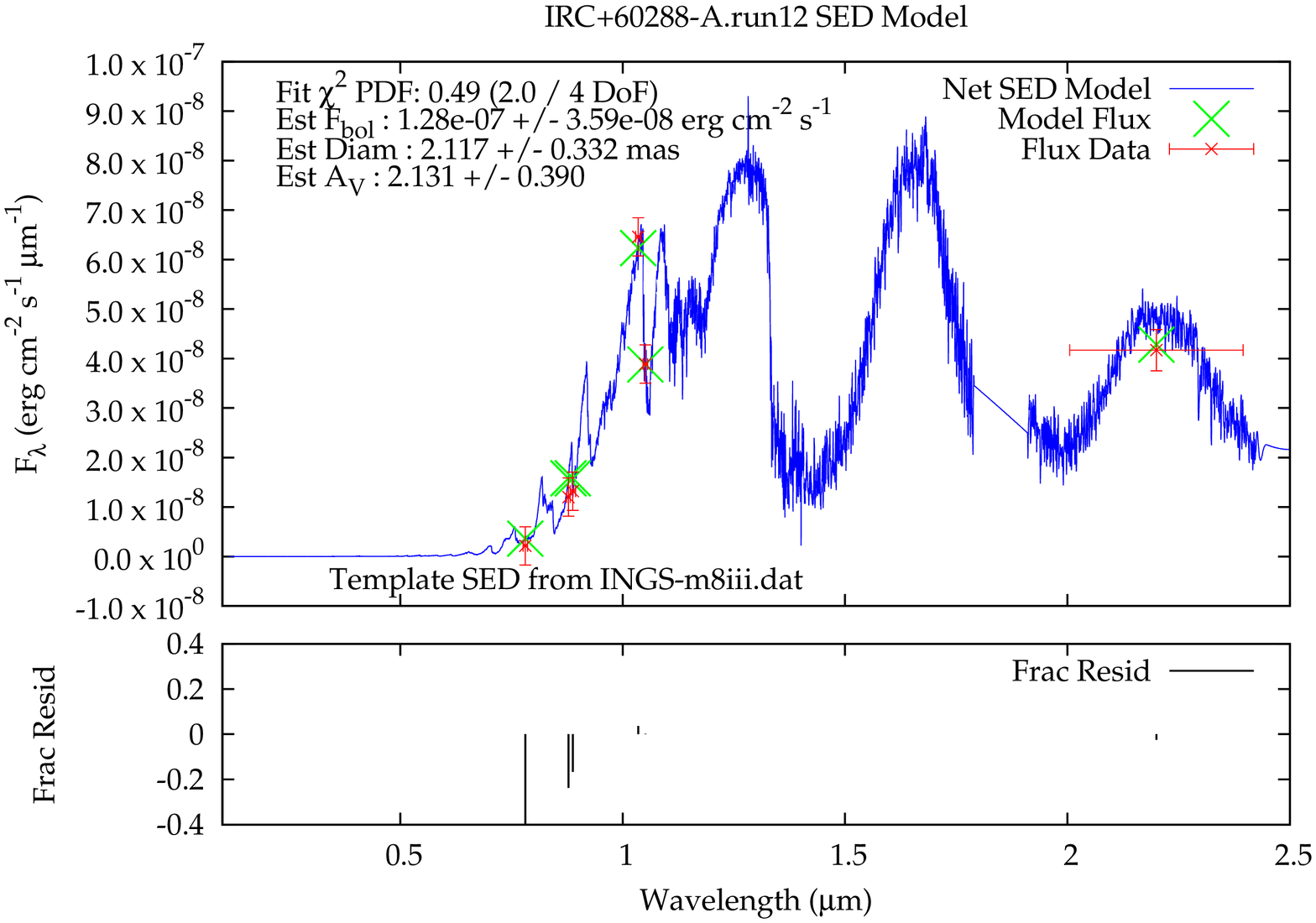}} \\
\subfigure[IRC+60288-B (M8III)]{\includegraphics[width = 2.35in,angle=270]{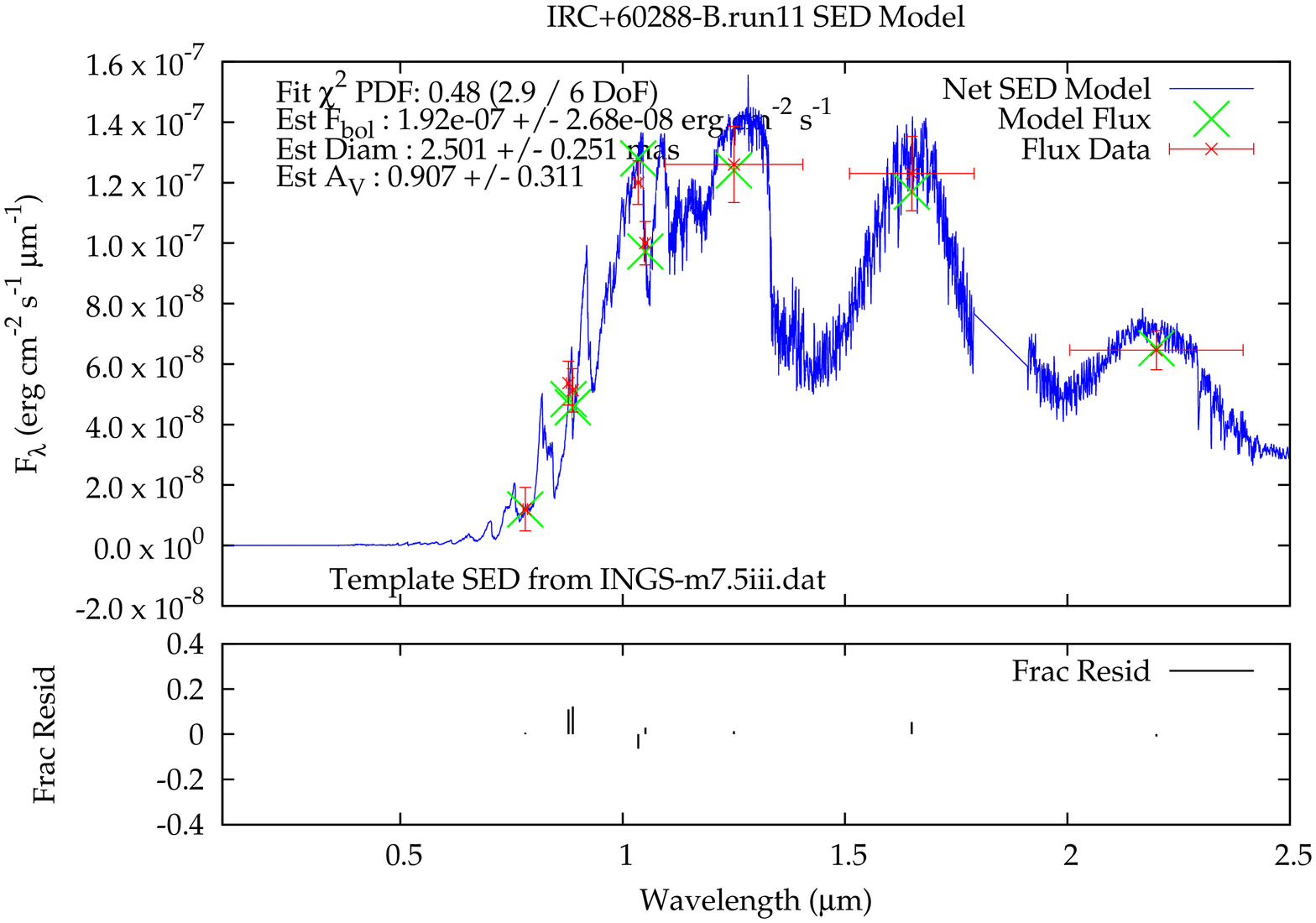}}
\subfigure[IRC+60288-C (M8III)]{\includegraphics[width = 2.35in,angle=270]{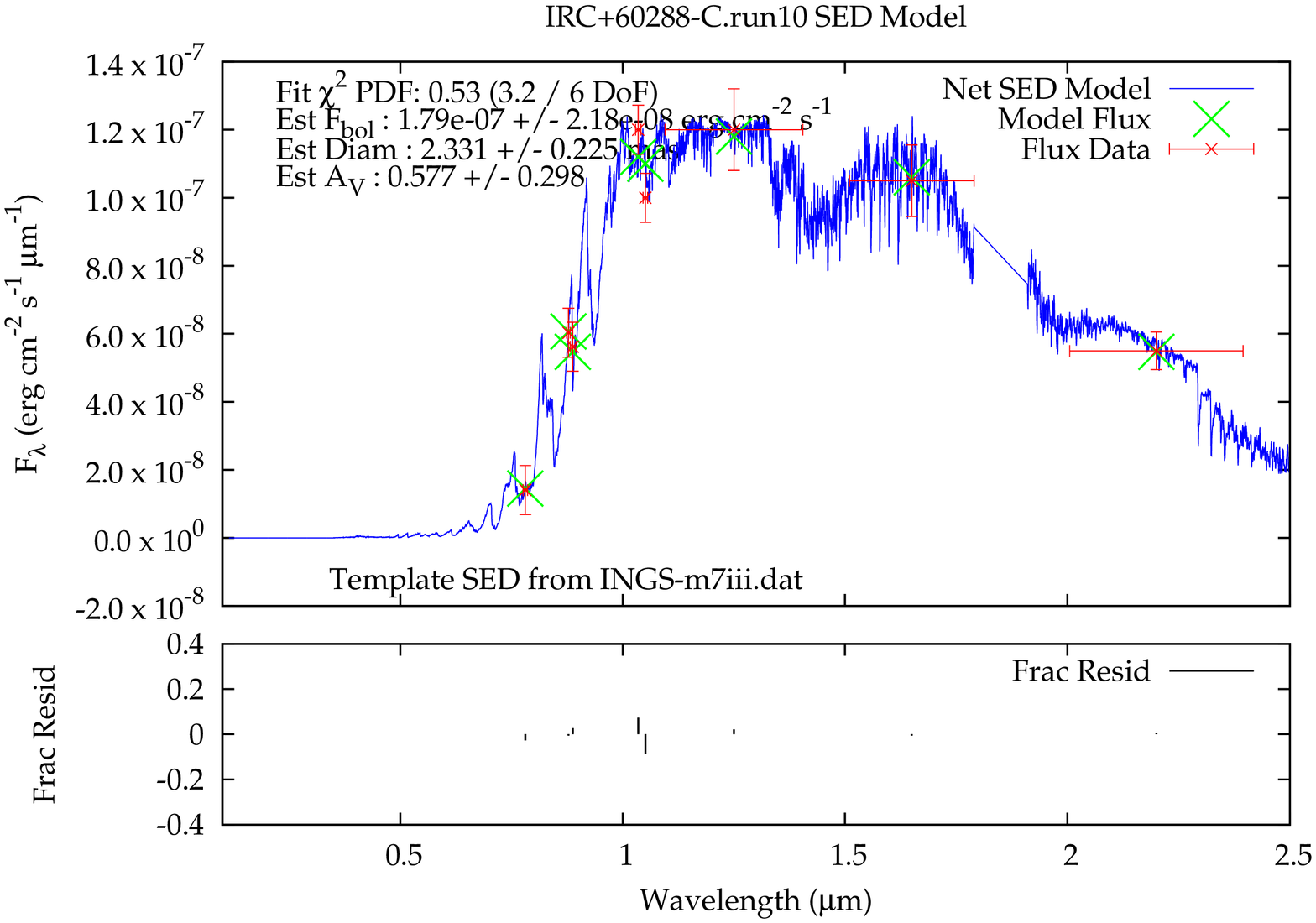}} \\
\subfigure[IRC+60289-A (M8III)]{\includegraphics[width = 2.35in,angle=270]{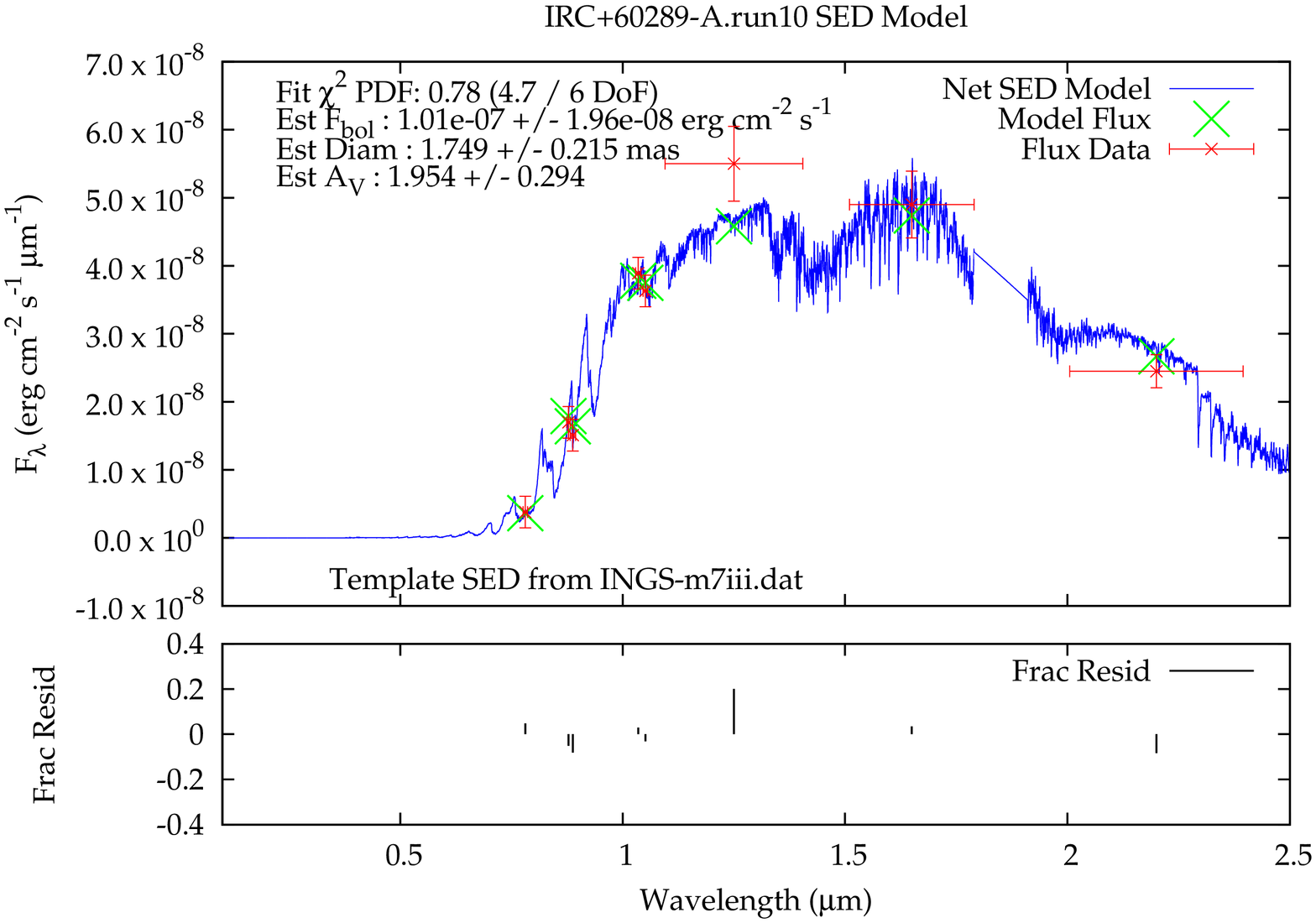}}
\subfigure[IRC+60316-A (M6III)]{\includegraphics[width = 2.35in,angle=270]{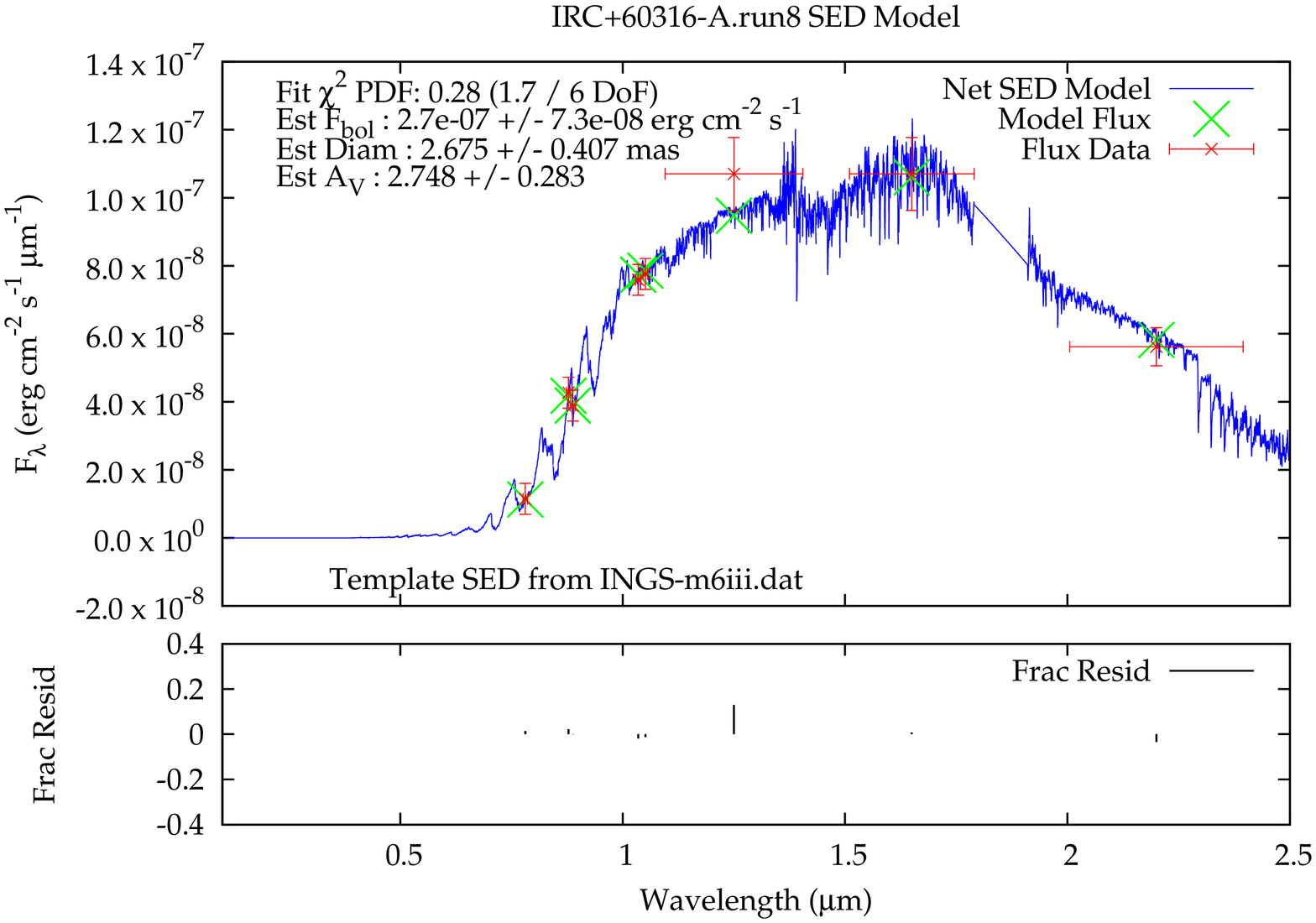}} \\
\caption{SED fits as described in \S 2.2.}
\end{figure}

\begin{figure}
\subfigure[IRC+60334-A (M9III)]{\includegraphics[width = 2.35in,angle=270]{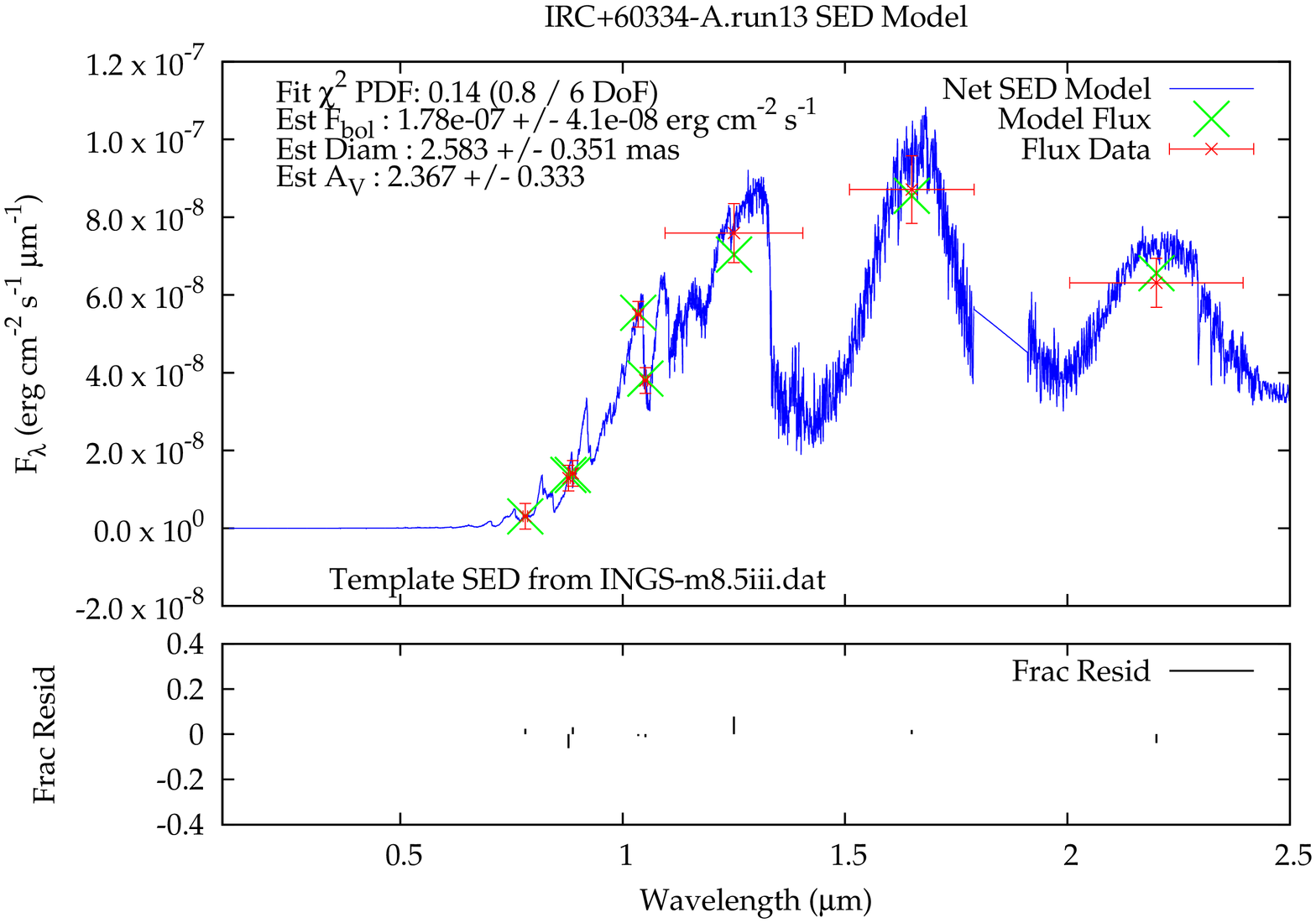}}
\subfigure[IRC+70102-A (M9III)]{\includegraphics[width = 2.35in,angle=270]{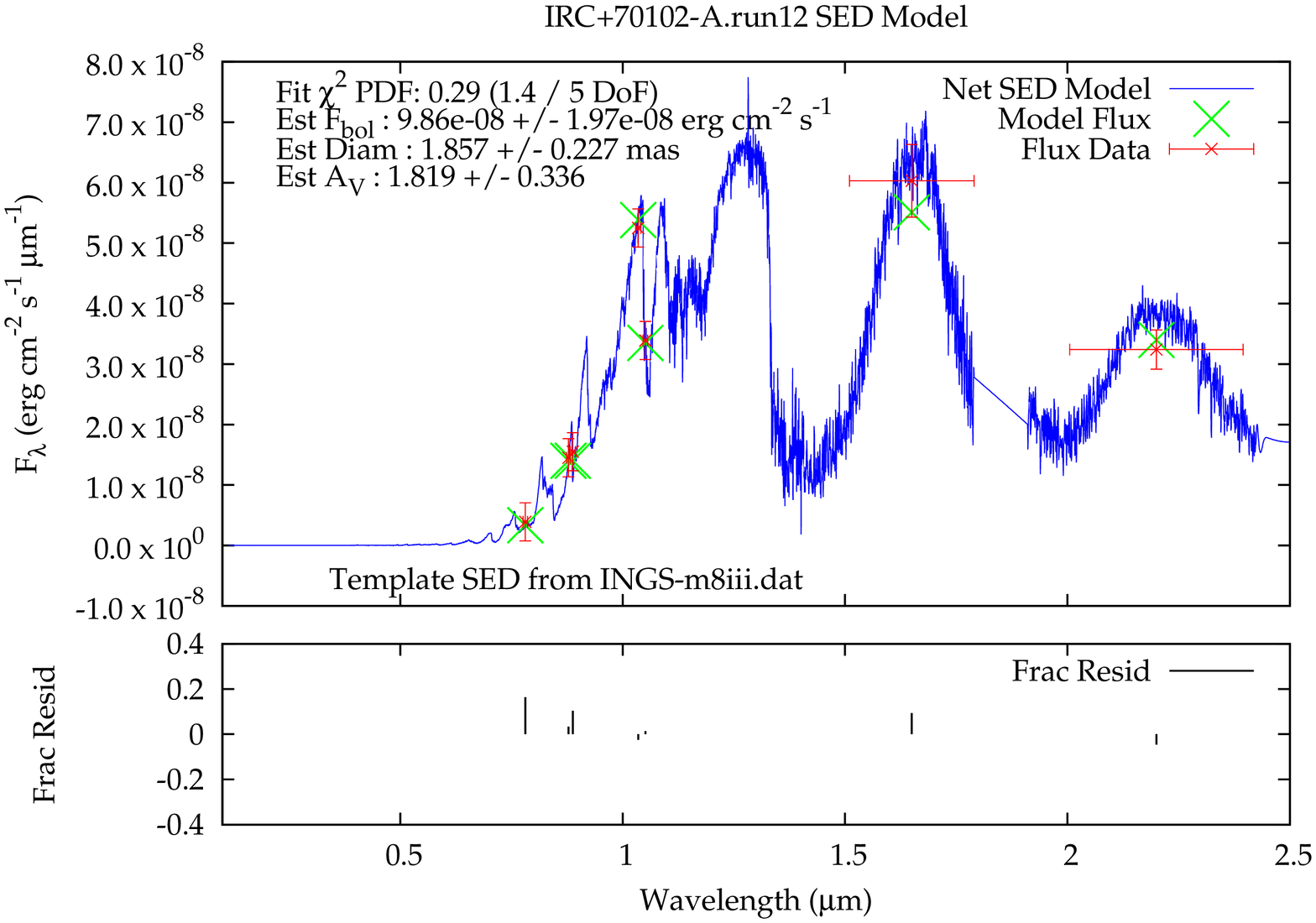}} \\
\subfigure[IRC+70171-A (M9.5III)]{\includegraphics[width = 2.35in,angle=270]{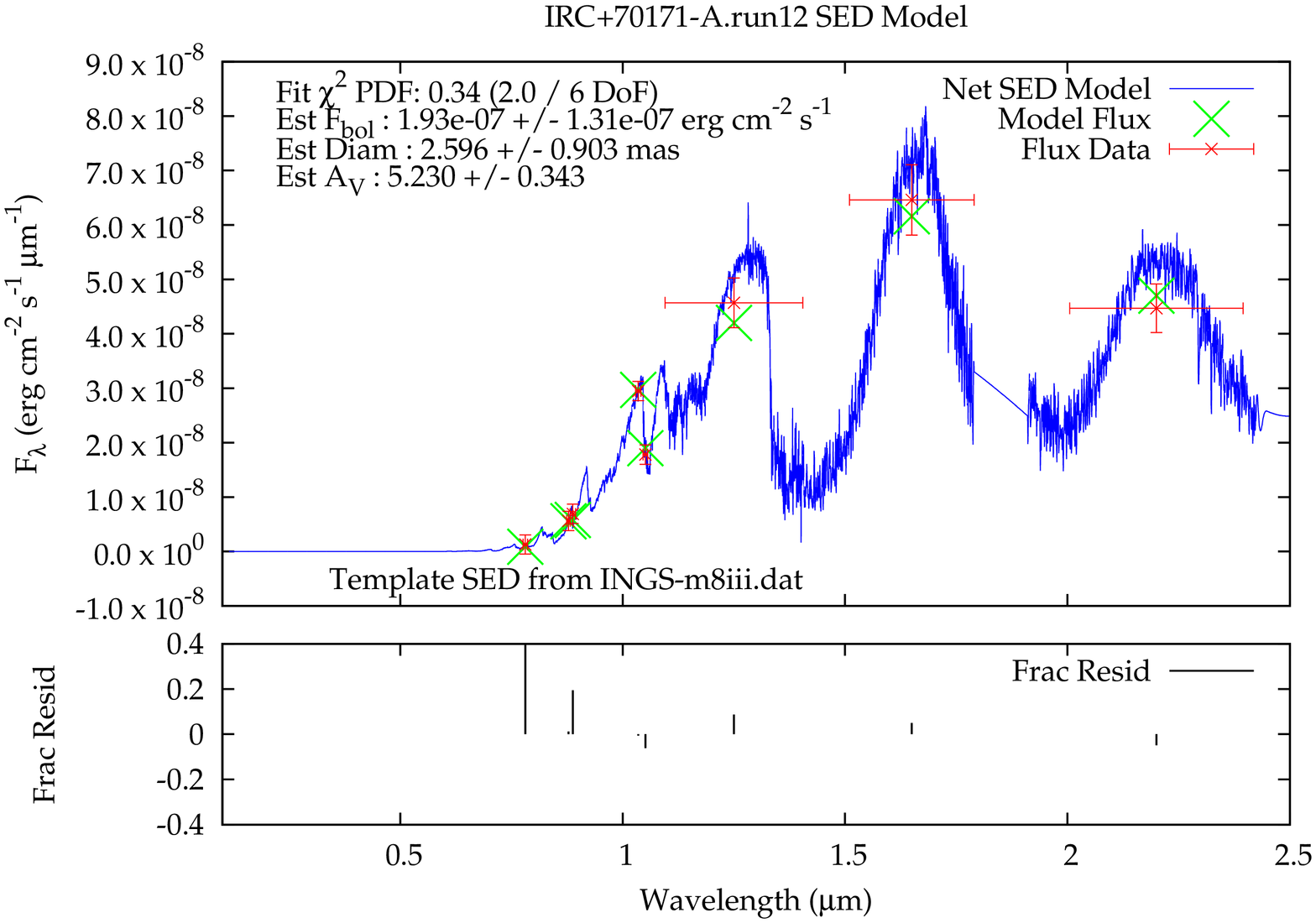}}
\subfigure[IRC+70171-B (M9III)]{\includegraphics[width = 2.35in,angle=270]{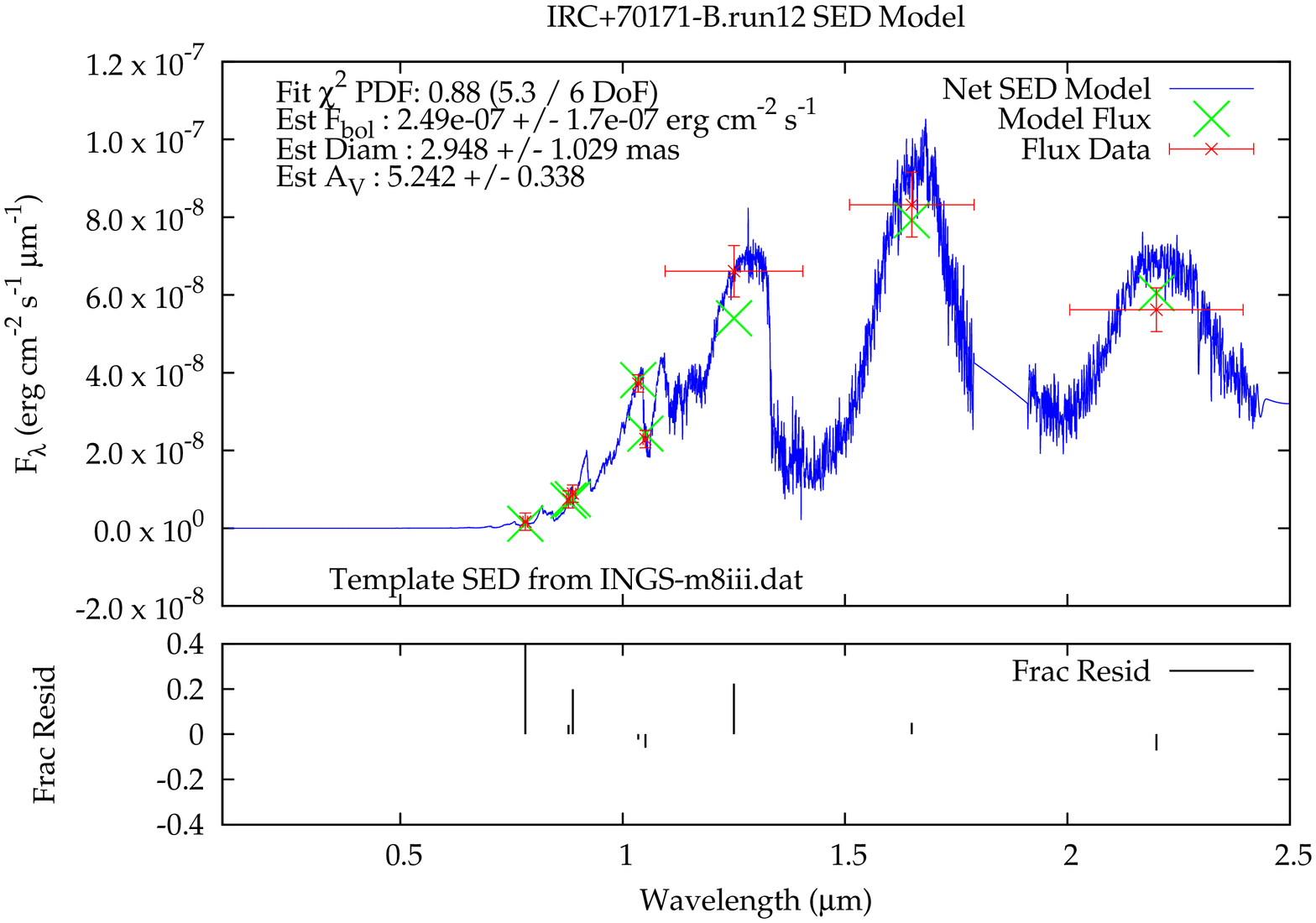}} \\
\subfigure[IRC+80005-A (M9III)]{\includegraphics[width = 2.35in,angle=270]{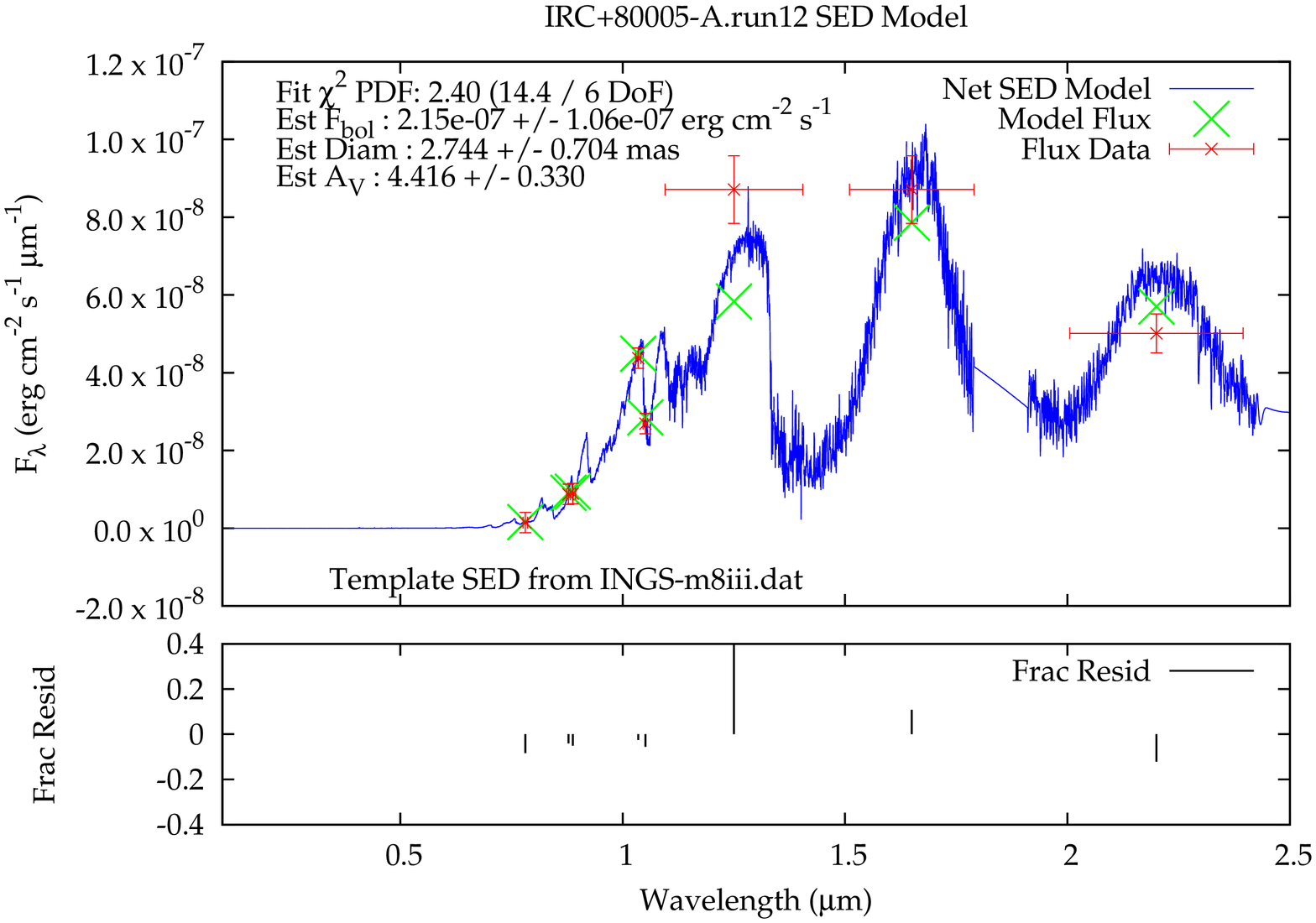}}
\subfigure[IRC+80005-B (M9.5III)]{\includegraphics[width = 2.35in,angle=270]{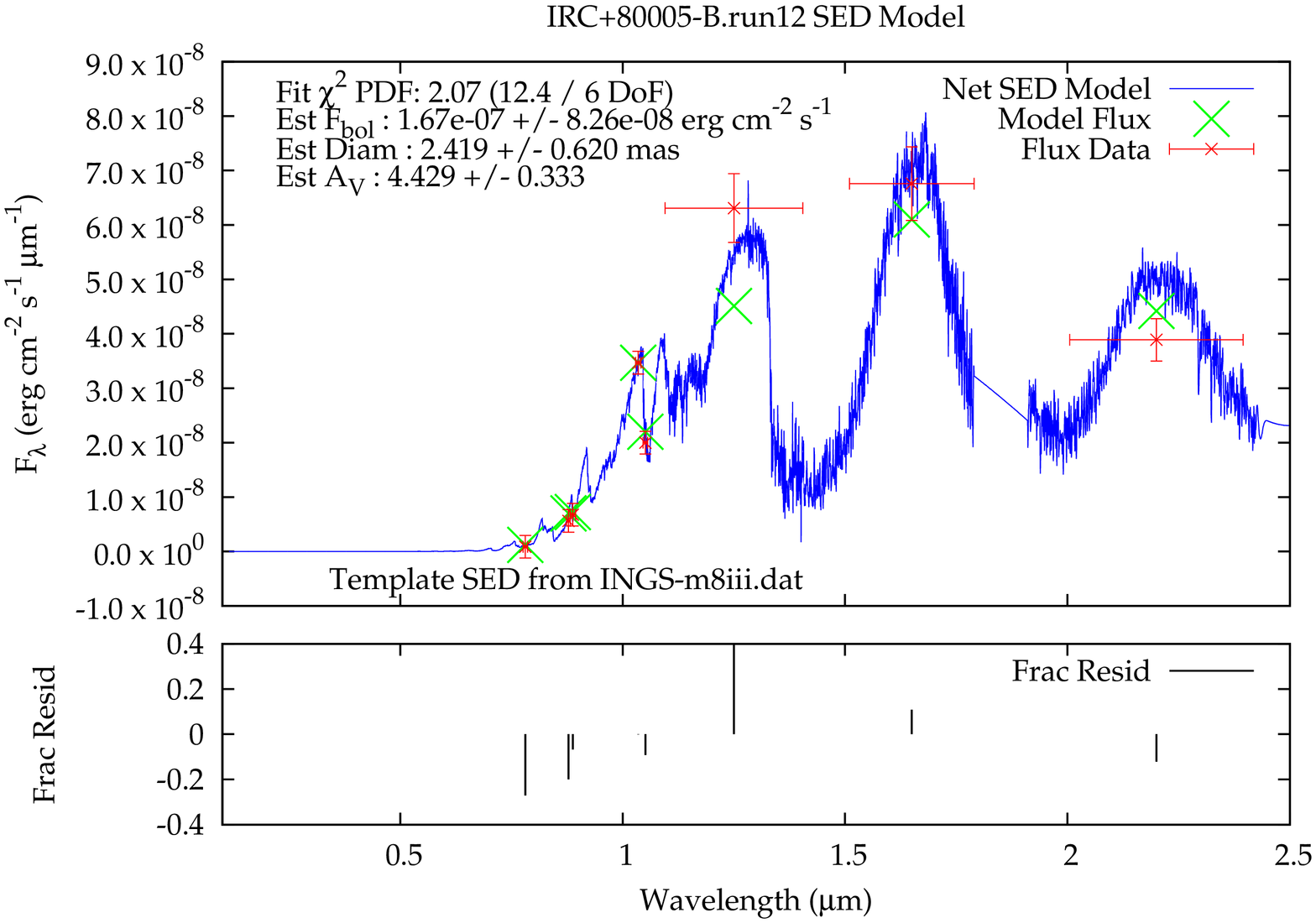}} \\
\caption{SED fits as described in \S 2.2.}
\end{figure}

\begin{figure}
\subfigure[IRC-10236-A (M10III)]{\includegraphics[width = 2.35in,angle=270]{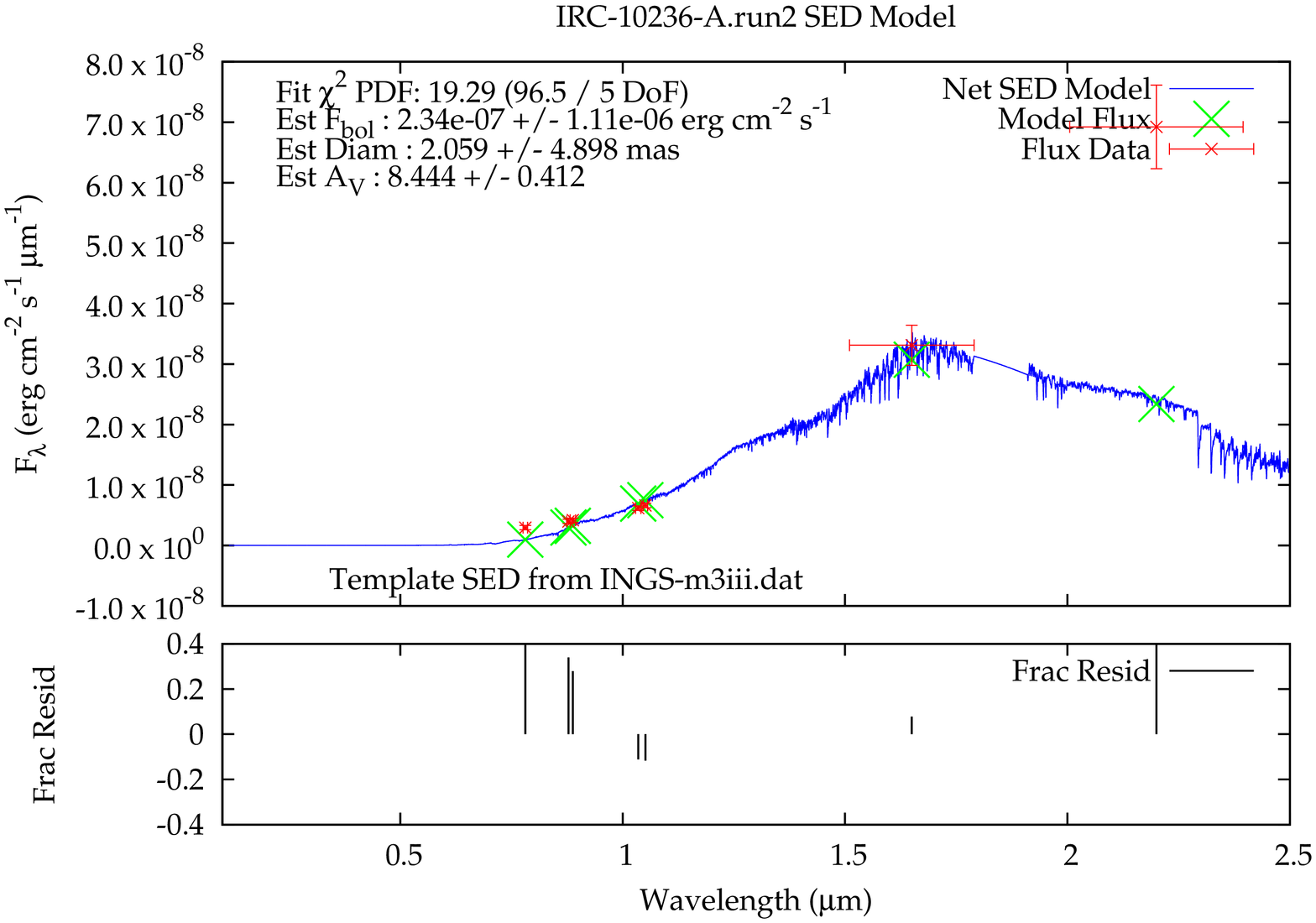}}
\subfigure[IRC-20293-A (M8III)]{\includegraphics[width = 2.35in,angle=270]{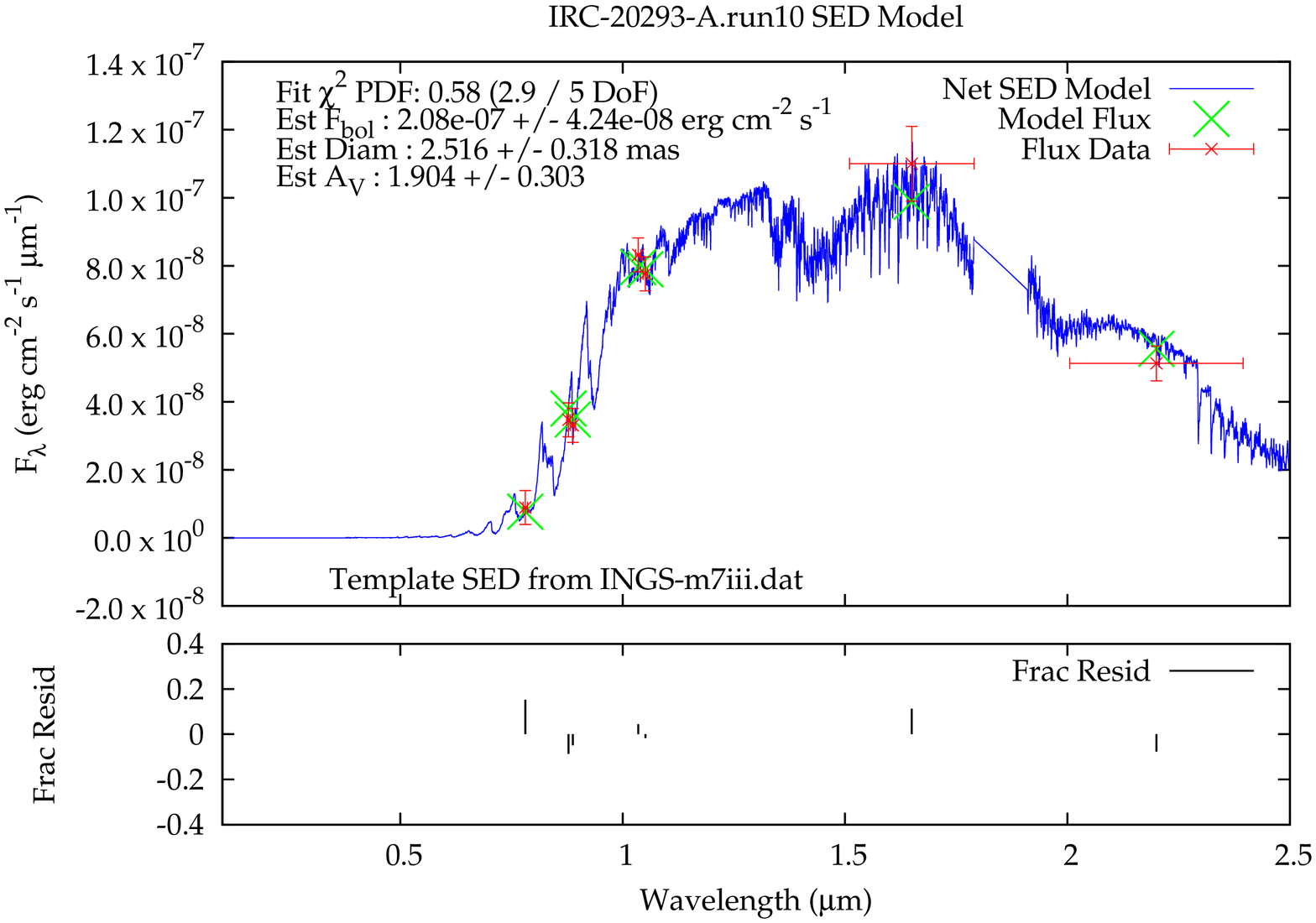}} \\
\subfigure[IRC-30217-A (M10III)]{\includegraphics[width = 2.35in,angle=270]{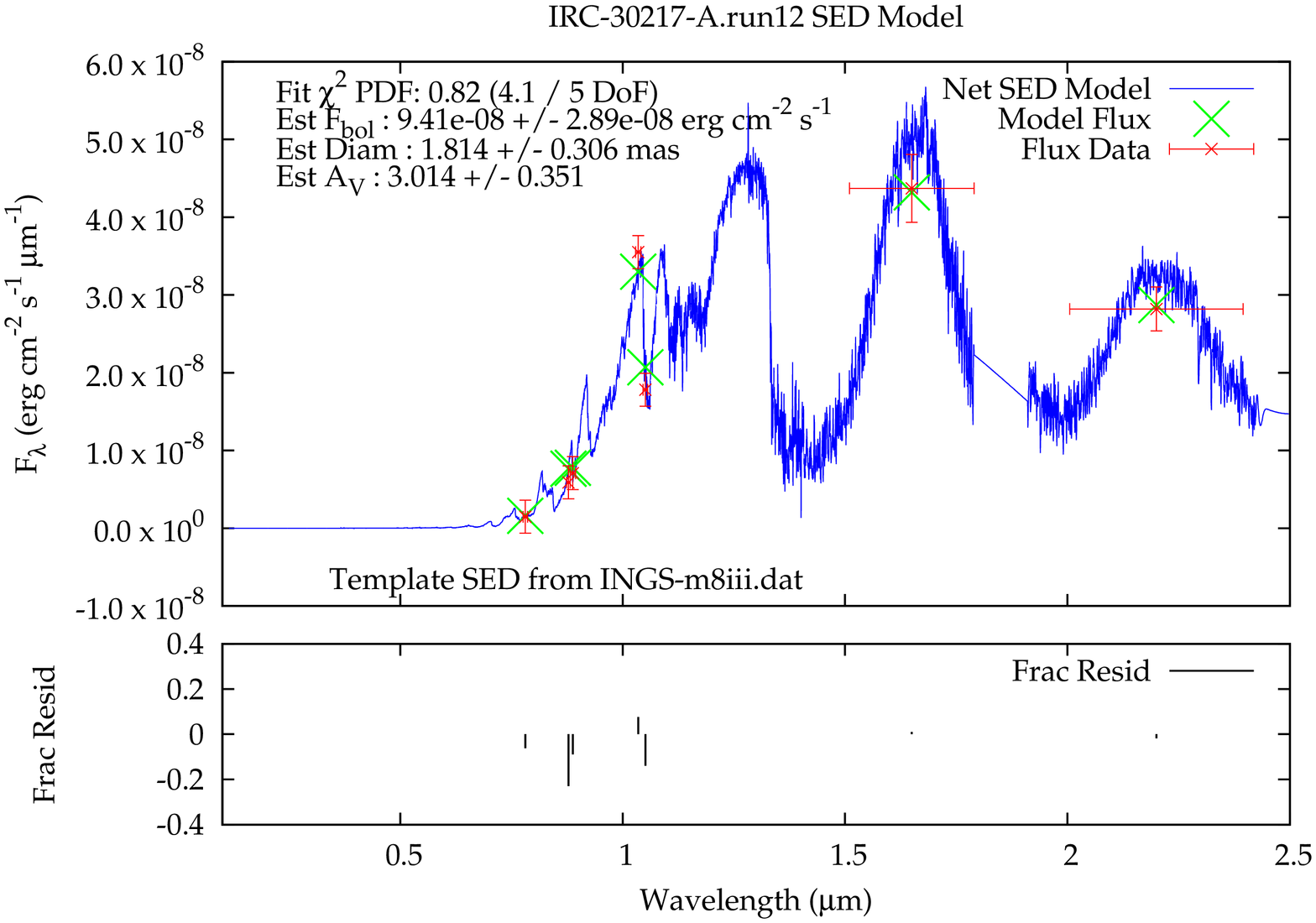}}
\subfigure[IRC-30217-B (M9.5III)]{\includegraphics[width = 2.35in,angle=270]{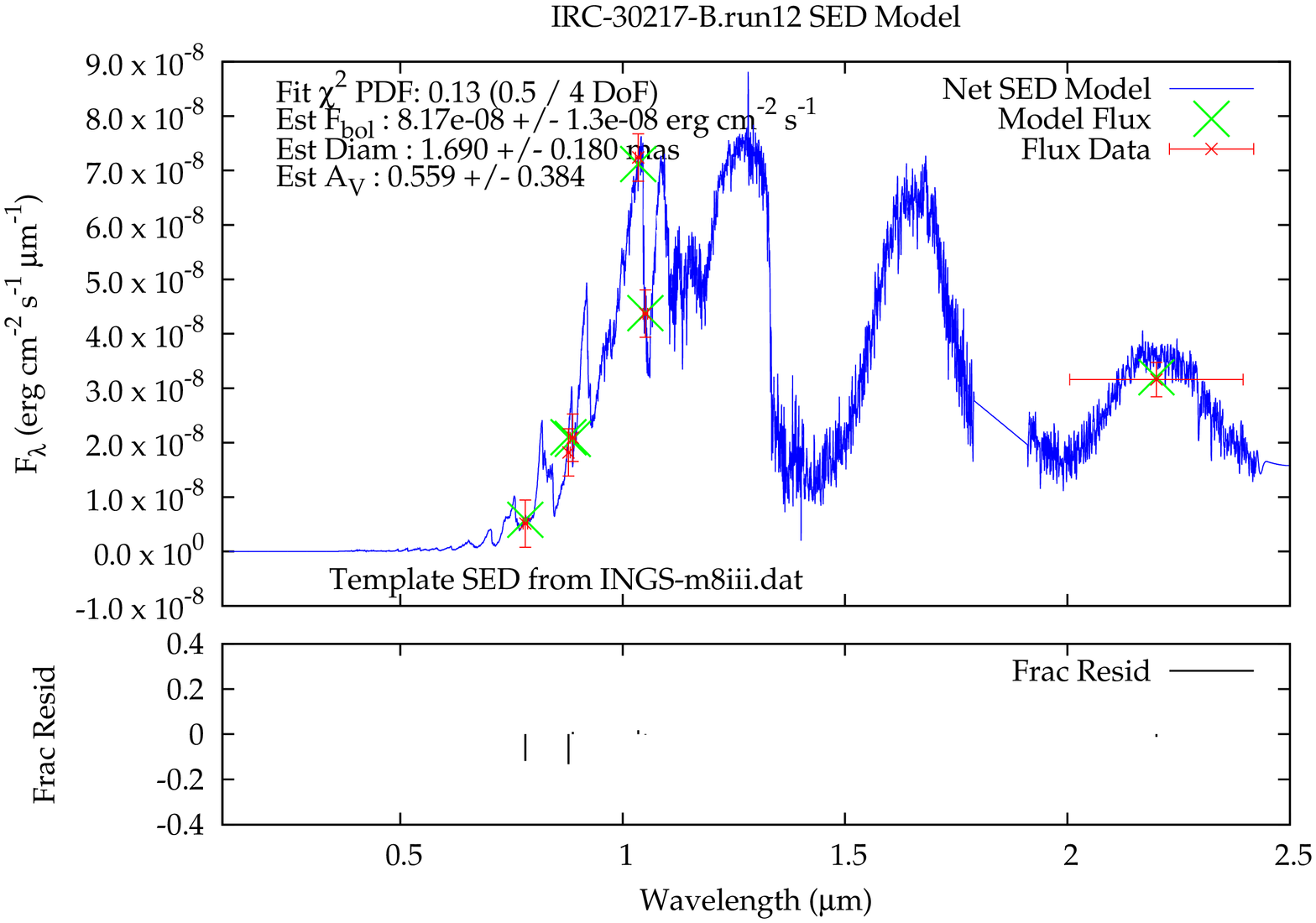}} \\
\subfigure[R-And-A (M10III)]{\includegraphics[width = 2.35in,angle=270]{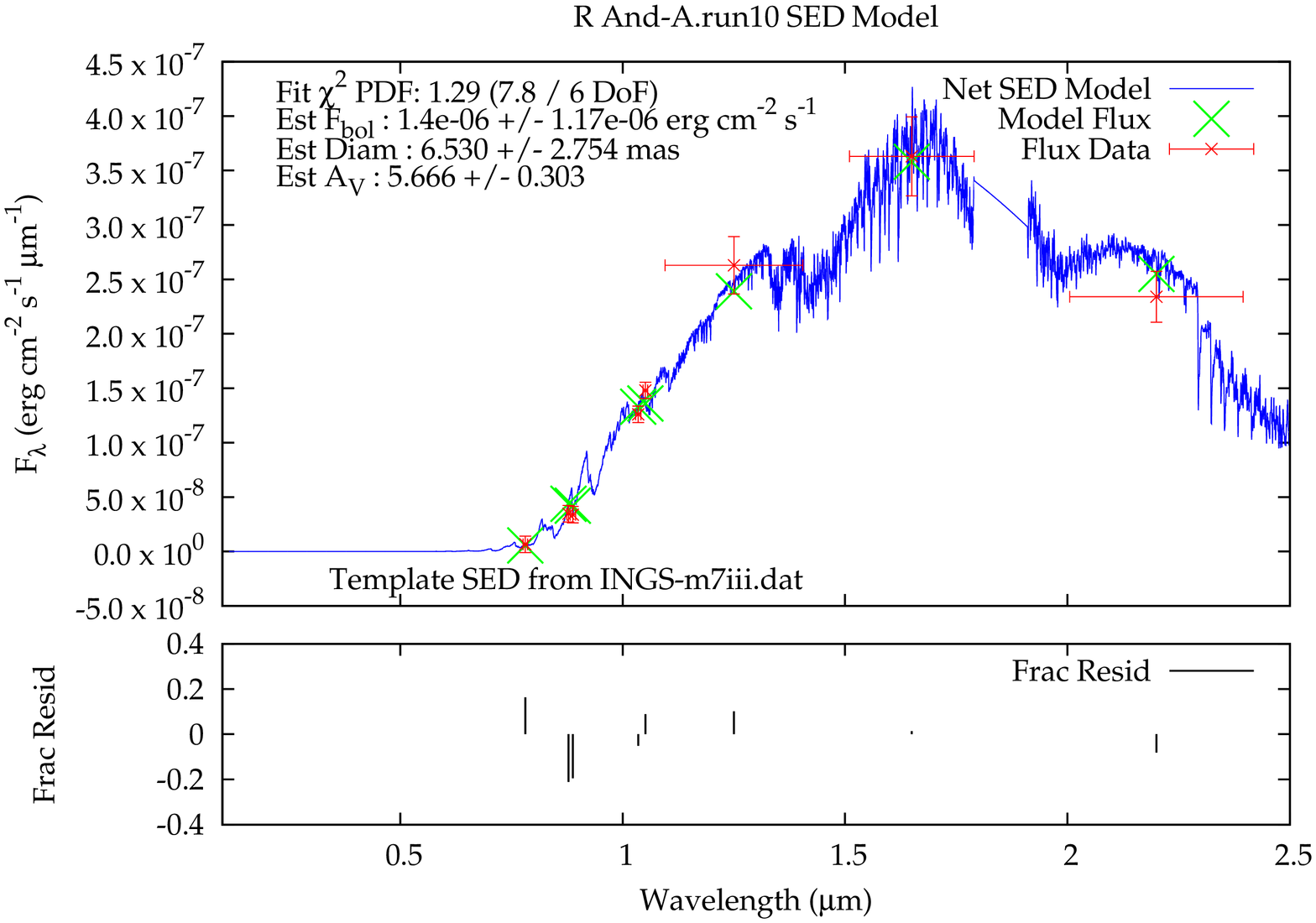}}
\subfigure[R-And-B (M10III)]{\includegraphics[width = 2.35in,angle=270]{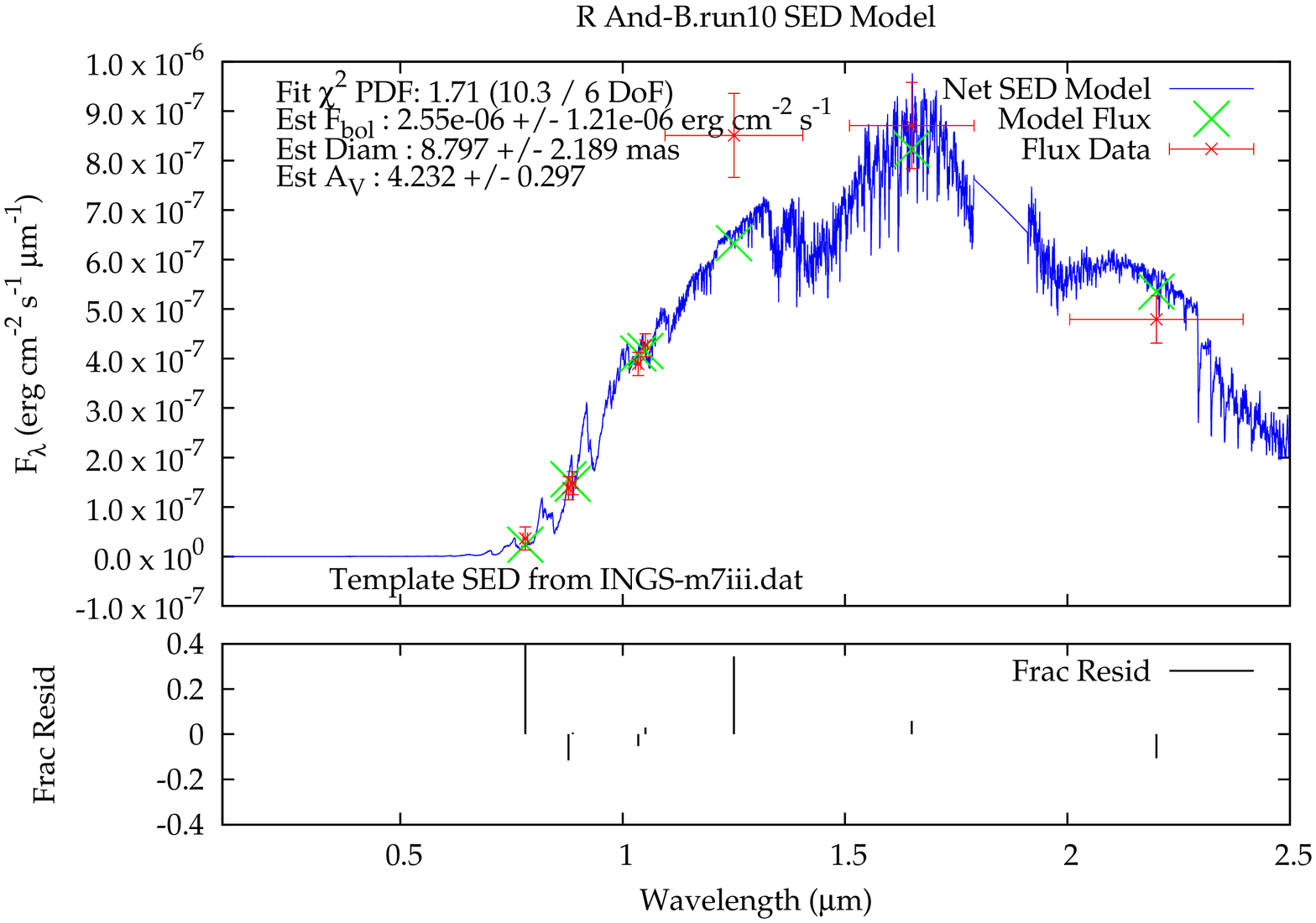}} \\
\caption{SED fits as described in \S 2.2.}
\end{figure}

\begin{figure}
\subfigure[R-Boo-A (M8III)]{\includegraphics[width = 2.35in,angle=270]{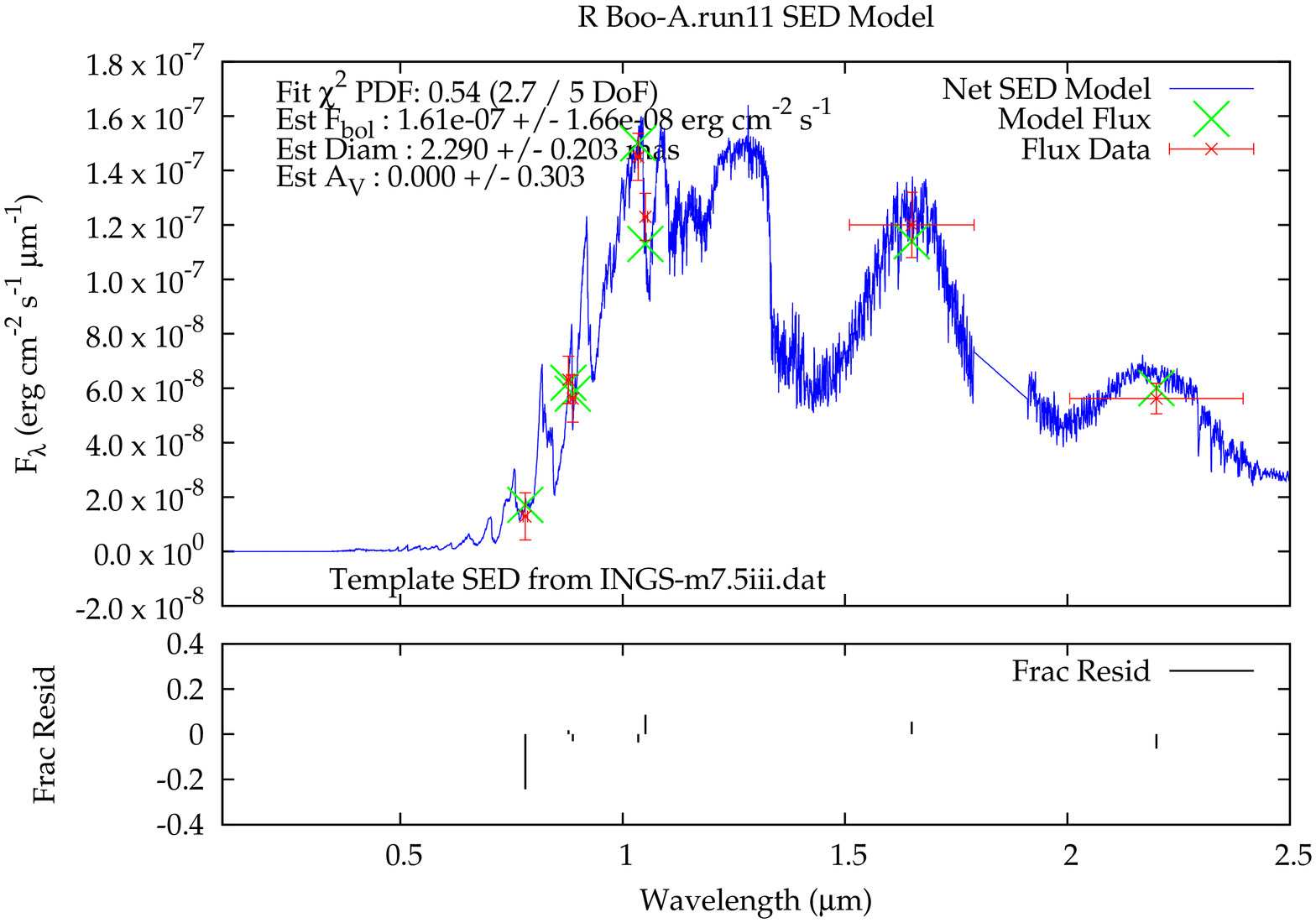}}
\subfigure[R-Cam-A (M10III)]{\includegraphics[width = 2.35in,angle=270]{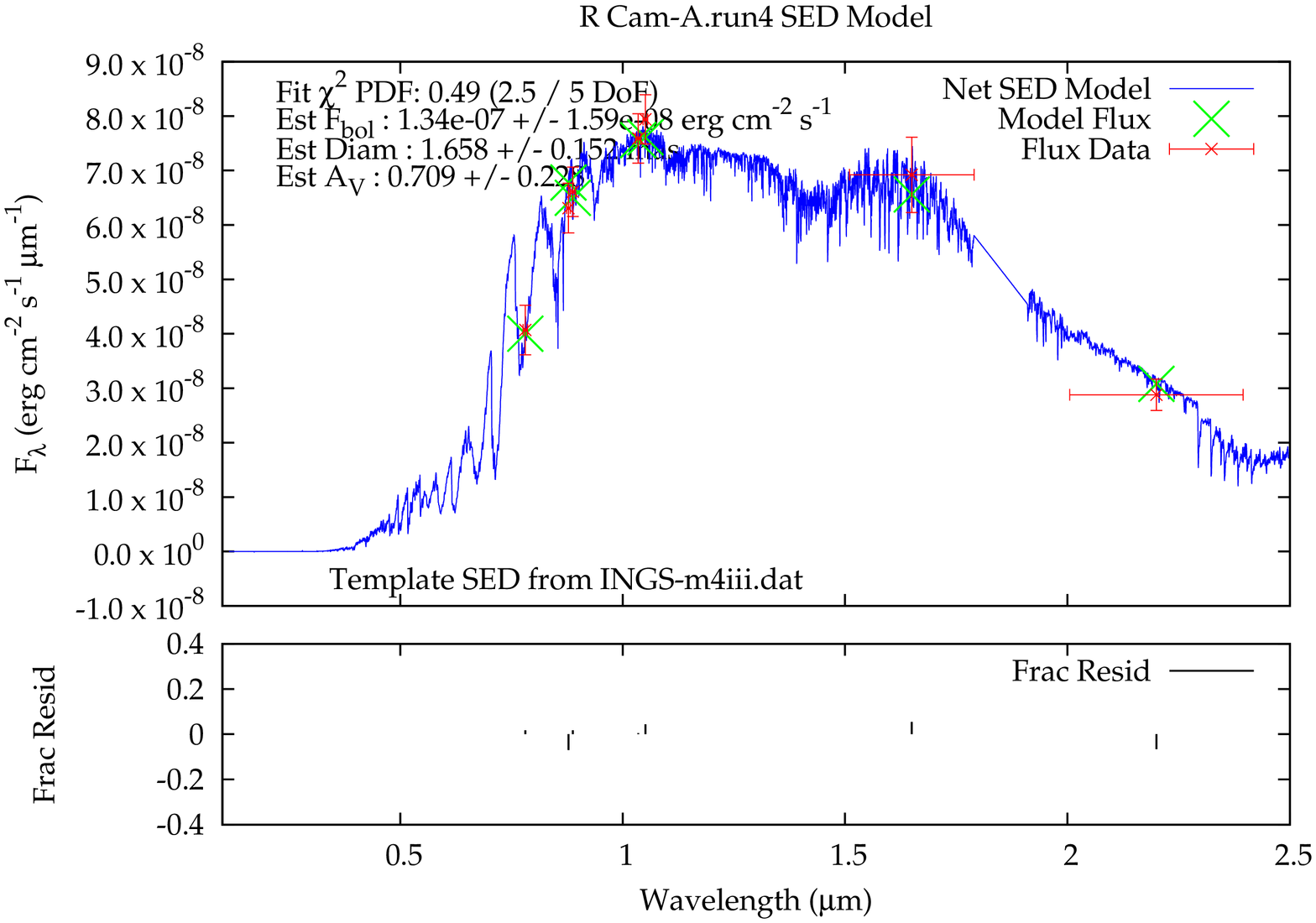}} \\
\subfigure[R-Cas-A (M9.5III)]{\includegraphics[width = 2.35in,angle=270]{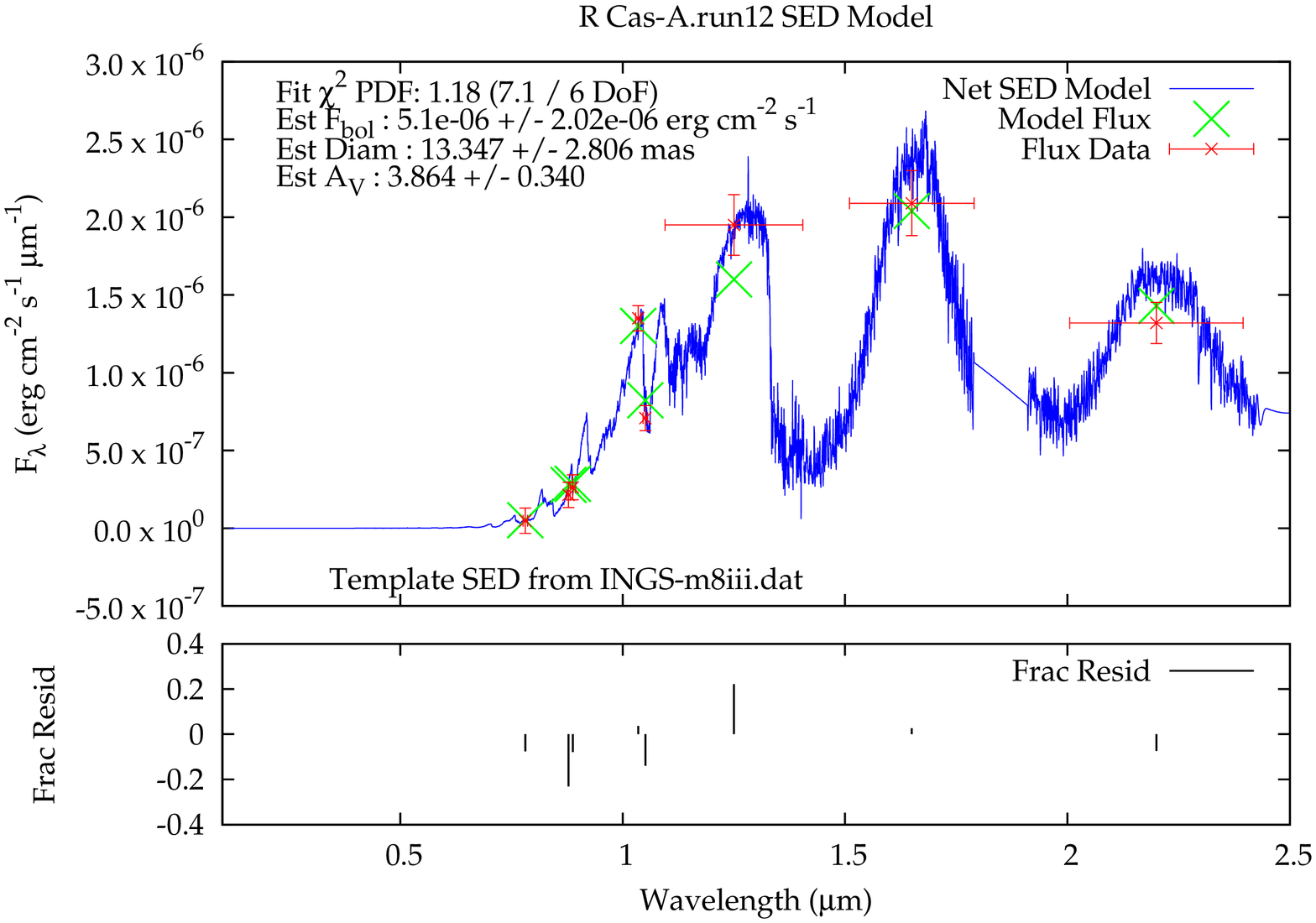}}
\subfigure[R-Cas-B (M10III)]{\includegraphics[width = 2.35in,angle=270]{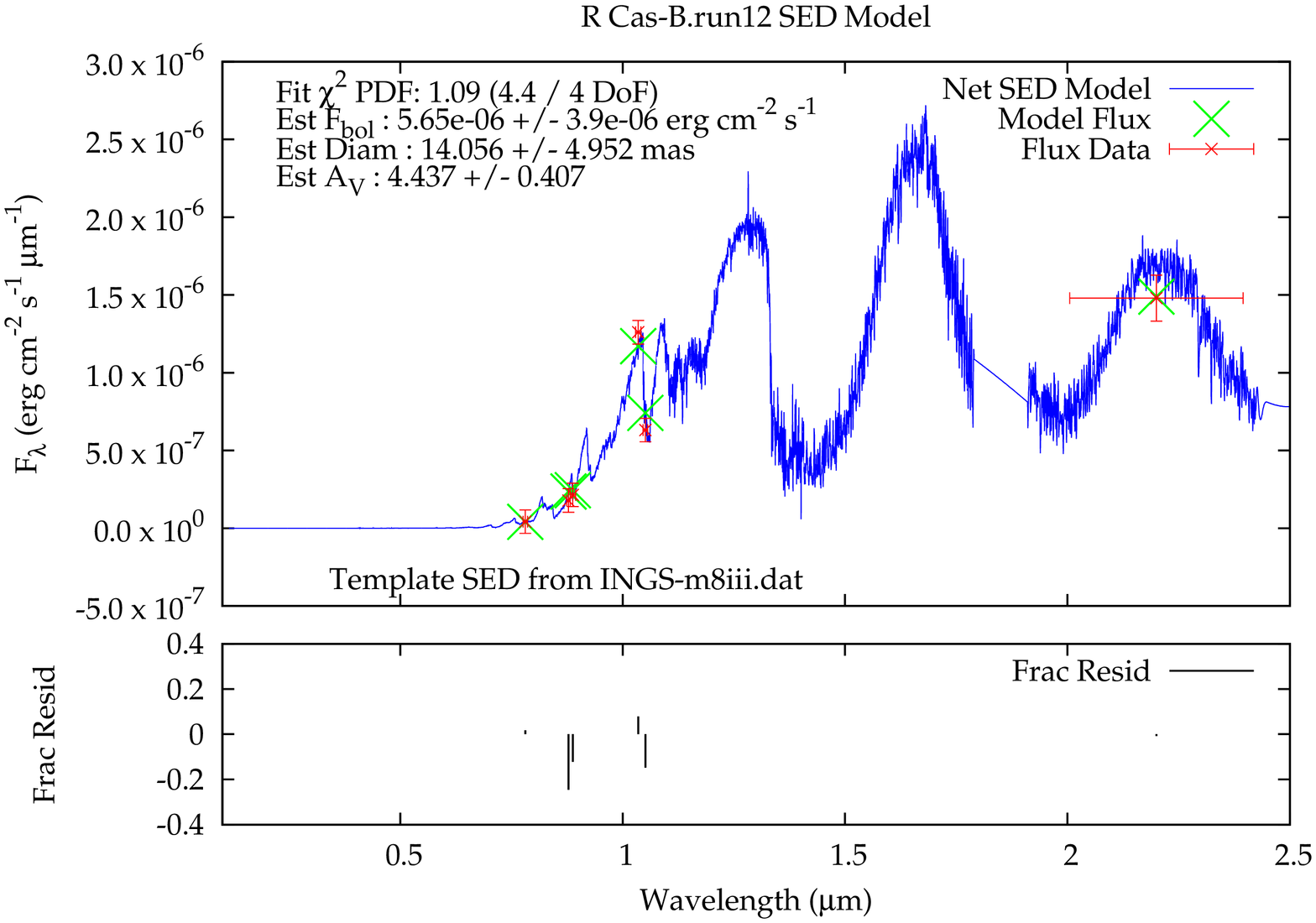}} \\
\subfigure[R-Cas-C (M9.5III)]{\includegraphics[width = 2.35in,angle=270]{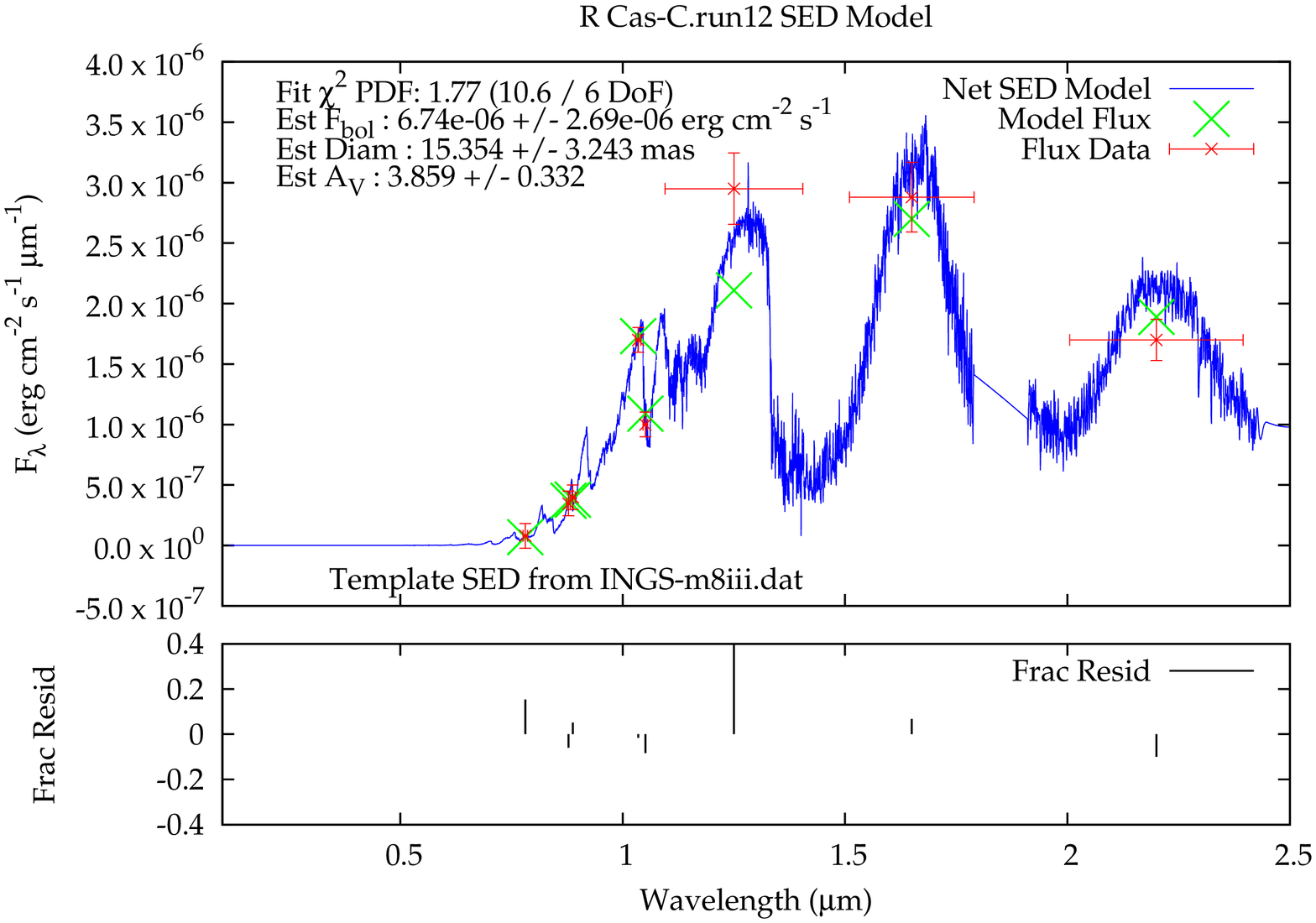}}
\subfigure[R-Cyg-A (M10III)]{\includegraphics[width = 2.35in,angle=270]{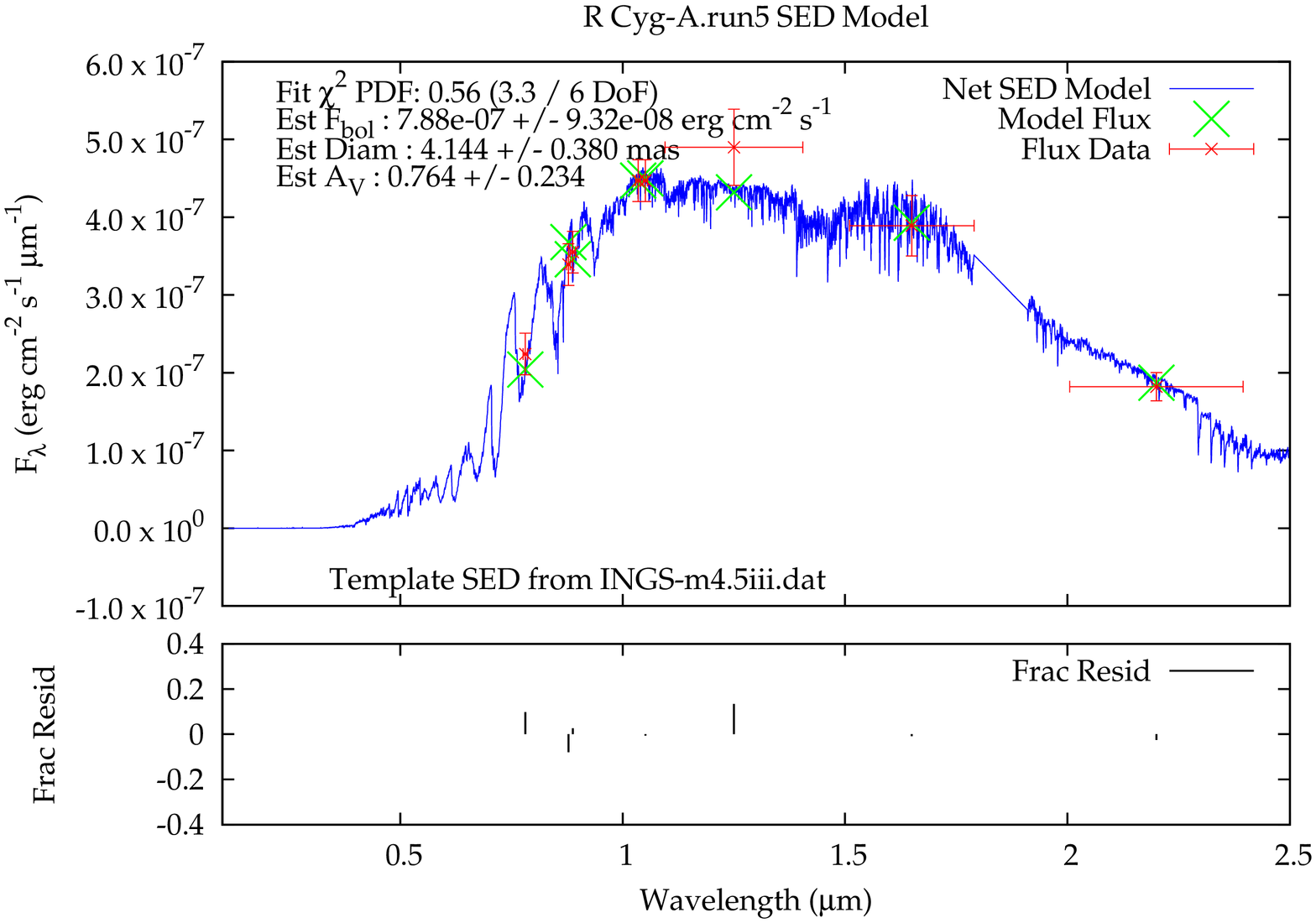}} \\
\caption{SED fits as described in \S 2.2.}
\end{figure}

\begin{figure}
\subfigure[R-Tri-A (M7III)]{\includegraphics[width = 2.35in,angle=270]{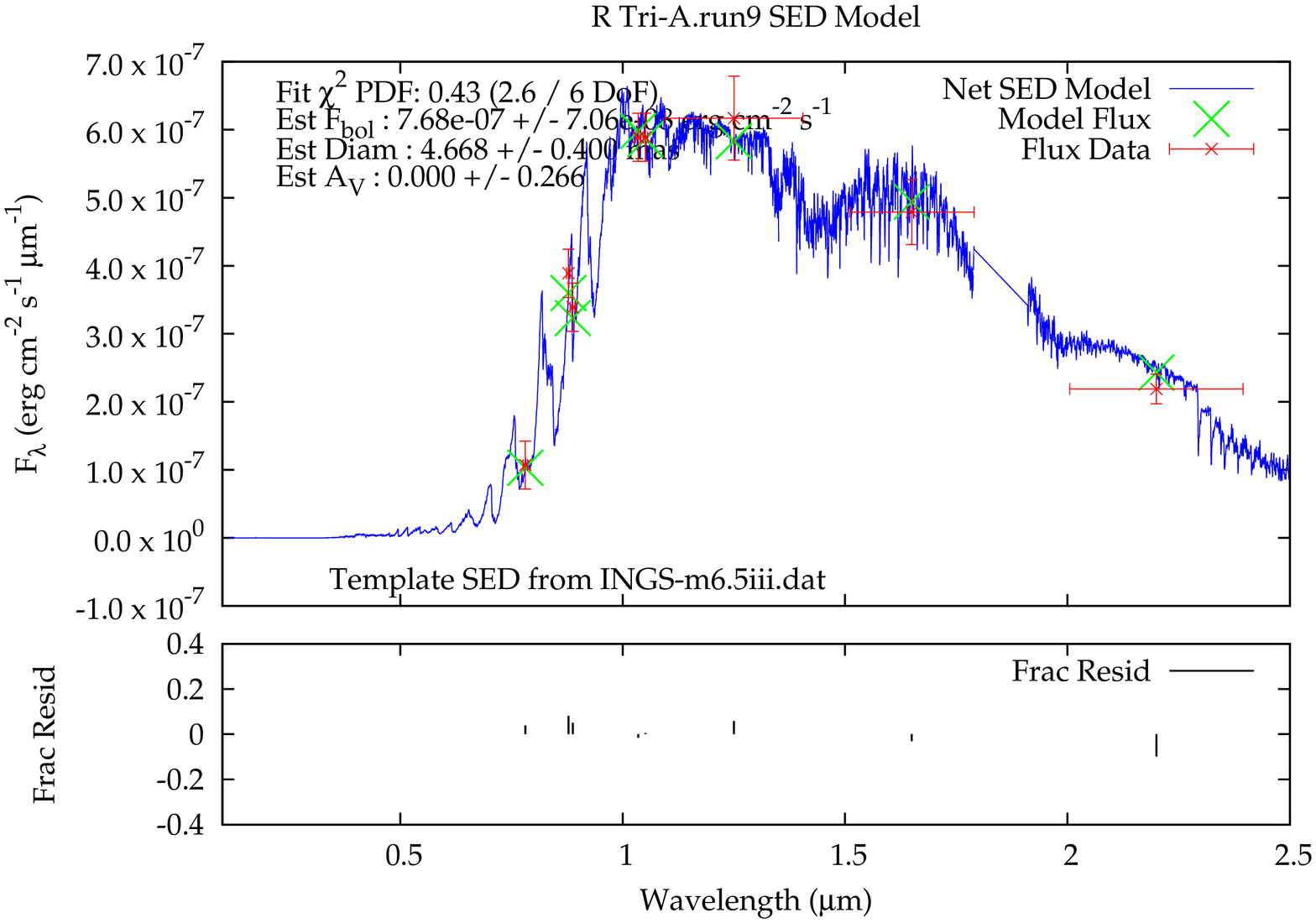}}
\subfigure[R-Tri-B (M5.5III)]{\includegraphics[width = 2.35in,angle=270]{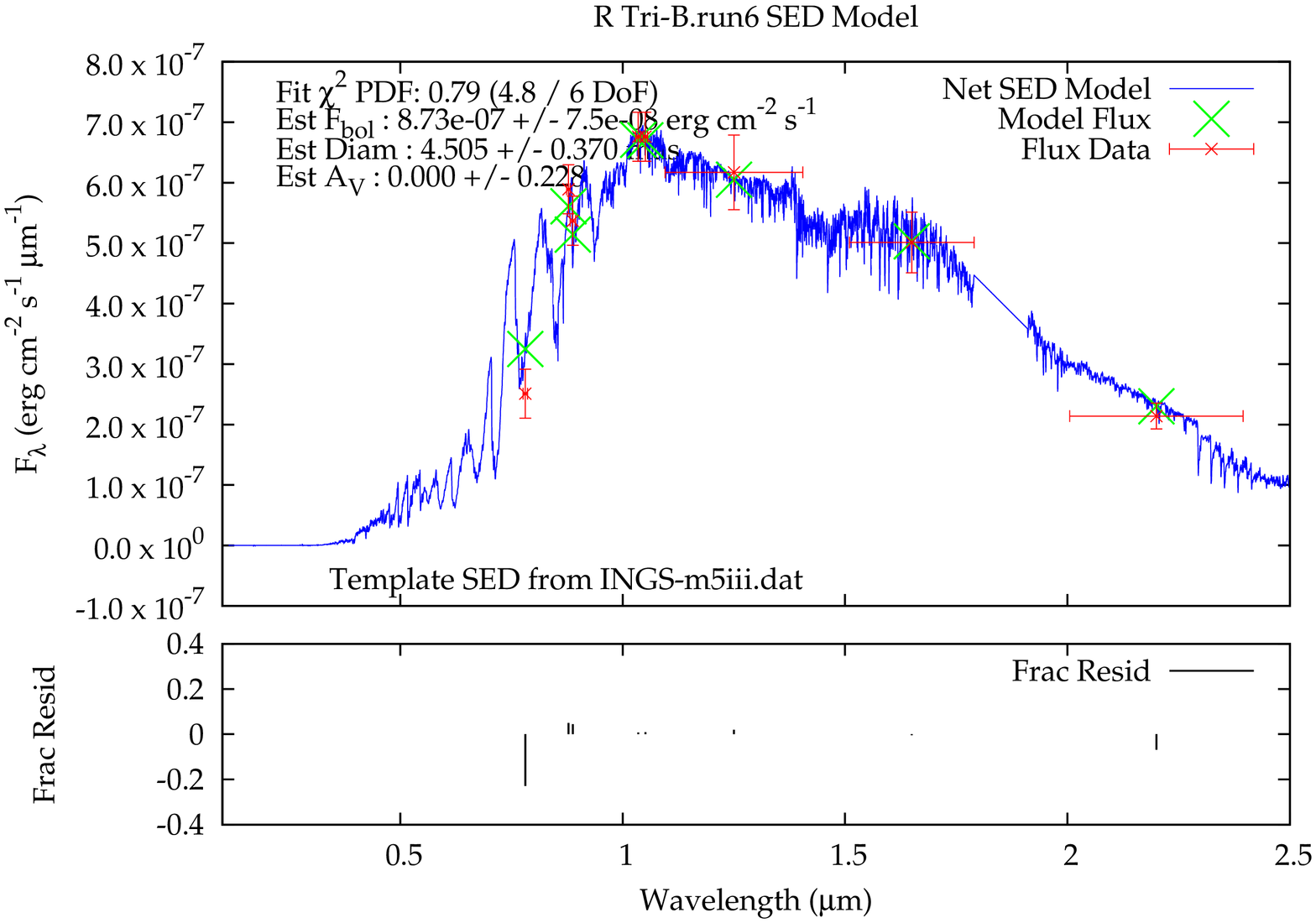}} \\
\subfigure[R-Tri-C (M5III)]{\includegraphics[width = 2.35in,angle=270]{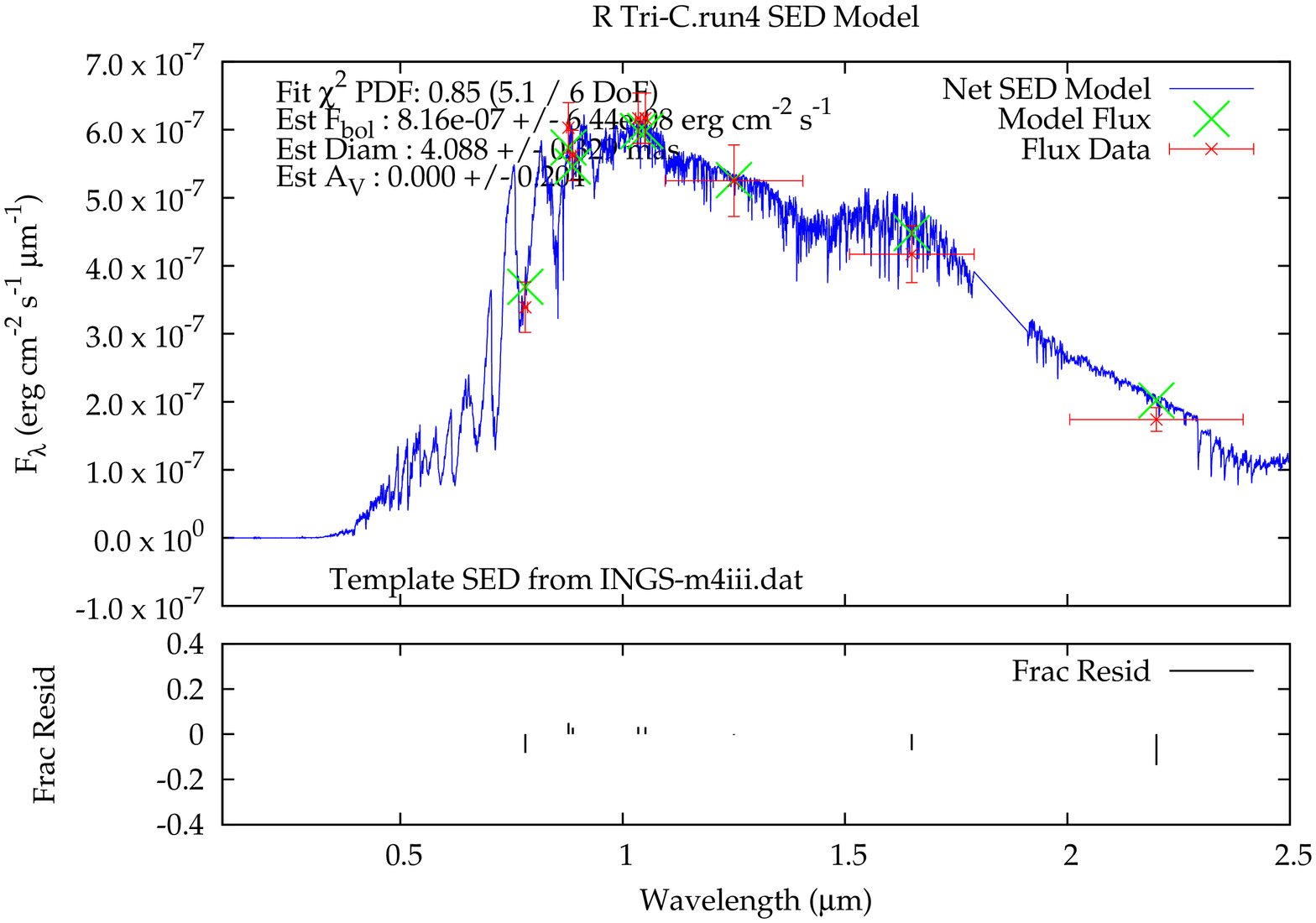}}
\subfigure[S-Lac-A (M8III)]{\includegraphics[width = 2.35in,angle=270]{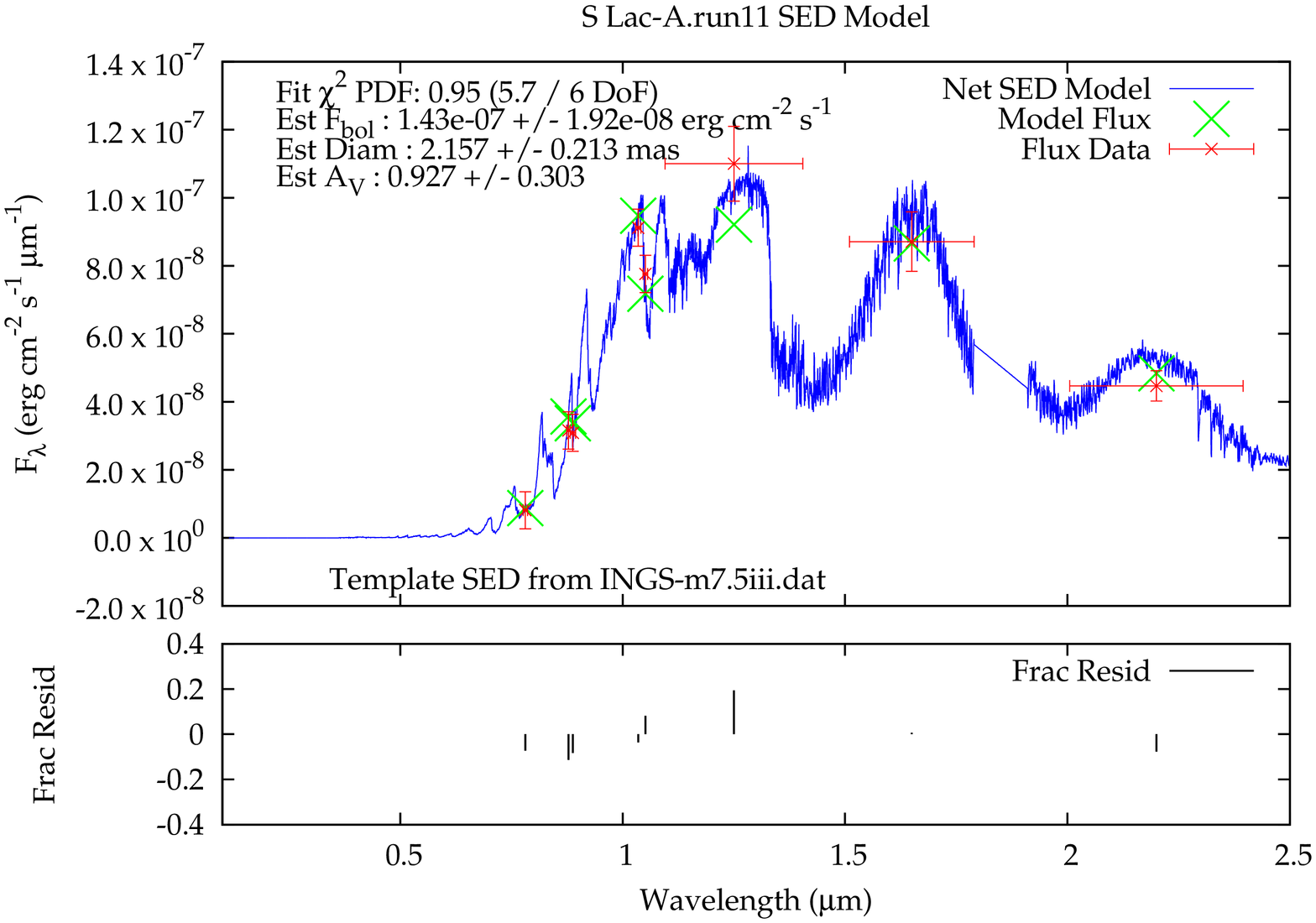}} \\
\subfigure[SS-Cas-A (M4III)]{\includegraphics[width = 2.35in,angle=270]{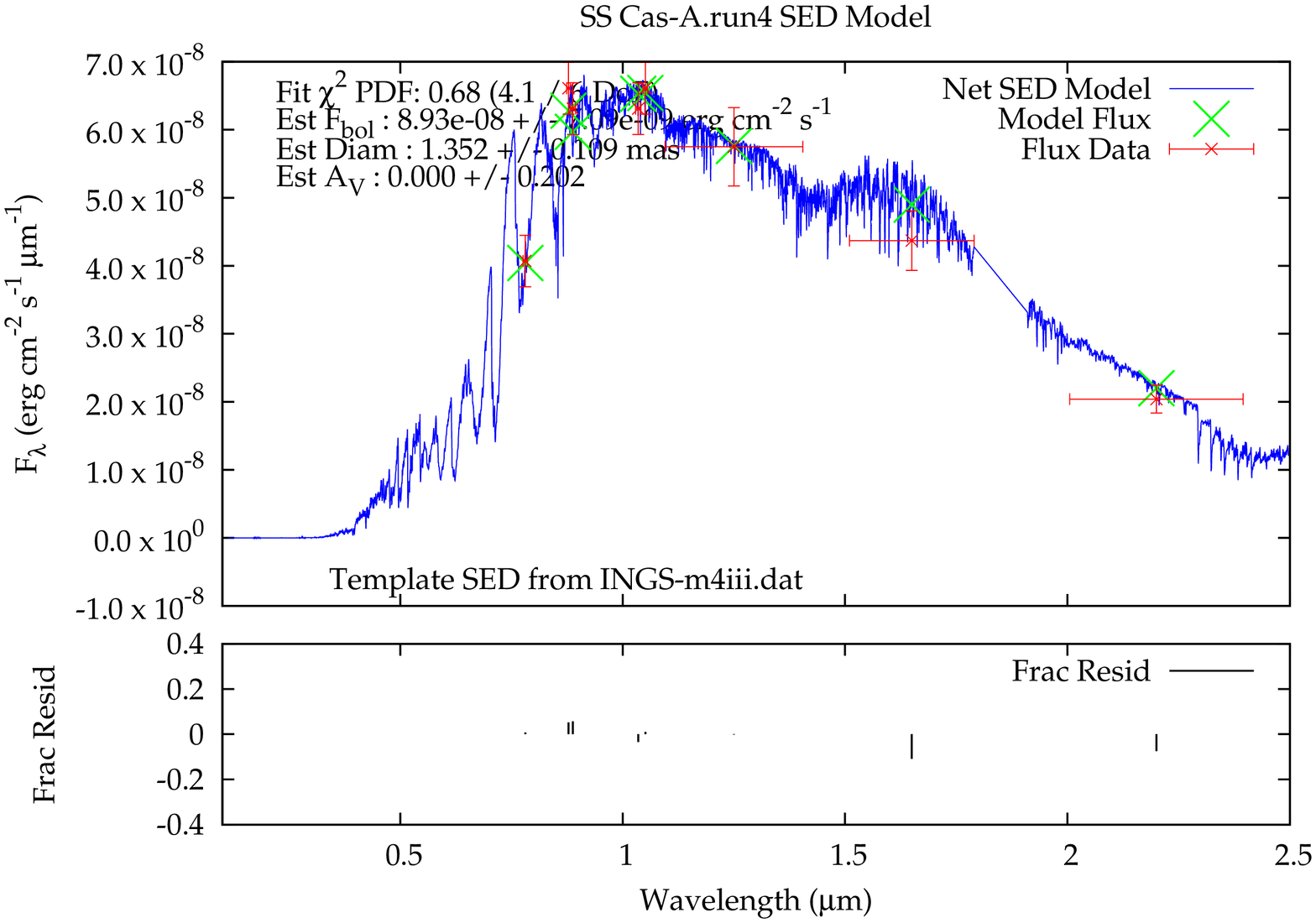}}
\subfigure[T-Aqr-A (M8III)]{\includegraphics[width = 2.35in,angle=270]{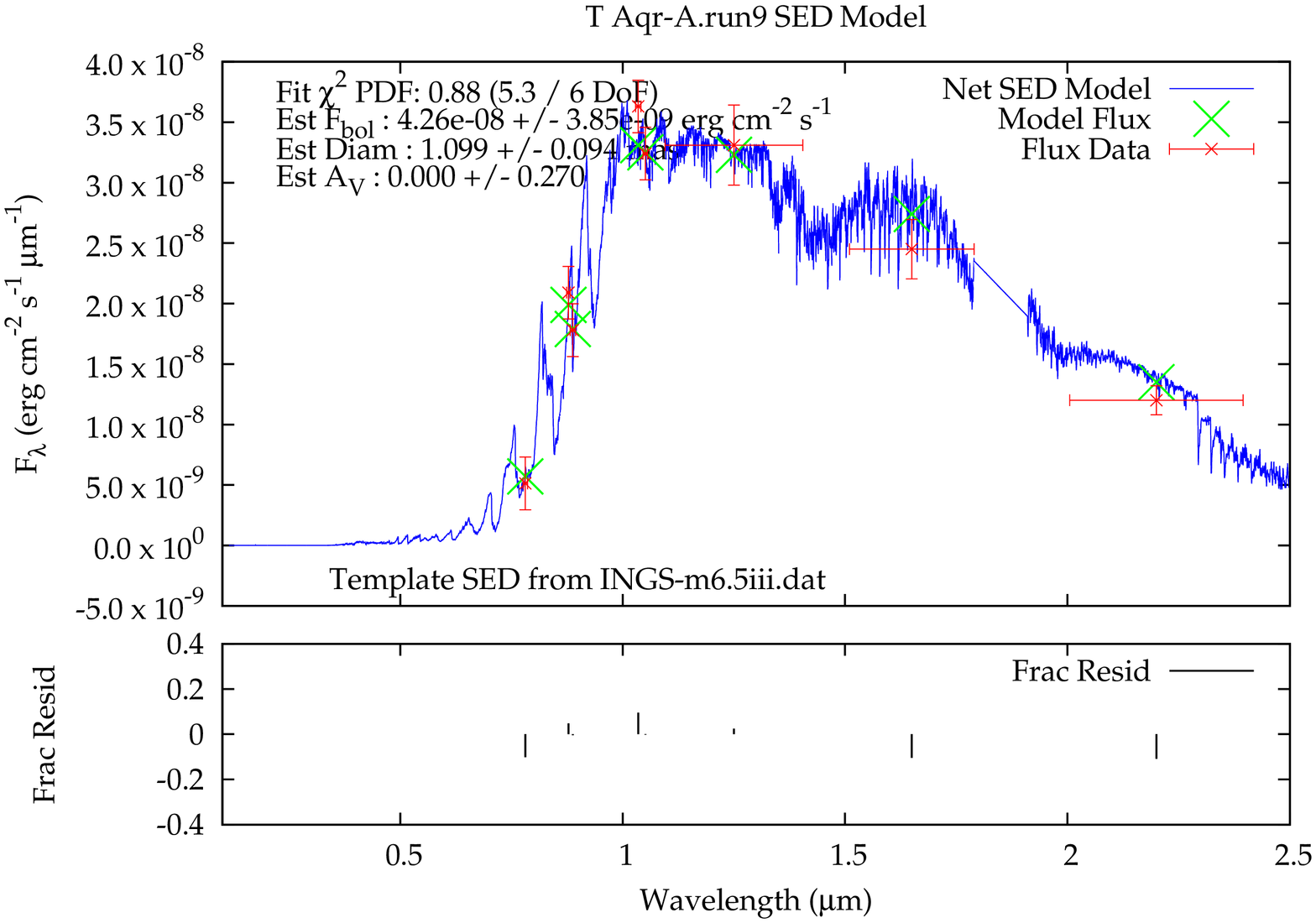}} \\
\caption{SED fits as described in \S 2.2.}
\end{figure}

\begin{figure}
\subfigure[T-Ari-A (M7III)]{\includegraphics[width = 2.35in,angle=270]{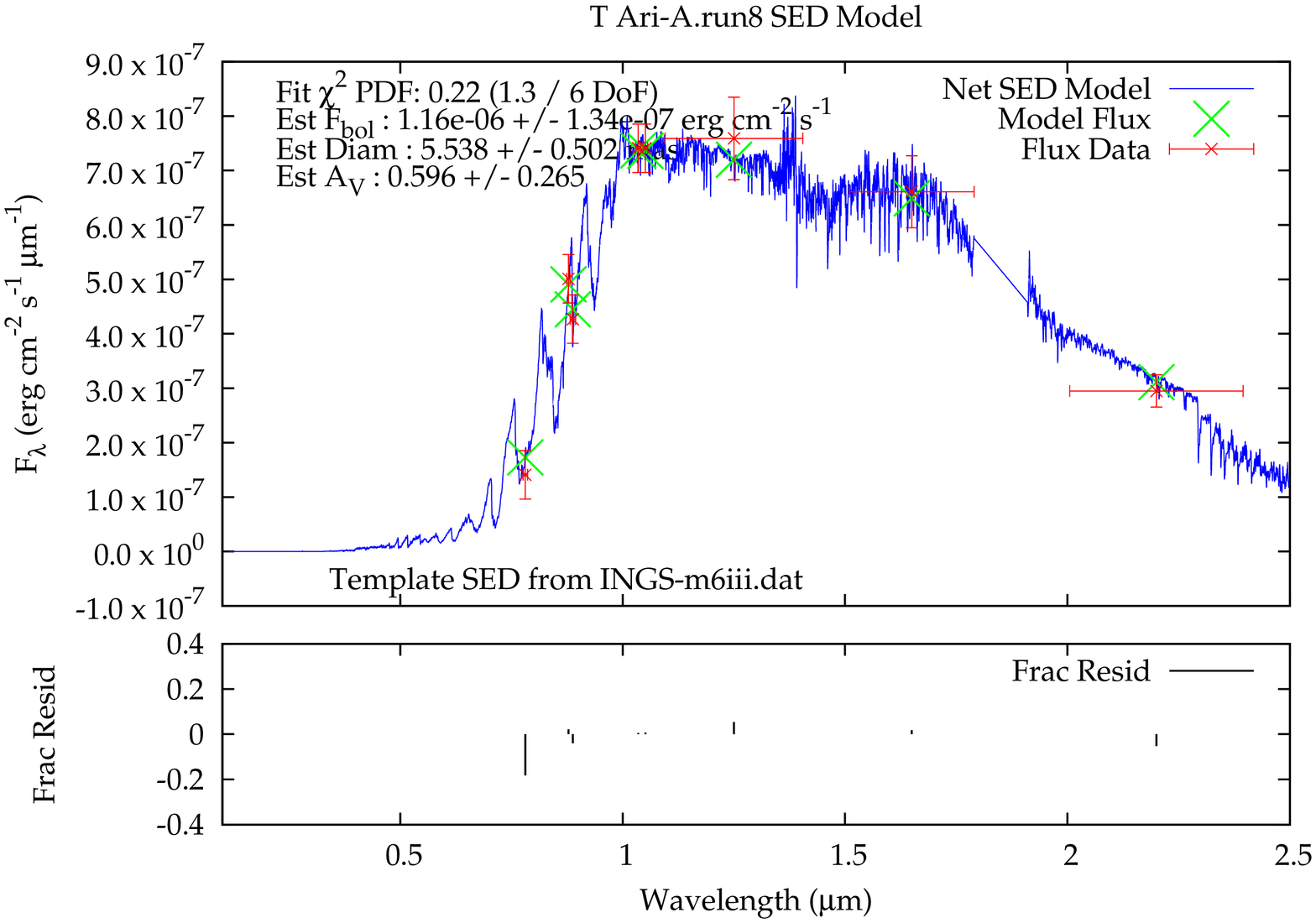}}
\subfigure[T-Cas-A (M9III)]{\includegraphics[width = 2.35in,angle=270]{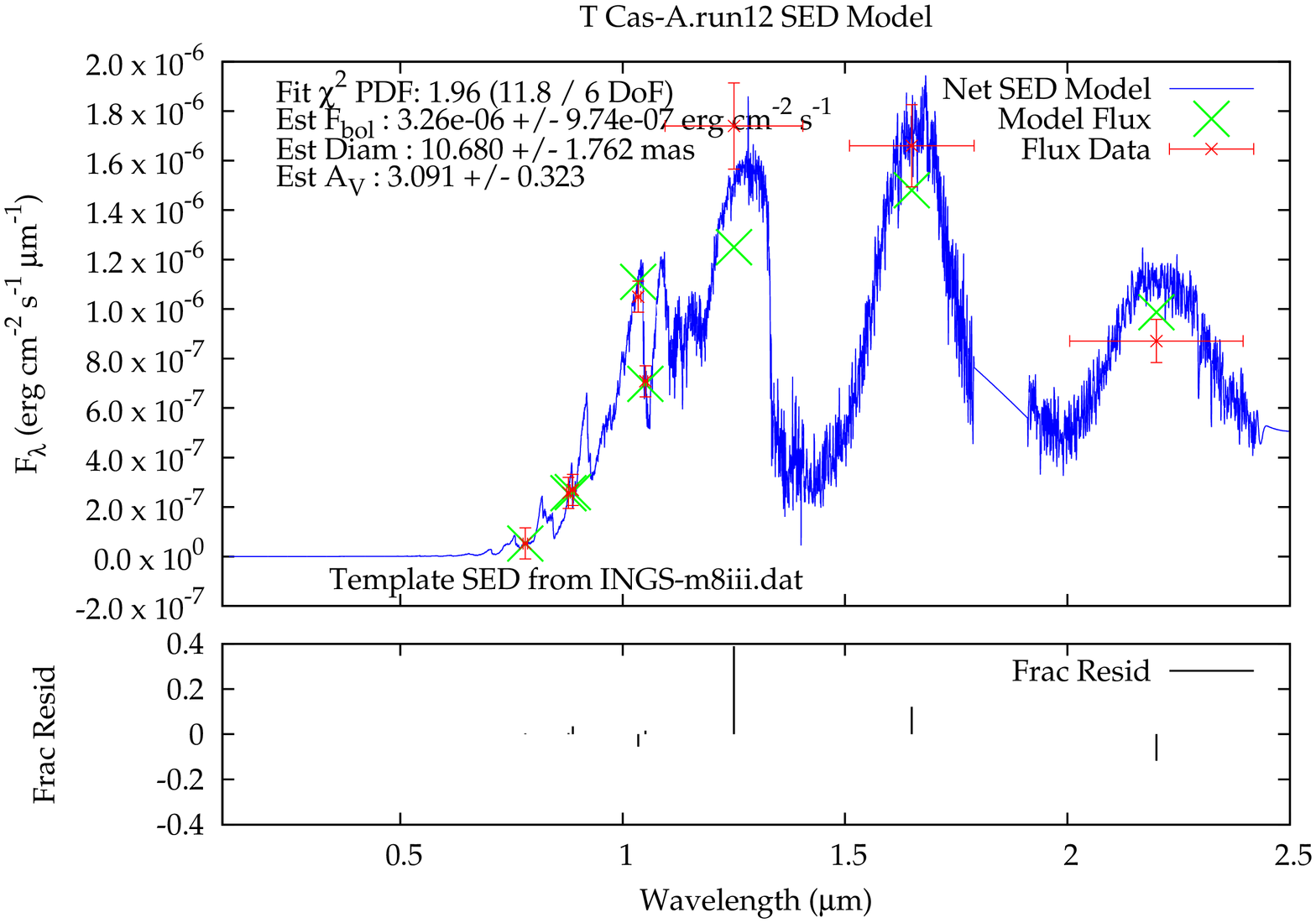}} \\
\subfigure[T-Cas-B (M8III)]{\includegraphics[width = 2.35in,angle=270]{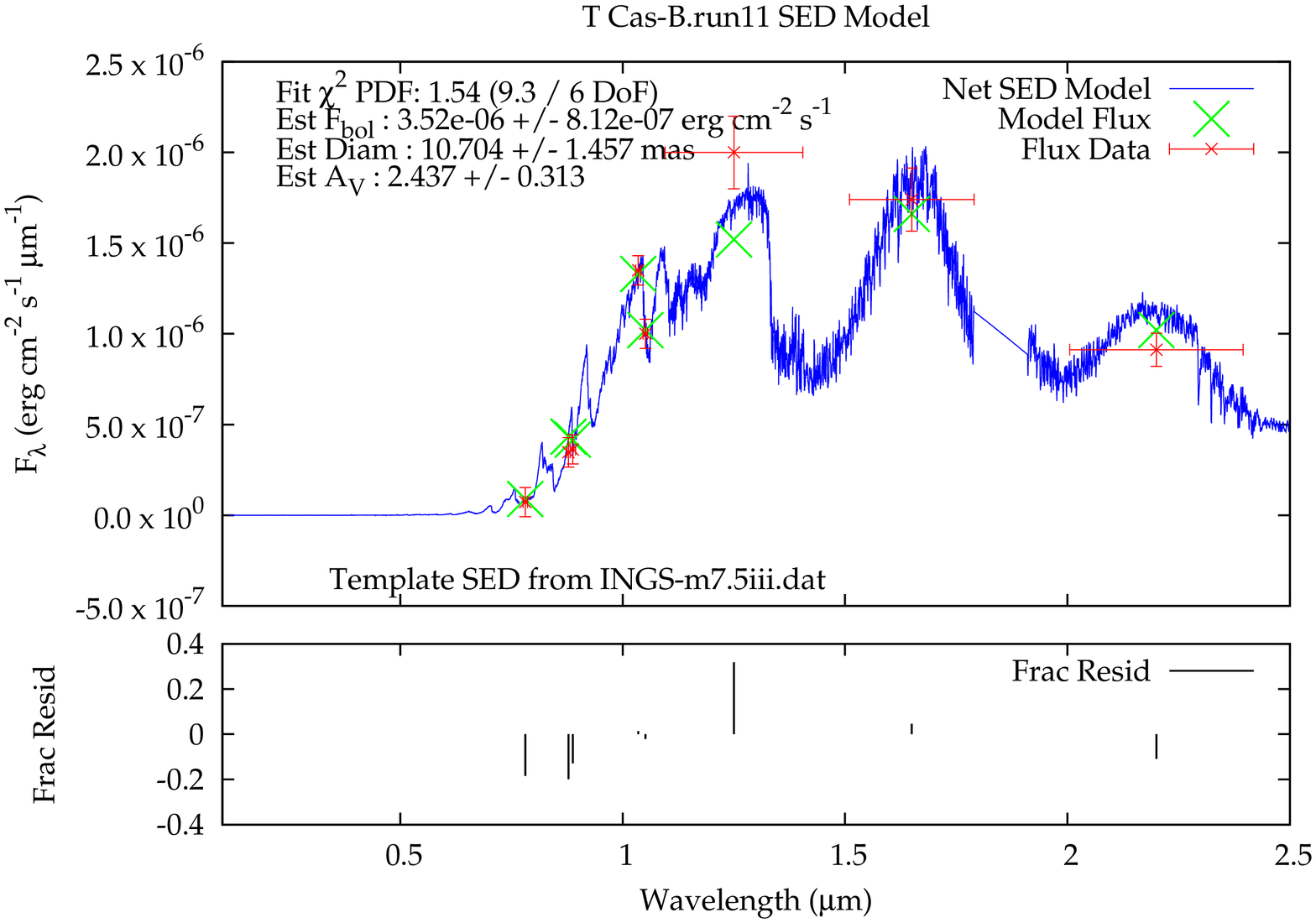}}
\subfigure[T-Cep-A (M8III)]{\includegraphics[width = 2.35in,angle=270]{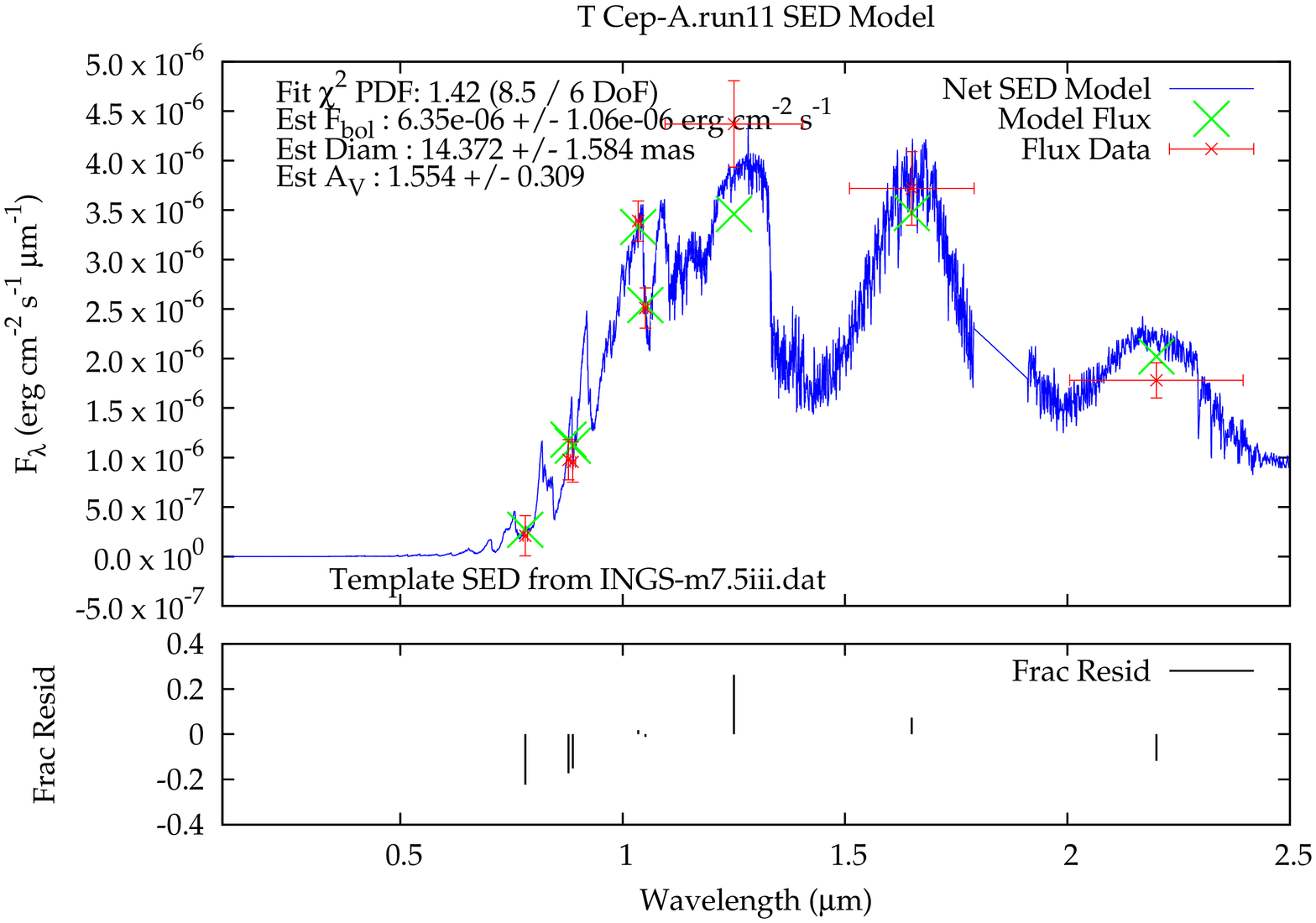}} \\
\subfigure[U-Ari-A (M9.5III)]{\includegraphics[width = 2.35in,angle=270]{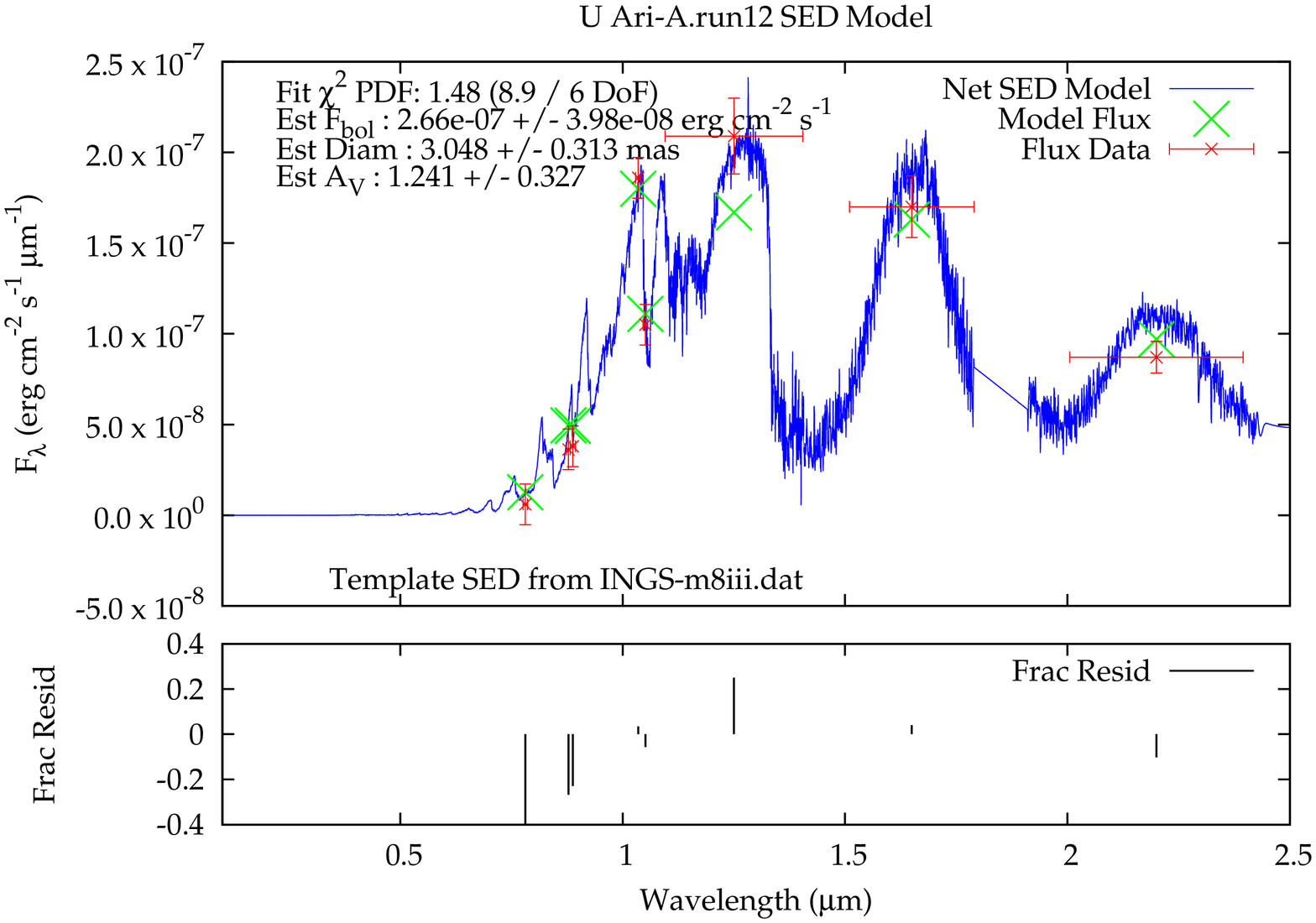}}
\subfigure[U-Ari-B (M9.5III)]{\includegraphics[width = 2.35in,angle=270]{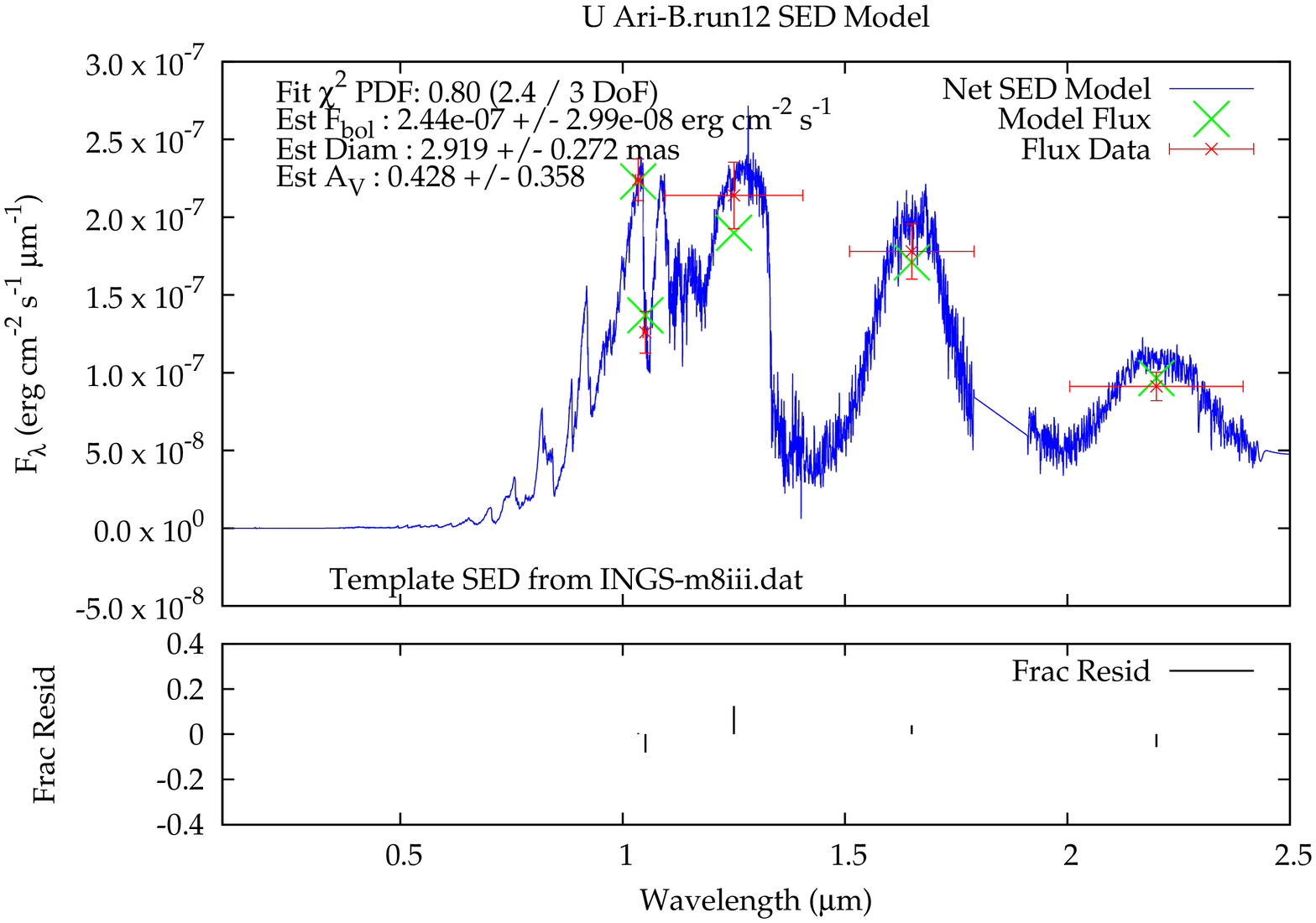}} \\
\caption{SED fits as described in \S 2.2.}
\end{figure}

\begin{figure}
\subfigure[V-Cas-A (M8III)]{\includegraphics[width = 2.35in,angle=270]{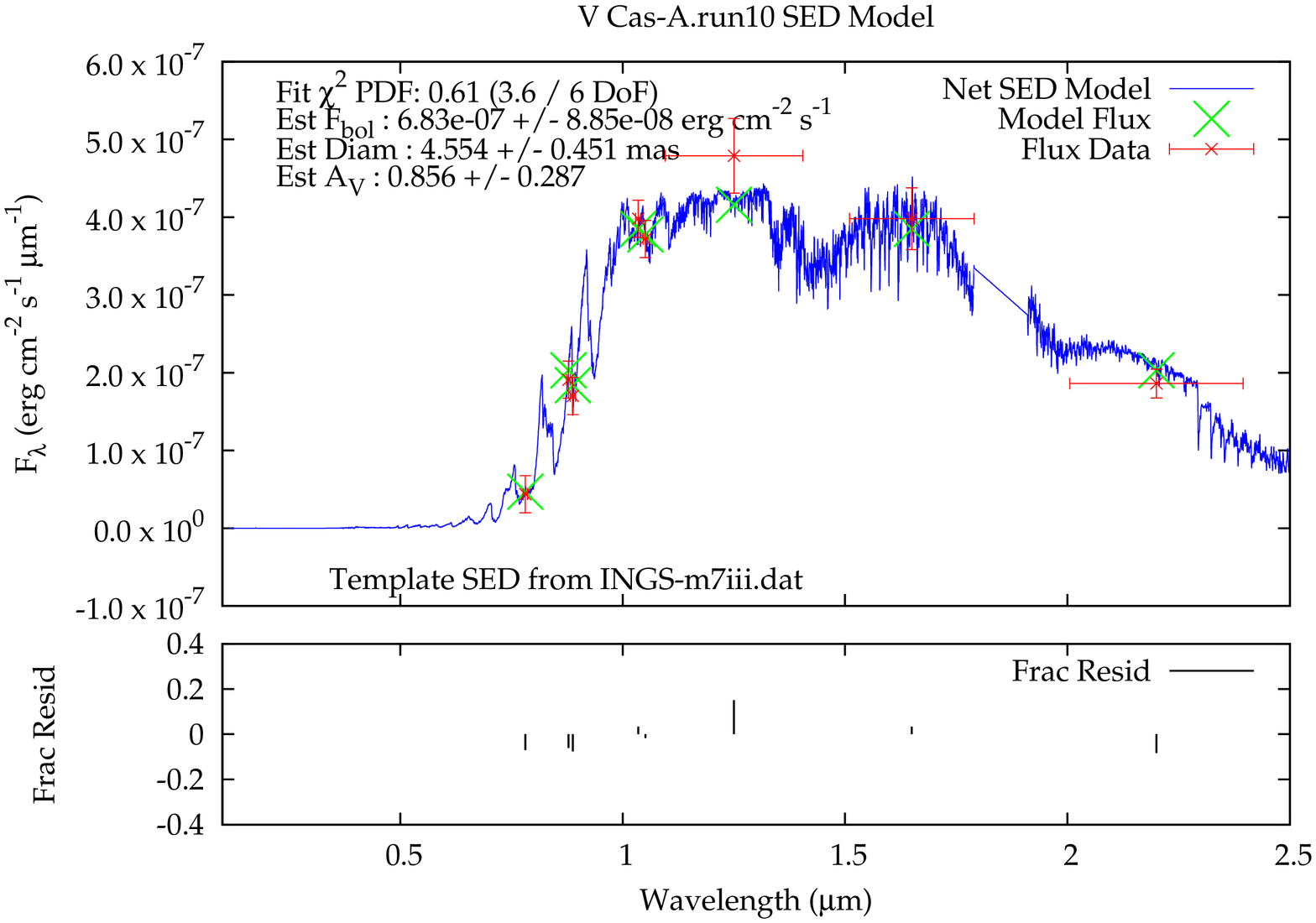}}
\subfigure[V-Mon-A (M5III)]{\includegraphics[width = 2.35in,angle=270]{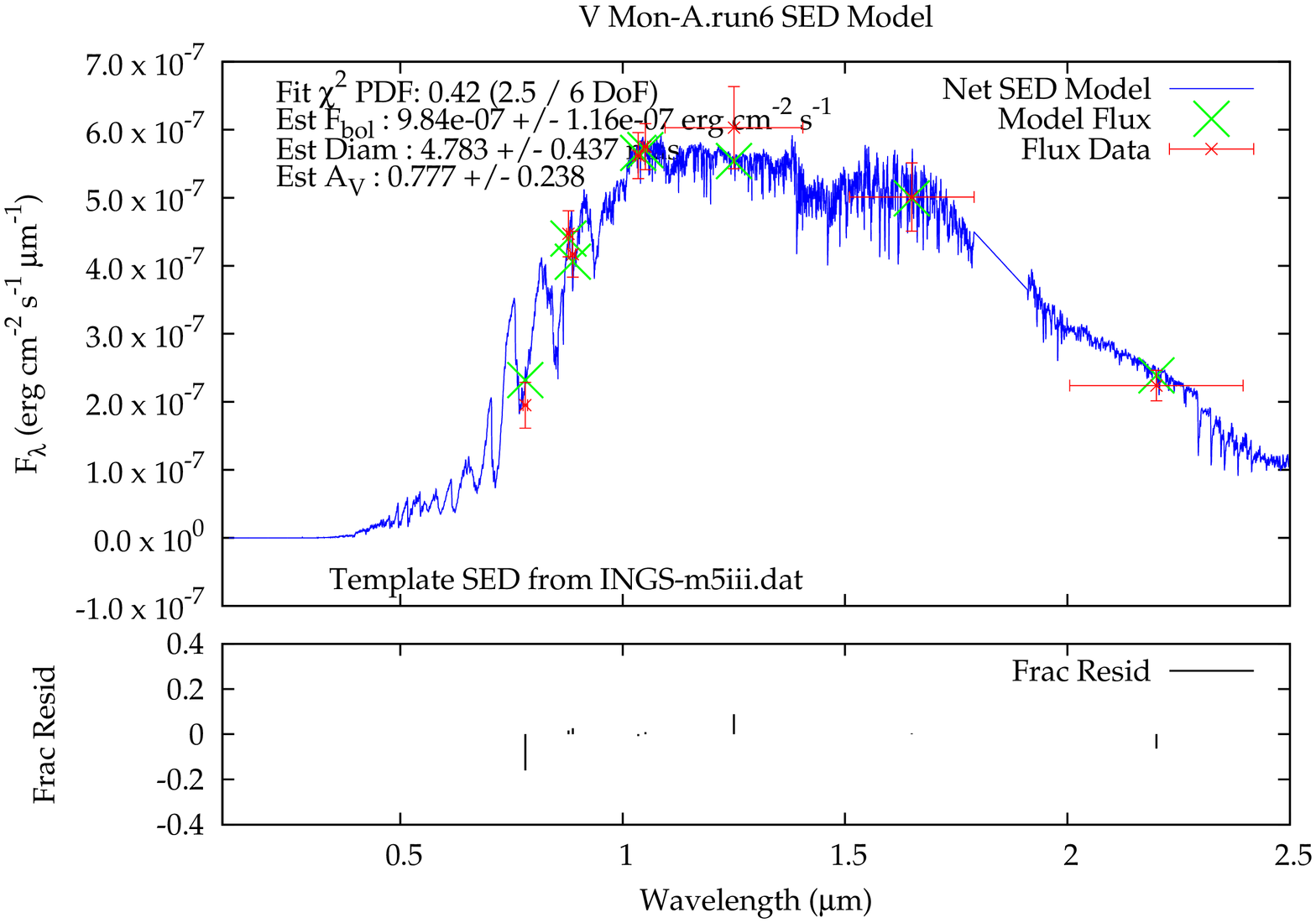}} \\
\subfigure[W-And-A (M9III)]{\includegraphics[width = 2.35in,angle=270]{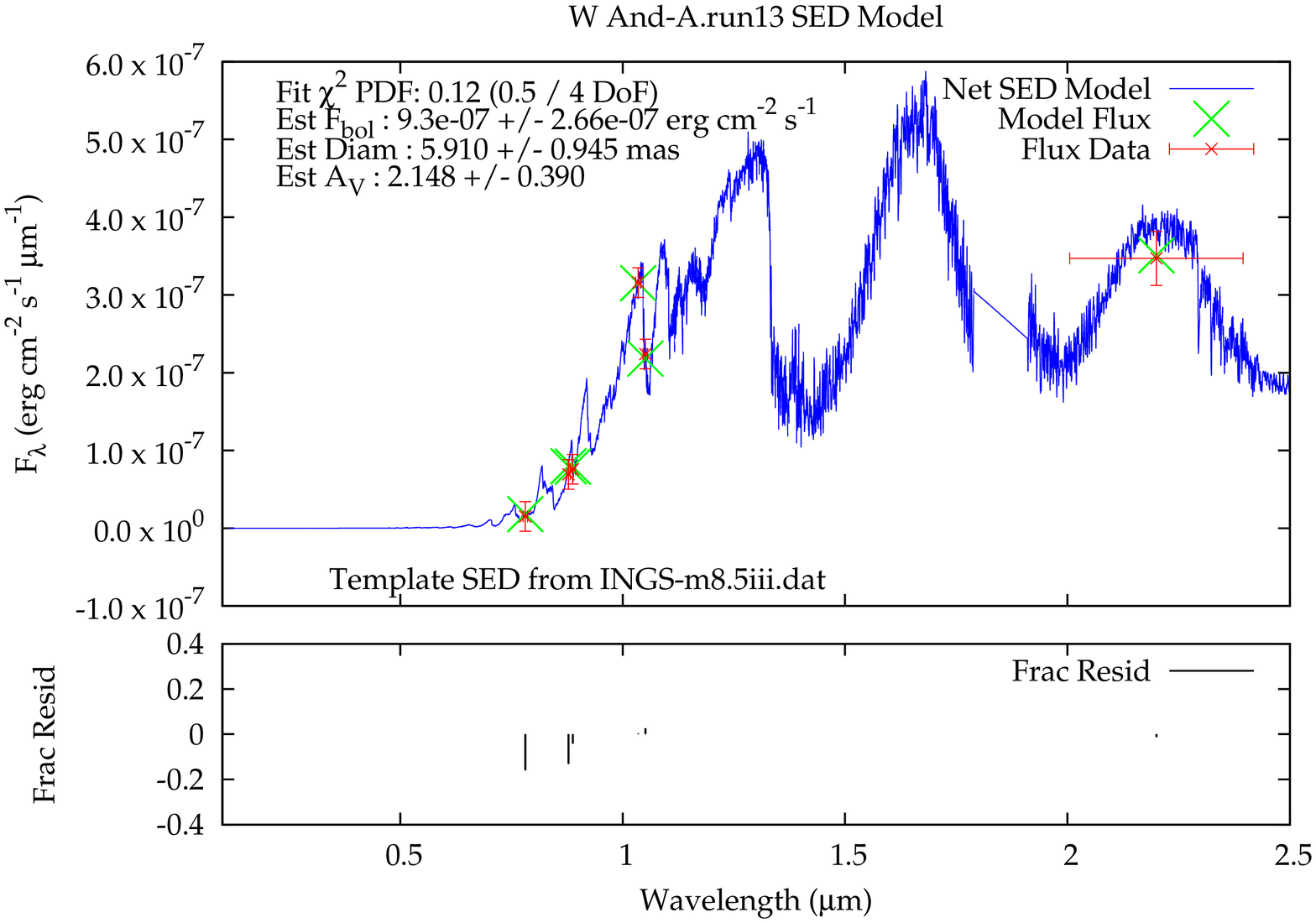}}
\subfigure[W-And-B (M8.5III)]{\includegraphics[width = 2.35in,angle=270]{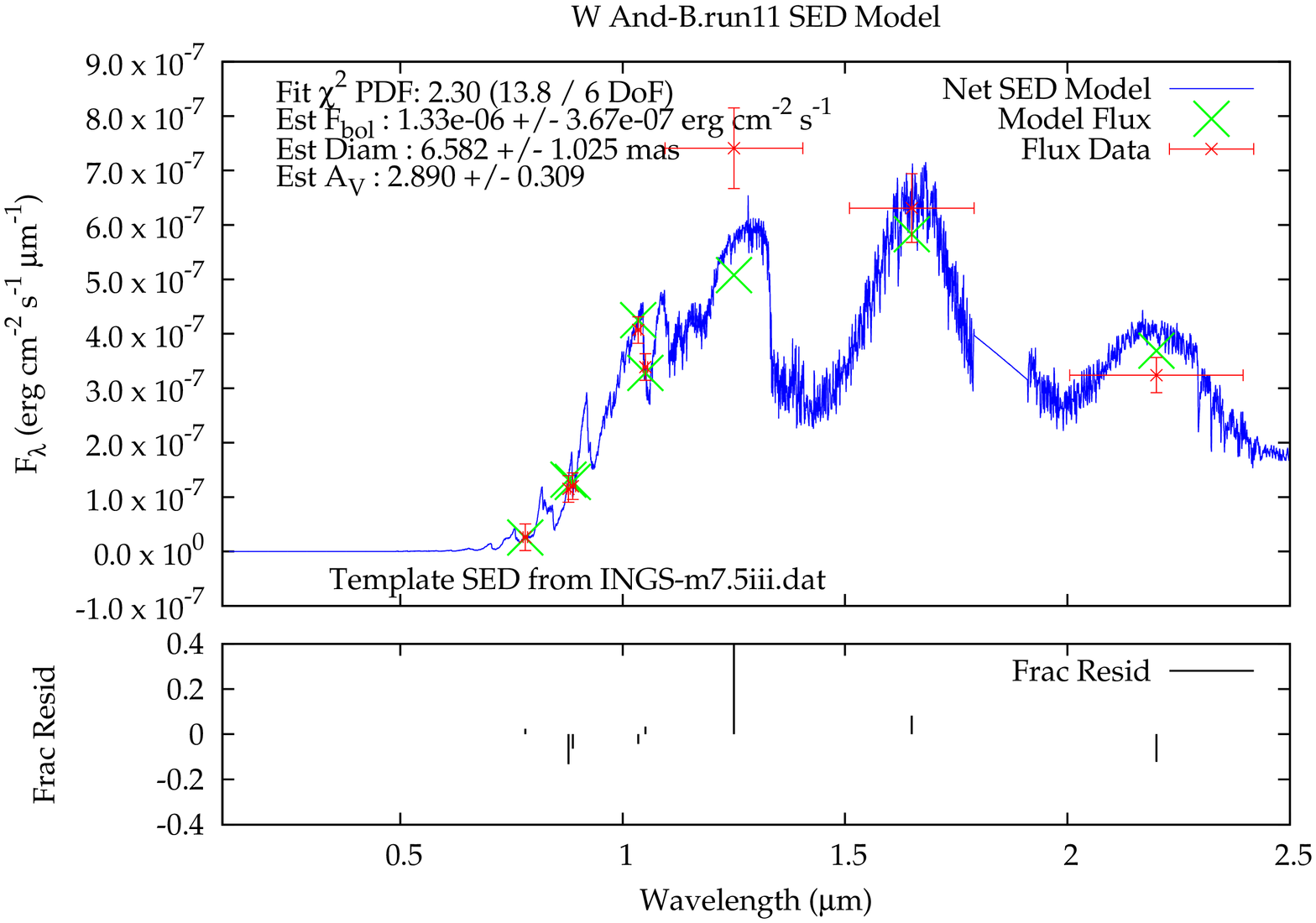}} \\
\subfigure[W-And-C (M8III)]{\includegraphics[width = 2.35in,angle=270]{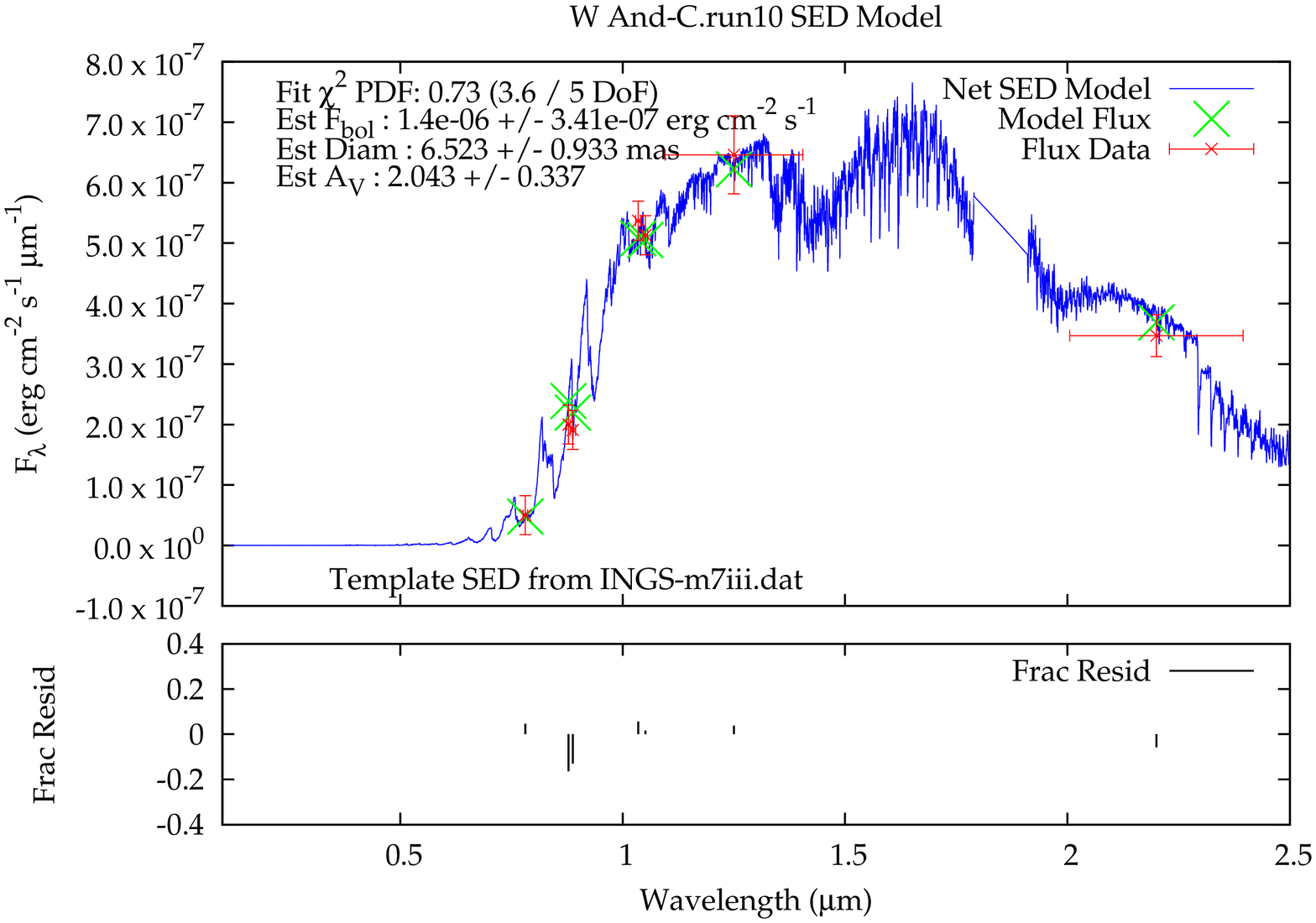}}
\subfigure[W-And-D (M6.5III)]{\includegraphics[width = 2.35in,angle=270]{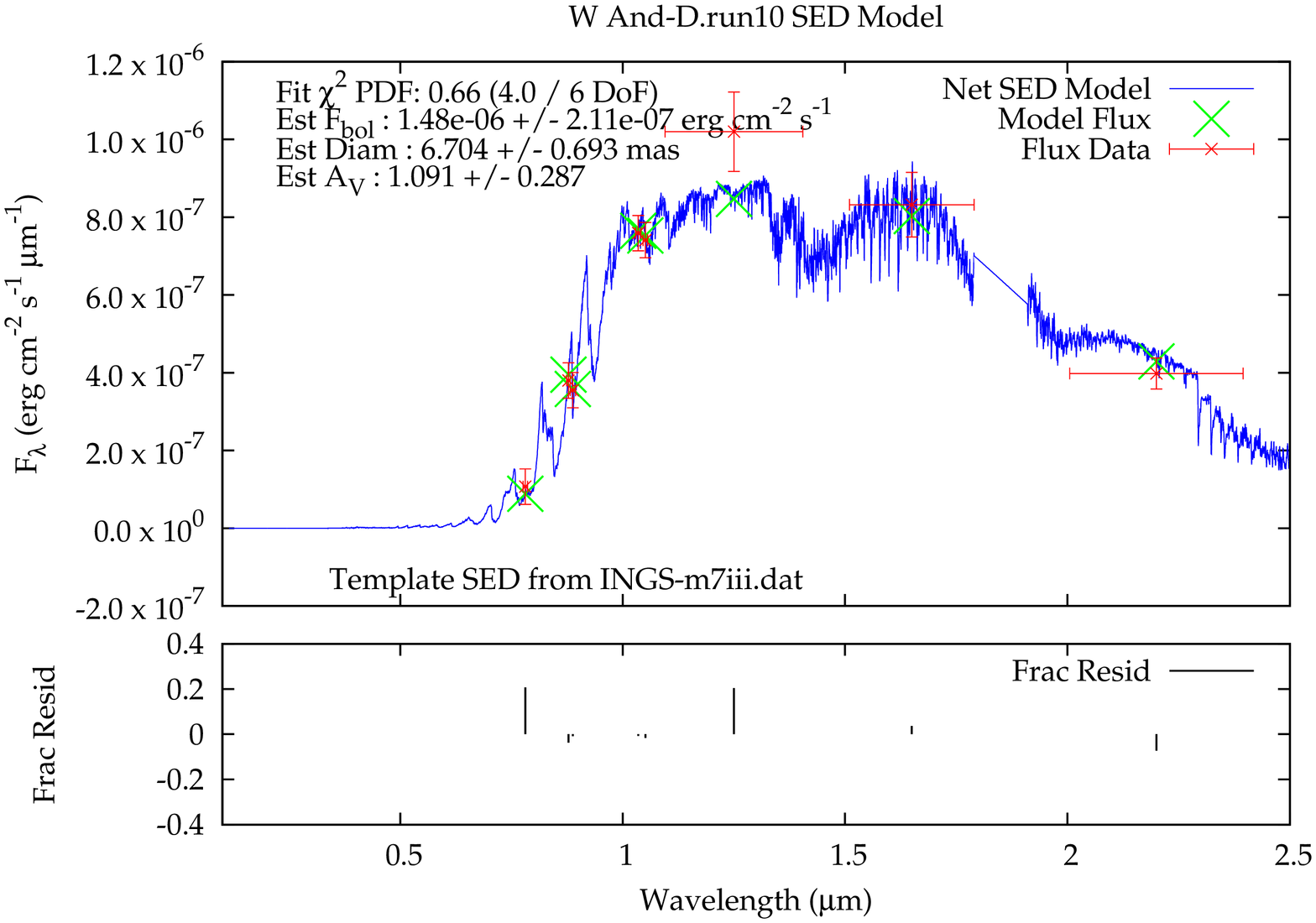}} \\
\caption{SED fits as described in \S 2.2.}
\end{figure}

\begin{figure}
\subfigure[W-Lyr-A (M6.5III)]{\includegraphics[width = 2.35in,angle=270]{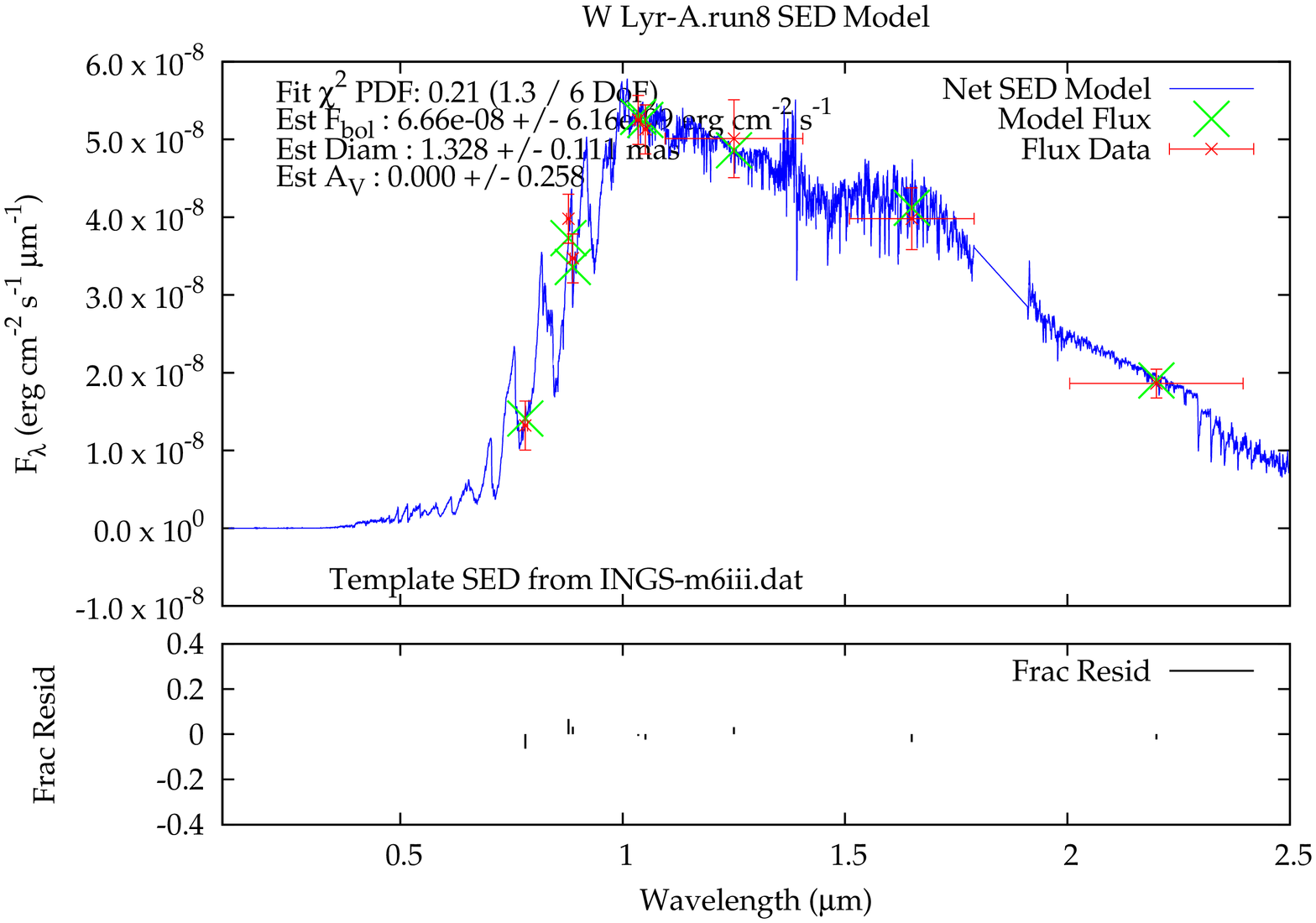}}
\subfigure[W-Peg-A (M9III)]{\includegraphics[width = 2.35in,angle=270]{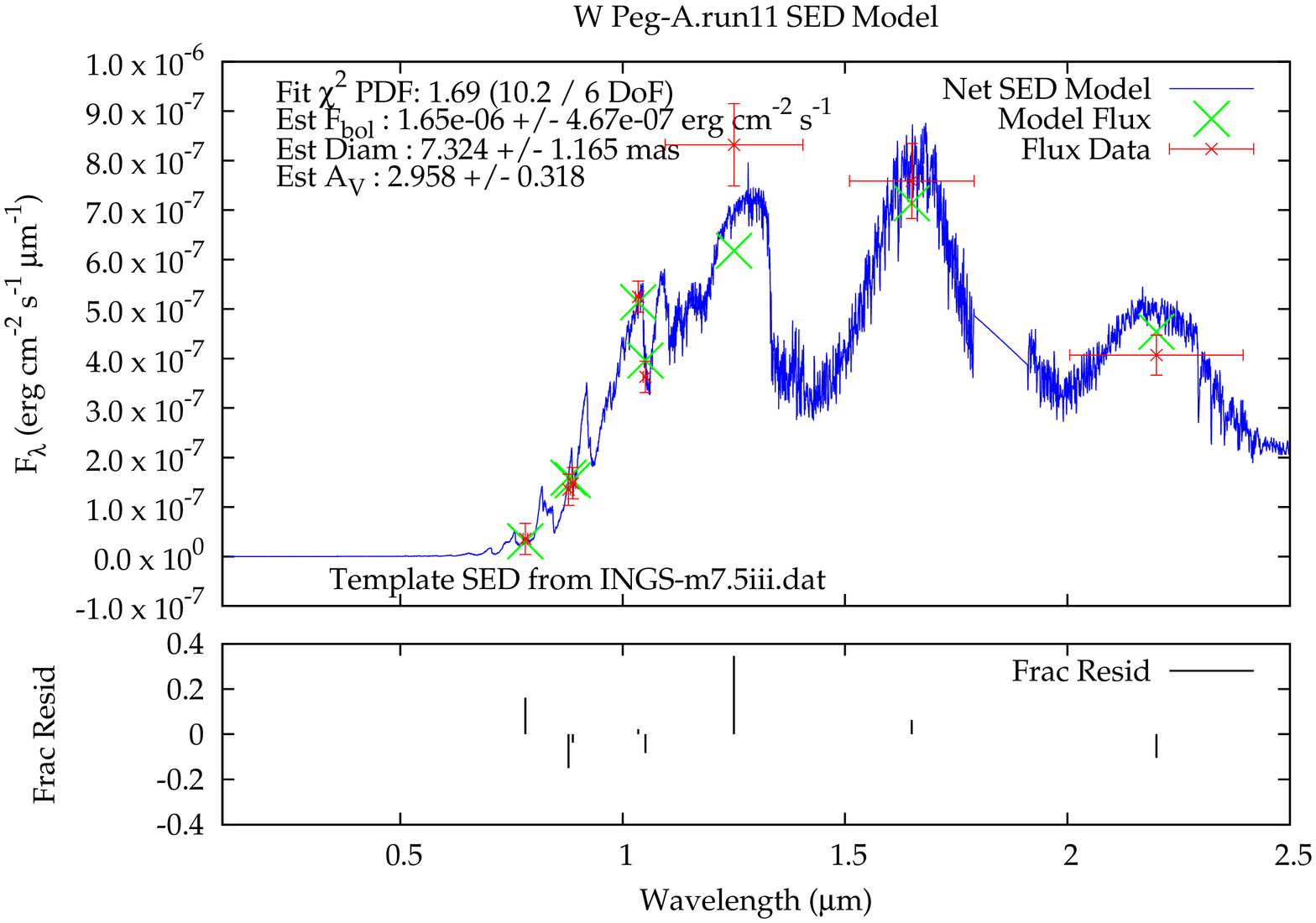}} \\
\subfigure[X-Cyg-A (M8III)]{\includegraphics[width = 2.35in,angle=270]{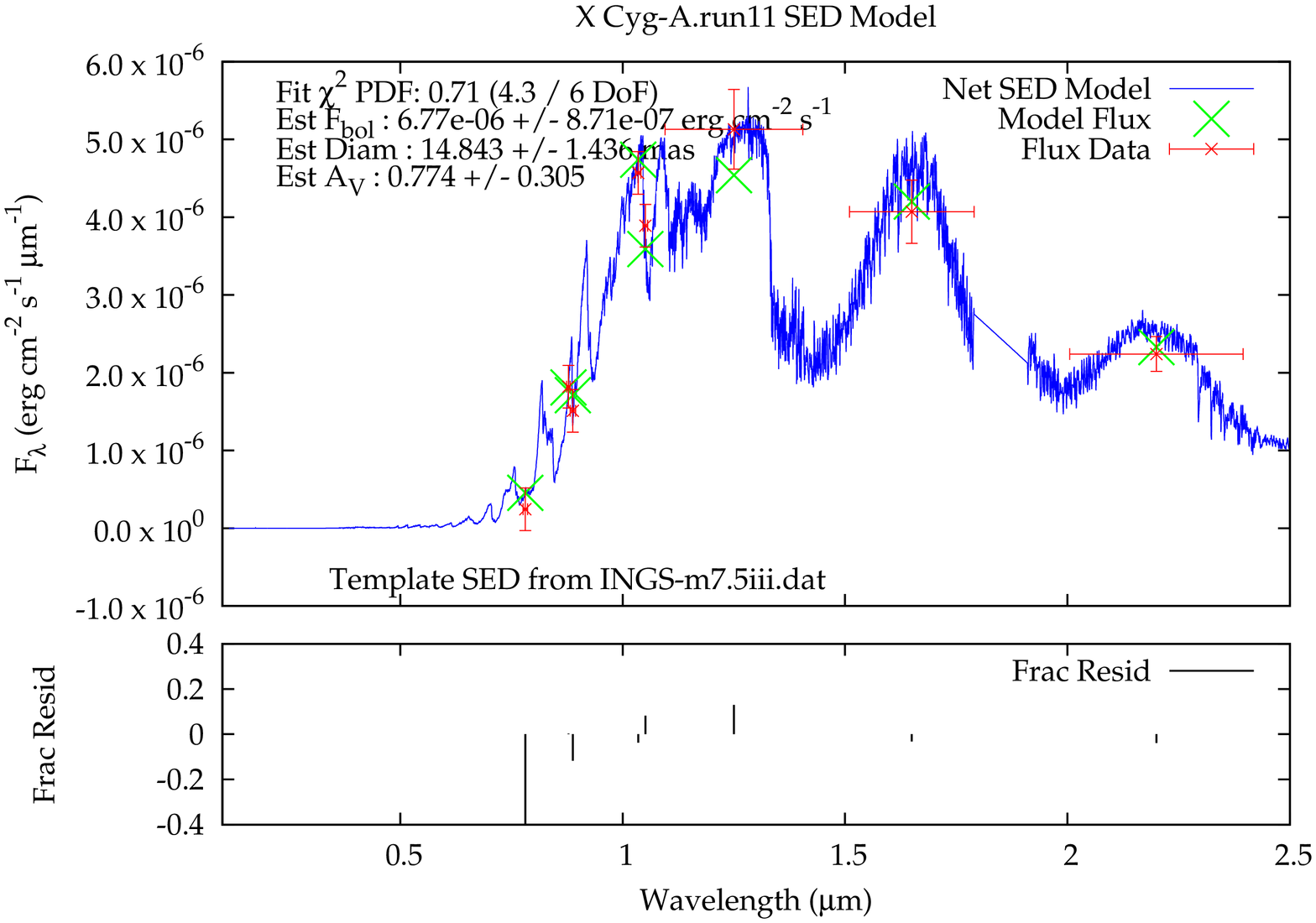}}
\subfigure[Y-Cas-A (M8III)]{\includegraphics[width = 2.35in,angle=270]{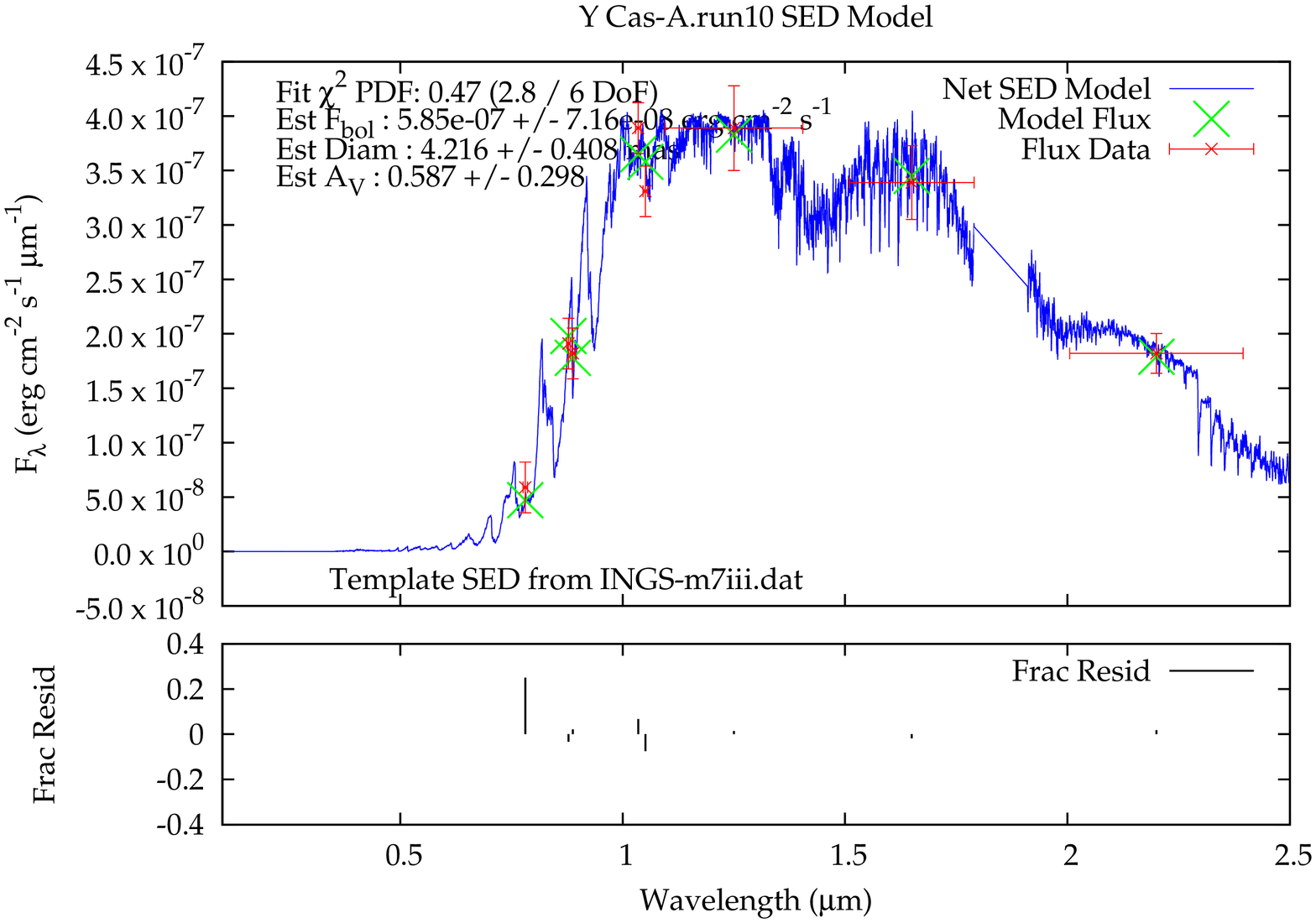}} \\
\subfigure[Z-Peg-A (M8.5III)]{\includegraphics[width = 2.35in,angle=270]{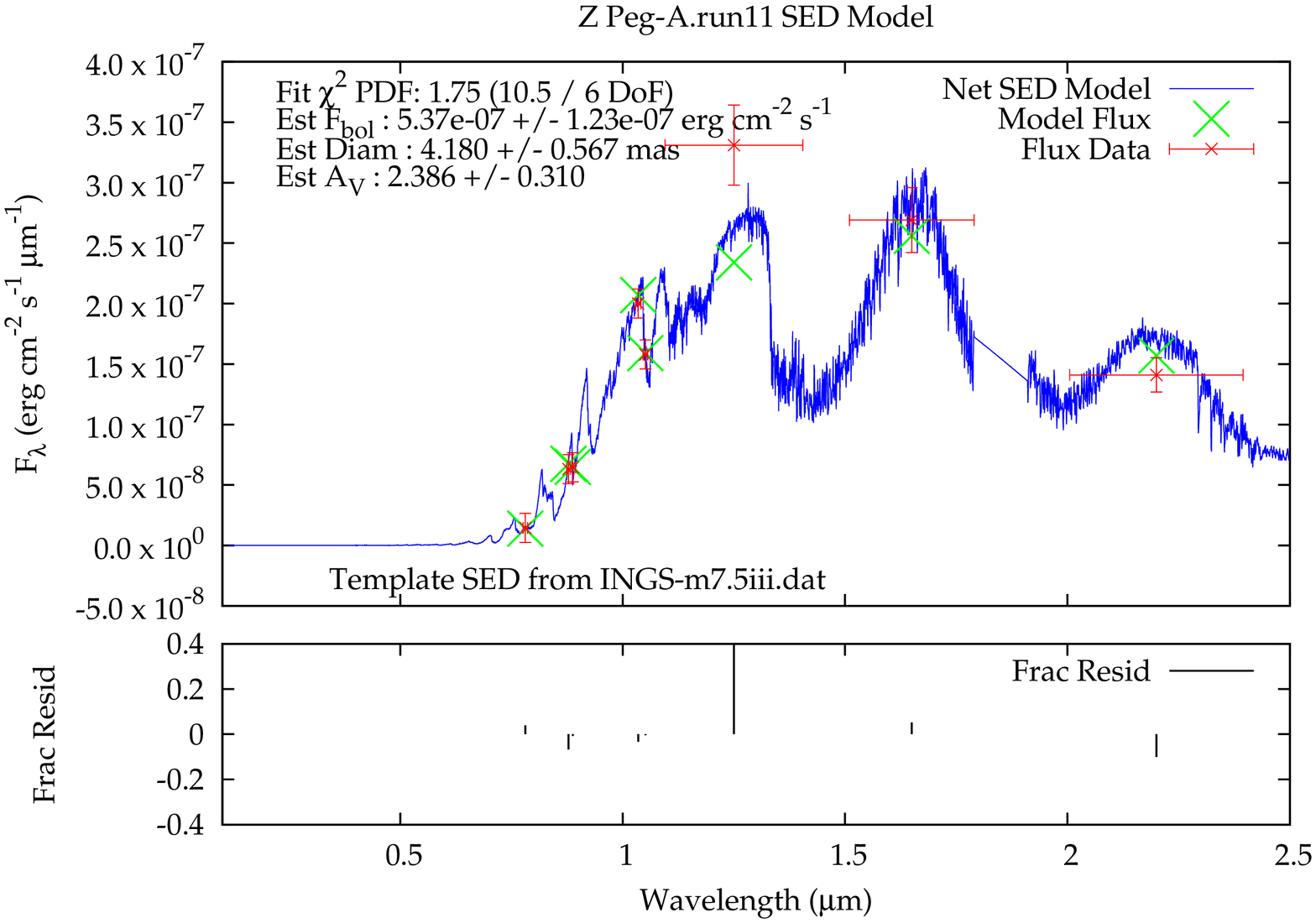}}

\caption{SED fits as described in \S 2.2.}
\end{figure}

\end{document}